\documentclass[fleqn,usenatbib,usedcolumn]{mnras}

\pdfoutput=1

\usepackage{newtxtext,newtxmath}
\renewcommand{\la}{\lesssim} % for less than similar from newtxmath, not \la from mnras.cls
\renewcommand{\ga}{\gtrsim} % for greater than similar from newtxmath, not \la from mnras.cls
\usepackage[T1]{fontenc}
\usepackage{graphicx}	% Including figure files
\usepackage{amsmath}	% Advanced maths commands
\usepackage{pdflscape}  % Landscape tables and figures
\usepackage[usenames,dvipsnames]{color}

\newcommand{\Sref}[1]{Section \ref{#1}}
\newcommand{\Tref}[1]{Table \ref{#1}}

\sfcode`\.=1001\sfcode`\?=1001\sfcode`\!=1001
\newcommand{\Fref}[1]{\ifhmode \ifnum\spacefactor=1001 Figure \ref{#1}\else Fig.\ \ref{#1}\fi \else Figure \ref{#1}\fi}
\newcommand{\Eref}[1]{\ifhmode \ifnum\spacefactor=1001 Equation (\ref{#1})\else equation (\ref{#1})\fi \else Equation (\ref{#1})\fi}

\newcommand{\kms}{\ensuremath{\textrm{km\,s}^{-1}}}

\newcommand{\pcmsq}{\ensuremath{\textrm{cm}^{-2}}}
\newcommand{\SN}{\ensuremath{\textrm{S/N}}}
\newcommand{\CN}{\ensuremath{\textrm{C/N}}}

\newcommand{\lya}{\ensuremath{\textrm{Ly-}\alpha}}
\newcommand{\lyb}{\ensuremath{\textrm{Ly-}\beta}}
\newcommand{\zem}{\ensuremath{z_\textrm{\scriptsize em}}}
\newcommand{\zab}{\ensuremath{z_\textrm{\scriptsize abs}}}
\newcommand{\NHI}{\ensuremath{N_\textsc{h\scriptsize{\,i}}}}

\newcommand{\lNHI}{\ensuremath{\log(N_\textsc{h\scriptsize{\,i}}/\textrm{cm}^{-2})}}
\newcommand{\tran}[3]{\ensuremath{\ion{#1}{#2}\,\lambda\textrm{#3}}}

\newcommand{\headsort}{\ensuremath{\textsc{uves\_headsort}}}
\newcommand{\popler}{\ensuremath{\textsc{uves\_popler}}}

\usepackage{array}
\newcolumntype{:}{>{\global\let\currentrowstyle\relax}}
\newcolumntype{;}{>{\currentrowstyle}}

\title[UVES SQUAD Data Release 1]{The UVES Spectral Quasar Absorption Database (SQUAD) Data Release 1: The first 10 million seconds}

\author[M. T. Murphy et al.]{Michael T. Murphy,$^{1}$\thanks{E-mail: mmurphy@swin.edu.au (MTM)} Glenn G. Kacprzak,$^{1}$ Giulia A. D. Savorgnan,$^{1}$\newauthor
Robert F. Carswell$^{2}$\\
  $^{1}$Centre for Astrophysics and Supercomputing, Swinburne University of Technology, Hawthorn, Victoria 3122, Australia\\
  $^{2}$Institute of Astronomy, University of Cambridge, Madingley Road, Cambridge, CB3 0HA, UK
}

\date{Accepted 2018 October 8. Received 2018 September 14; in original form 2018 August 10}

\pubyear{2018}

\volume{{\rm in press}}

\begin{document}
\label{firstpage}
\pagerange{\pageref{firstpage}--\pageref{lastpage}}
\maketitle

% Single paragraph, not more than 250 words (200 for Letters), no references.
\begin{abstract}
We present the first data release (DR1) of the UVES Spectral Quasar Absorption Database (SQUAD), comprising 467 fully reduced, continuum-fitted high-resolution quasar spectra from the Ultraviolet and Visual Echelle Spectrograph (UVES) on the European Southern Observatory's Very Large Telescope. The quasars have redshifts $z=0$--5, and a total exposure time of 10 million seconds provides continuum-to-noise ratios of 4--342 (median 20) per 2.5-\kms\ pixel at 5500\,\AA. The SQUAD spectra are fully reproducible from the raw, archival UVES exposures with open-source software, including our \popler\ tool for combining multiple extracted echelle exposures which we document here. All processing steps are completely transparent and can be improved upon or modified for specific applications. A primary goal of SQUAD is to enable statistical studies of large quasar and absorber samples, and we provide tools and basic information to assist three broad scientific uses: studies of damped Lyman-$\alpha$ systems (DLAs), absorption-line surveys and time-variable absorption lines. For example, we provide a catalogue of 155 DLAs whose Lyman-$\alpha$ lines are covered by the DR1 spectra, 18 of which are reported for the first time. The \ion{H}{i} column densities of these new DLAs are measured from the DR1 spectra. DR1 is publicly available and includes all reduced data and information to reproduce the final spectra.
\end{abstract}

% Select between one and six entries from the list of approved keywords: http://oxfordjournals.org/our_journals/mnrasl/for_authors/mnraskey.pdf

%%%%%%%%%%%%%%%%%%%%%%%%%%%%%%%%%%%%%%%%%%%%%%%%%%

%%%%%%%%%%%%%%%%% BODY OF PAPER %%%%%%%%%%%%%%%%%%

\section{Introduction}\label{s:intro}

The era of 8-and-10-metre telescopes has revolutionised the study of quasar absorption spectra. Before the Keck I 10-metre telescope's first light with the High Resolution Echelle Spectrometer in 1993 \citep[HIRES;][]{Vogt:1994:362}, few quasars were bright enough to be studied with reasonable signal-to-noise ratio (\SN) at resolving powers $R\ga40000$ with smaller telescopes. This new reach was extended to the southern hemisphere in 1999 with the Ultraviolet and Visual Echelle Spectrograph \citep[UVES;][]{Dekker:2000:534} on the European Southern Observatory's (ESO's) 8-metre Very Large Telescope (VLT).

Since its commissioning, UVES has contributed to a wide variety of extragalactic discoveries and studies, particularly using absorption lines arising in gas clouds along quasar sight-lines. For example, UVES spectra have been used to trace the metallicity, power-spectrum and thermal history of the intergalactic medium via \lya\ forest absorption lines \citep[e.g.][]{Schaye:2003:768,Kim:2004:355,Boera:2014:1916}. The chemical abundances of circumgalactic environments, traced by the highest-column density clouds -- the damped \lya\ systems (DLAs) and sub-DLAs -- have been studied in detail with UVES \citep[e.g][]{Molaro:2000:54,Pettini:2002:21,Pettini:2008:2011,Dessauges-Zavadsky:2003:447}. UVES spectra have also been used to discover and analyse molecular hydrogen and carbon monoxide in (sub-)DLAs \citep[e.g.][]{Ledoux:2003:209,Noterdaeme:2008:327,Srianand:2008:L39} and, recently, in likely examples of the high-redshift interstellar medium \citep[e.g.][]{Noterdaeme:2015:A24,Noterdaeme:2017:A82}. Measurements of key cosmological parameters have been made with UVES quasar spectra; for example, deuterium abundance constraints on the total energy density of baryons \citep[e.g.][]{Pettini:2008:1499,Pettini:2012:2477,Riemer-Sorensen:2017:3239} and the redshift evolution of the cosmic microwave background temperature \citep[e.g.][]{Noterdaeme:2011:L7}. UVES quasar spectra have even been used to constrain cosmological variations in the fundamental constants of nature \citep[e.g.][]{Quast:2004:L7,King:2008:251304,King:2012:3370,Rahmani:2013:861,Molaro:2013:A68,Murphy:2016:2461}.

It is notable that most of the above studies utilised a UVES spectrum of a single quasar. While this demonstrates the high scientific value of such spectra, large samples are often required to enable some scientific projects, to make a meaningful measurement or improvement over previous ones\footnote{A crude illustration of the latter point is that, according to NASA's Astrophysical Data System, all but two of the 15 most cited UVES quasar absorption papers used considerable samples of spectra.}. One difficulty is that reducing raw high-resolution (i.e.\ echelle) spectroscopic data is challenging and can require considerable experience, even with observatory-supplied data reduction pipelines. Combining the spectra from many quasar exposures and continuum-fitting the final spectrum are almost always required, but these steps are not straight-forward and usually fall outside the scope of reduction pipelines. Therefore, most studies using UVES quasar spectra have not made general-purpose, combined spectra publicly available. Doing so can be time-consuming and low priority compared to the immediate, specific scientific purpose for which the UVES observations were proposed. This has severely limited the availability and use of large samples of high-resolution quasar spectra.

Considerable efforts have already been invested to address these limitations: \citet{Zafar:2013:A140} presented a database of 250 UVES quasar spectra and \citet{OMeara:2015:111,OMeara:2017:114} has provided 300 HIRES quasar spectra. To further assist, we provide here the UVES Spectral Quasar Absorption Database (SQUAD) first data release (DR1): 467 ``science-ready'' UVES quasar spectra at redshifts $\zem=0$--5. Importantly, the processing steps for each quasar spectrum in DR1 are fully transparent and repeatable. That is, all the steps to reduce and combine the multiple exposures of a quasar, and ``clean'' and continuum fit its combined spectrum, are fully visible and can be repeated by executing a few commands using public, open-source software. This end-to-end transparency and reproducibility ensures that scientific applications for which certain aspects of the data are important (e.g.\ the wavelength calibration accuracy) have an unbroken record of their treatment, from raw data to final, combined spectrum. Our public software ensures that each spectrum can be improved by its users as it is employed for different purposes (each with a different scientific focus), or modified to suit a particular scientific application, and that all changes can easily be made transparent and reproducible to others. We have also attempted to process the DR1 spectra as uniformly as possible so they may be most useful for statistical studies of large quasar samples.

The UVES SQUAD differs in several ways from the database of \citet{Zafar:2013:A140}. The latter drew on ESO's Advanced Data Products (ADP) archive: automatic reductions of point-source exposures, in settings for which standardised (``master'') calibration files were available. This used the original {\sc eso-midas} data reduction pipeline which has now been superseded by a pipeline with superior spectral extraction quality and which we optimise for better wavelength calibration accuracy. \citeauthor{Zafar:2013:A140} combined the ADP-reduced exposures of a quasar using custom software. Our experience suggests that this processes can be very important, even critical, for some scientific applications, which further motivates our fully transparent and reproducible approach, and the ability for users to modify the parameters of the reduction and/or combination process easily. Finally, the ADP-reduced exposures are redispersed (re-gridded) onto a linear wavelength grid. Given that UVES is a grating cross-dispersed echelle, the resolving power does not vary strongly with wavelength, so a linear wavelength grid is inappropriate: it will undersample the resolution element at the bluest wavelengths and/or oversample it at the reddest wavelengths. Further, combining multiple exposures onto a common wavelength grid entails redispersing them again (e.g.\ to accommodate different heliocentric velocities). This introduces further correlations between the flux (and uncertainty) in neighbouring pixels, and slightly lowers the resolving power. We avoid these problems in the UVES SQUAD by reducing the raw data and redispersing all extracted exposures once to a common log-linear, vacuum--heliocentric grid. Combining all (extracted) exposures in this way provides the highest \SN, highest-resolution, appropriately sampled final spectrum of each quasar.

This paper is organised as follows. \Sref{s:selection} describes how DR1 is defined, how the quasars were identified in ESO's UVES data archive, and presents the DR1 quasar catalogue (\Tref{t:cat}). \Sref{s:reduction} details how appropriate calibration data were identified for each quasar exposure, and the data reduction process. \Sref{s:popler} documents our \popler\ software for combining multiple (extracted) UVES exposures of a quasar to produce the ``science-ready'', final DR1 spectra. \Sref{s:database} describes the basic properties of the DR1 spectra and the main remaining artefacts that most or all spectra contain. In \Sref{s:use} we illustrate three examples of the many applications for the DR1 sample: DLA studies, absorption-line surveys and studies of time-variable absorption lines. In particular, we present a catalogue of 155 DLAs where the \lya\ line is covered by the DR1 spectra, 18 of which have not been reported before. We measure \ion{H}{i} column densities for these new DLAs directly from the DR1 spectra. \Sref{s:conclusion} summarises the paper and discusses future SQUAD data releases.

The DR1 database, including all reduced data and files required to produce the final DR1 spectra, are publicly available in \citet{Murphy:2018:UVESSQUADDR1}.

\section{Quasar selection and catalogue}\label{s:selection}

\Tref{t:cat} catalogues the DR1 quasar and spectrum properties. This first data release is defined as containing the 475 quasars in the ESO UVES archive whose first exposure (longer than 100\,s) was observed before 30th June 2008. All exposures of these quasars (longer than 100\,s) observed before 17th November 2016 were included in the final, combined spectra in DR1. In total, 3088 exposures were selected and successfully processed, with a total exposure time of $10.09\times10^6$\,s (2803\,h, an average of 5.9\,h per quasar).

The quasar candidates satisfying these date criteria were selected by cross-matching the coordinates of all ``science'' observations in the ESO UVES archive (i.e.\ with DPR.CATG set to ``SCIENCE'') with the MILLIQUAS quasar catalogue \citep[][updated to version 5.2\footnote{See {\urlstyle{rm}\url{http://quasars.org/milliquas.htm}}}]{Flesch:2015:e010}. While this catalogue aims to include all quasars from the literature (up to August 2017), it will not include unpublished quasars. To identify such cases, we checked the ESO proposal titles and observed object names (as labelled by the observers) for all programs that observed any MILLIQUAS quasar with UVES and searched for any objects observed in those programs that may be quasars (and not already reported in MILLIQUAS). This approach identified 9 of the final 475 quasars selected for DR1, and a further 18 objects that, upon data reduction and exposure combination, were clearly not quasars (17 stars and one galaxy). While it is possible that some quasars were not selected by our approach, our manual checking of the proposal titles and object names should ensure this number is very small or zero.

\begin{figure*}
\begin{center}
\centerline{\hbox{
    \includegraphics[width=1.10\columnwidth]{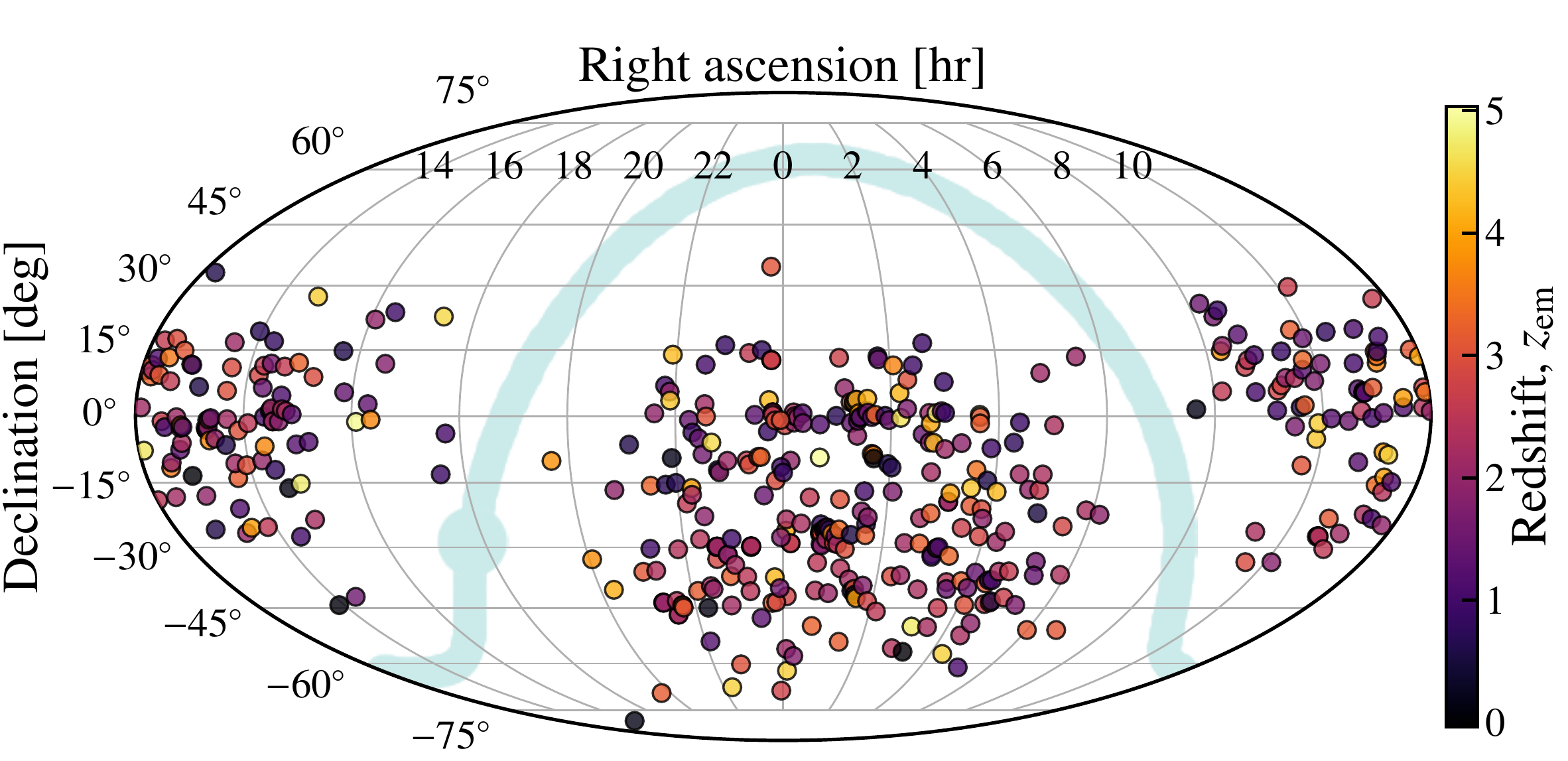}
    \hspace{0.01\columnwidth}
    \includegraphics[width=0.78\columnwidth]{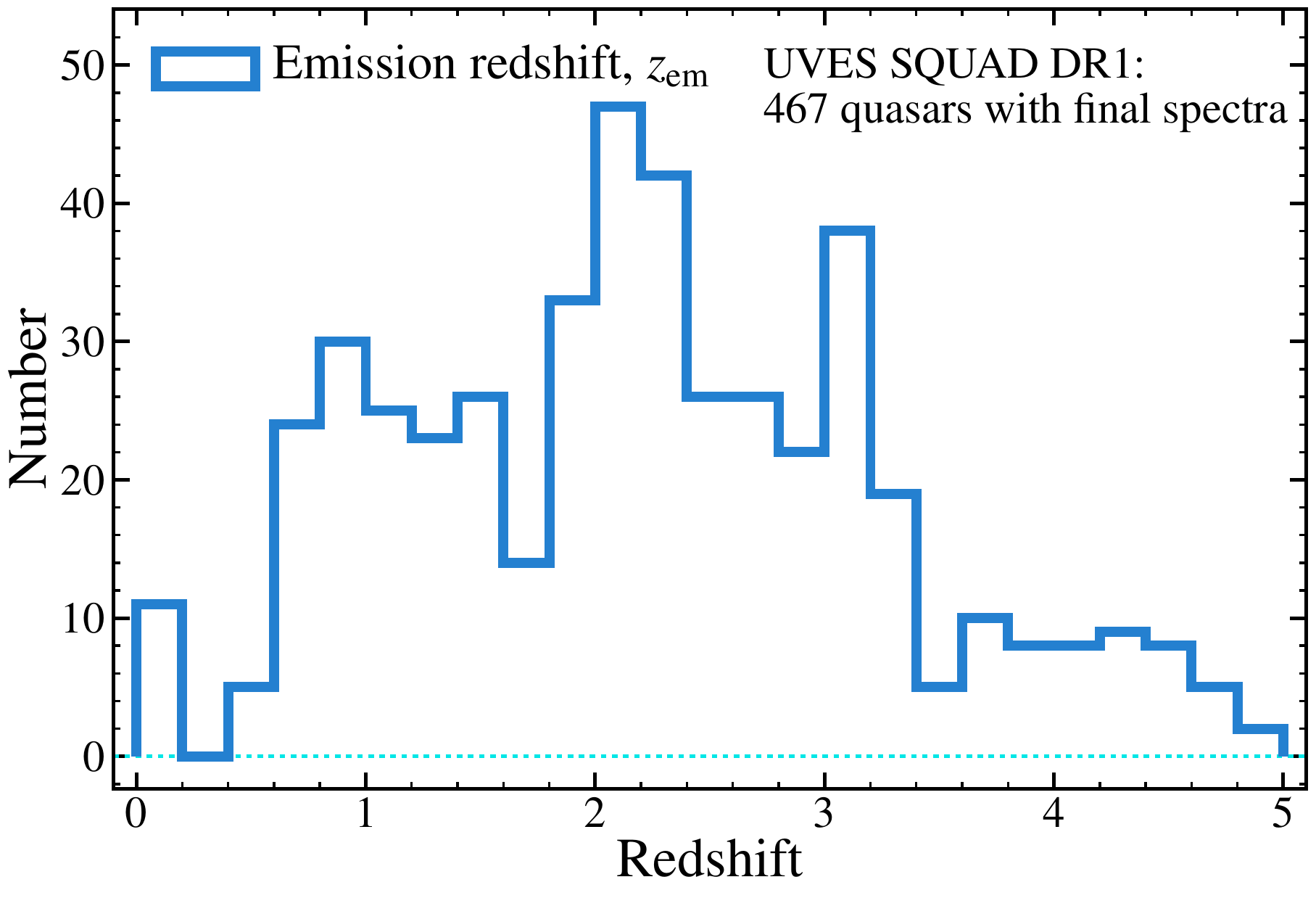}
}}\vspace{-0.5em}
\caption{Left panel: Sky distribution of the 467 DR1 quasars in \Tref{t:cat} for which final spectra were produced. The colour/shading of the points encodes the quasar redshift as indicated. The light blue shaded stripe and bulge represent the Galactic plane and centre. Right panel: Distribution of emission redshifts ($z_{\rm em,Adopt}$ in \Tref{t:cat}) for the DR1 quasars with final spectra.}
\label{f:sky+zem}
\end{center}
\end{figure*}

All quasar candidates were identified in the SuperCosmos Sky Survey database \citep{Hambly:2001:1279} to determine a complete set of J2000 coordinates. In cases where a spectrum is available from the Sloan Digital Sky Survey
\newcommand\oldtabcolsep{\tabcolsep}
\setlength{\tabcolsep}{0.27em}
\begin{landscape}
\begin{table}
  \caption{The UVES SQUAD DR1 sample of 475 quasars, of which we provide final spectra for 467. The first 20 columns provide the quasar names, coordinates, redshifts and optical/infra-red photometry sourced from several databases: SDSS, SuperCosmos, NED and SIMBAD (see text). The next seven columns specify the important observational information: the number of UVES exposures, their total duration, the ESO Program IDs, the UVES wavelength settings, slit widths and on-chip binnings used, and the prevailing seeing values reported in the ESO Science Archive (when available, the minimum, median and maximum seeing are reported, and ``NA'' is reported when seeing information is not available). The final five columns summarise the final spectrum properties (see text for full explanations): the ``Spec.\ status'' flag indicates whether we produced a final spectrum (values 0, 1 and 2), or whether multiple objects in the slit, a lack of calibration exposures, or a very high redshift precluded this (values 3, 4 and 5); the dispersion of the final spectrum (\kms\ per pixel); its wavelength coverage (\Sref{s:database}); the continuum-to-noise ratio and nominal resolving power calculated at five representative wavelengths (see \Sref{s:database}). The full table is available as Supporting Information online and contains additional information, including the DLA catalogue discussed in \Sref{ss:dlas}.}
\label{t:cat}
{\footnotesize
\begin{tabular}{cccclllccccccccc}
\hline
  \multicolumn{1}{c}{DR1 Name} &
  \multicolumn{1}{c}{RA$_{\rm Adopt}$} &
  \multicolumn{1}{c}{Dec$_{\rm Adopt}$} &
  \multicolumn{1}{c}{$z_{\rm em,Adopt}$} &
  \multicolumn{1}{c}{SDSS Name} &
  \multicolumn{1}{c}{NED Name} &
  \multicolumn{1}{c}{SIMBAD Name} &
  \multicolumn{1}{c}{$z_{\rm em,SDSS}$} &
  \multicolumn{1}{c}{$z_{\rm em,NED}$} &
  \multicolumn{1}{c}{$z_{\rm em,SIMBAD}$} &
  \multicolumn{1}{c}{B$_{\rm SSS}$} &
  \multicolumn{1}{c}{R1$_{\rm SSS}$} &
  \multicolumn{1}{c}{R2$_{\rm SSS}$} &
  \multicolumn{1}{c}{I$_{\rm SSS}$} \\
  &
  \multicolumn{2}{c}{(J2000)} &
  &
  &
  &
  &
  &
  &
  &
  \multicolumn{1}{c}{[mag]} &
  \multicolumn{1}{c}{[mag]} &
  \multicolumn{1}{c}{[mag]} &
  \multicolumn{1}{c}{[mag]} \\
\hline
  J000149$-$015939 & 00:01:49.94 & $-$01:59:39.4 & 2.815 & SDSS J000149.94$-$015939.4 & LBQS 2359$-$0216B          & [LE2003] Q2359$-$02A       & 2.815 & 2.817 & 0     & 18.69 & 18.31 & 18.52 & 18.59 \\
  J000322$-$260318 & 00:03:22.94 & $-$26:03:18.3 & 4.098 &                            & [HB89] 0000$-$263          & QSO B0000$-$26             &       & 4.098 & 4.111 & 19.55 & 17.12 & 16.94 & 16.74 \\
  J000344$-$232355 & 00:03:44.91 & $-$23:23:55.3 & 2.280 &                            & HE 0001$-$2340             & QSO B0001$-$2340           &       & 2.280 & 2.280 & 16.75 & 16.69 & 16.47 & 15.93 \\
  J000443$-$555044 & 00:04:43.28 & $-$55:50:44.6 & 2.100 &                            &                            &                            &       &       &       & 18.37 & 18.01 & 17.50 & 17.06 \\
  J000448$-$415728 & 00:04:48.27 & $-$41:57:28.1 & 2.760 &                            & TOLOLO 0002$-$422          & QSO B0002$-$4214           &       & 2.760 & 2.760 & 17.91 & 17.45 & 17.14 & 16.86 \\
  J000651$-$620803 & 00:06:51.62 & $-$62:08:03.3 & 4.455 &                            & BR J0006$-$6208            & QSO B0004$-$6224           &       & 4.455 & 4.455 & 21.62 & 18.70 & 19.04 & 18.06 \\
  J000815$-$095854 & 00:08:15.33 & $-$09:58:54.3 & 1.955 & SDSS J000815.33$-$095854.3 & SDSS J000815.33$-$095854.0 & SDSS J000815.33$-$095854.3 & 1.955 & 1.95  & 1.951 & 18.33 & 18.24 & 17.92 & 17.53 \\
  J000852$-$290043 & 00:08:52.70 & $-$29:00:43.7 & 2.645 &                            & 2QZ J000852.7$-$290044     & QSO B0006$-$2917           &       & 2.645 & 2.645 & 19.16 & 18.27 & 18.17 & 17.93 \\
  J000857$-$290126 & 00:08:57.72 & $-$29:01:26.4 & 2.607 &                            & 2QZ J000857.7$-$290126     & QSO B0006$-$2918           &       & 2.607 & 2.607 & 19.93 & 19.34 & 19.41 & 19.06 \\
  J001130$+$005550 & 00:11:30.55 & $+$00:55:50.8 & 2.290 & SDSS J001130.55$+$005550.8 & [HB89] 0008$+$006          & [HHB2004] A1               & 2.290 & 2.309 & 2.291 & 19.16 &       & 18.43 & 17.69 \\
\hline
\end{tabular}

\bigskip

\begin{tabular}{cccccccllccccc}
\hline
  \multicolumn{1}{c}{$u_{\rm SDSS}$} &
  \multicolumn{1}{c}{$g_{\rm SDSS}$} &
  \multicolumn{1}{c}{$r_{\rm SDSS}$} &
  \multicolumn{1}{c}{$i_{\rm SDSS}$} &
  \multicolumn{1}{c}{$z_{\rm SDSS}$} &
  \multicolumn{1}{c}{Num.} &
  \multicolumn{1}{c}{Exp.\,Time} &
  \multicolumn{1}{c}{ESO Program IDs} &
  \multicolumn{1}{c}{Wavelength settings} &
  \multicolumn{1}{c}{Slit widths} &
  \multicolumn{1}{c}{Binnings} &
  \multicolumn{1}{c}{Seeing} &
  \multicolumn{1}{c}{Spec.} &
  \multicolumn{1}{c}{Disper-} \\
  &
  &
  &
  &
  &
  \multicolumn{1}{c}{Exp.} &
  &
  &
  &
  &
  \multicolumn{1}{c}{(spec.\,$\times$\,spat.)}&
  \multicolumn{1}{c}{(min.,\,med.,\,max.)} &
  \multicolumn{1}{c}{status} &
  \multicolumn{1}{c}{sion} \\
  \multicolumn{1}{c}{[mag]} &
  \multicolumn{1}{c}{[mag]} &
  \multicolumn{1}{c}{[mag]} &
  \multicolumn{1}{c}{[mag]} &
  \multicolumn{1}{c}{[mag]} &
  &
  \multicolumn{1}{c}{[s]} &
  &
  \multicolumn{1}{c}{[nm]} &
  \multicolumn{1}{c}{[arcsec]} &
  &
  \multicolumn{1}{c}{[arcsec]} &
  &
  \multicolumn{1}{c}{[\kms]} \\
\hline
  20.26 & 18.95 & 18.79 & 18.82 & 18.63 & 6  & 21300  & 073.B-0787(A),\,66.A-0624(A)                  & 346,\,390,\,437,\,580,\,760,\,860             & 0.9,\,1.0 & 2$\times$2              & 0.56,\,1.50,\,2.03 & 0     & 2.5 \\
        &       &       &       &       & 4  & 16100  & 60.A-9022(A)                                  & 437,\,860                                     & 0.9       & 2$\times$2              & 0.34,\,0.49,\,1.97 & 1,\,2 & 2.5 \\
        &       &       &       &       & 26 & 108482 & 083.A-0733(A),\,083.A-0733(I),\,166.A-0106(A) & 346,\,390,\,420,\,437,\,580,\,700,\,760,\,860 & 0.7,\,1.0 & 1$\times$1,\,2$\times$2 & 0.40,\,0.94,\,2.20 & 0     & 1.3 \\
        &       &       &       &       & 7  & 21288  & 074.A-0473(A)                                 & 390,\,564                                     & 0.8       & 2$\times$2              & 0.61,\,1.24,\,2.28 & 0     & 2.5 \\
        &       &       &       &       & 46 & 178664 & 166.A-0106(A),\,185.A-0745(C),\,185.A-0745(F) & 346,\,390,\,437,\,564,\,580,\,760,\,860       & 0.8,\,1.0 & 1$\times$1,\,2$\times$2 & 0.37,\,0.80,\,2.81 & 0     & 2.0 \\
        &       &       &       &       & 2  & 7199   & 69.A-0613(A)                                  & 580                                           & 1.0       & 2$\times$2              & 0.83,\,1.06,\,1.10 & 0     & 2.5 \\                                
  19.17 & 18.85 & 18.36 & 18.00 & 17.90 & 6  & 21600  & 076.A-0376(A)                                 & 390,\,760                                     & 0.9,\,1.0 & 2$\times$2              & 0.86,\,1.30,\,2.20 & 0     & 2.5 \\
        &       &       &       &       & 6  & 24000  & 075.A-0617(A),\,70.A-0031(A)                  & 390,\,437,\,760,\,860                         & 1.1,\,1.2 & 2$\times$2              & 0.62,\,0.80,\,1.09 & 0     & 2.5 \\
        &       &       &       &       & 6  & 17700  & 075.A-0617(A),\,70.A-0031(A)                  & 390,\,437,\,760,\,860                         & 1.1,\,1.2 & 2$\times$2              & 0.58,\,0.95,\,1.50 & 0     & 2.5 \\
  20.02 & 19.06 & 18.69 & 18.29 & 17.99 & 2  & 7200   & 267.B-5698(A)                                 & 437,\,860                                     & 1.0       & 2$\times$2              & NA,\,NA,\,NA       & 0     & 2.5 \\
\hline
\end{tabular}

\bigskip

\begin{tabular}{lcc}
\hline
  \multicolumn{1}{c}{Wavelength coverage} &
  \multicolumn{1}{c}{Cont.-to-noise ratio at $\lambda_{\rm obs}=$} &
  \multicolumn{1}{c}{Nom.\ Res.\ Power at $\lambda_{\rm obs}=$} \\
  &
  \multicolumn{1}{c}{3500,\,4500,\,5500,\,6500,\,7500\,\AA} &
  \multicolumn{1}{c}{3500,\,4500,\,5500,\,6500,\,7500\,\AA} \\
  \multicolumn{1}{c}{[\AA]} &
  &
  \multicolumn{1}{c}{[1000]} \\
\hline
  3231--5760,\,5761--\ldots--10429                                         & 6,\,12,\,12,\,15,\,11      & 49.8,\,49.8,\,52.0,\,49.9,\,52.0 \\
  3952--4981,\,4986--5000,\,6731--8521                                     & 0,\,65,\,0,\,0,\,48        & 0,\,54.5,\,0,\,0,\,52.0          \\
  3064--4894,\,4898--4902,\,4905--4949,\,4954--5235,\,5241--\ldots--10257  & 36,\,66,\,97,\,106,\,99    & 60.1,\,60.1,\,53.7,\,53.7,\,50.8 \\
  3290--4520,\,4622--5600,\,5675--6616,\,6619--6652                        & 22,\,16,\,42,\,40,\,0      & 59.8,\,59.8,\,56.9,\,56.9,\,0    \\
  3049--\ldots--10430                                                      & 49,\,113,\,122,\,144,\,101 & 49.8,\,55.8,\,47.8,\,50.6,\,47.8 \\
  4809--4855,\,4858--5755,\,5841--6802                                     & 0,\,0,\,18,\,28,\,0        & 0,\,0,\,47.8,\,47.8,\,0          \\
  3297--4519,\,5685--5742,\,5759--5826,\,5829--7521,\,7665--9466           & 11,\,26,\,0,\,40,\,42      & 49.8,\,49.8,\,0,\,52.0,\,52.0    \\
  3296--4984,\,5678--5921,\,5931--5982,\,5988--\ldots--10429               & 16,\,165,\,0,\,46,\,41     & 41.5,\,41.5,\,0,\,41.1,\,41.1    \\
  3324--4984,\,5679--5745,\,5760--7966,\,7967--\ldots--10434               & 5,\,16,\,0,\,16,\,16       & 41.5,\,41.5,\,0,\,41.1,\,41.1    \\
  3760--4984,\,6692--6770,\,6779--6921,\,6929--\ldots--10432               & 0,\,8,\,0,\,0,\,17         & 0,\,49.8,\,0,\,0,\,47.8          \\
\hline
\end{tabular}
}
%\begin{minipage}{\textwidth}
%{\footnotesize
%Notes: 
%}
%\end{minipage}

\end{table}
\end{landscape}
\setlength{\tabcolsep}{\oldtabcolsep}
\noindent  \citep[SDSS DR14;][]{Abolfathi:2018:42,Paris:2018:A51}, the SDSS coordinates were used in preference. These coordinates were used to name all quasars in DR1; this results in the unique ``DR1 Name'' field for each quasar in \Tref{t:cat}. Quasar emission redshifts were taken from a hierarchy of cross-matched databases: SDSS, NED, SIMBAD and, if the quasar appeared in none of these databases, our own approximate measurement from our final spectrum. The latter was required in 13 cases, but in one of these (J031257$-$563912), no emission line could be identified from which a redshift could be estimated (its redshift is set to zero in \Tref{t:cat}). This provides a nominal, adopted redshift for each quasar, named ``$z_{\rm em,Adopt}$'' in \Tref{t:cat}. The sky position and redshift distributions of the quasars are plotted in \Fref{f:sky+zem}.

DR1 contains final spectra for 467 of the 475 quasars in \Tref{t:cat}. Known shortcomings of the final spectrum, or the reason why one could not be produced, are encoded for each quasar in \Tref{t:cat} in the ``Spec.\ status'' flag, which can have the following values:
\begin{itemize}
\item ${\rm Spec.\ status} = 0$: A final spectrum was produced with no known problems.
\item ${\rm Spec.\ status} = 1$: Not all available exposures could be successfully processed for lack of appropriate calibration exposures. This applies to two quasars (J000322$-$260318 and J030722$-$494548). The wavelength coverage of their final spectra is significantly reduced as a result.
\item ${\rm Spec.\ status} = 2$: The wavelength calibration of at least some exposures are very obviously distorted. This applies to one quasar spectrum, J000322$-$260318; this should be used with caution.
\item ${\rm Spec.\ status} = 3$: Quasars at redshifts $\zem\ge5.8$, so the extremely thick \lya\ forest left very little flux in individual exposures. This applies to three quasars (J130608$+$035626, J103027$+$052455, J104433$-$012502). Combining the exposures using our approach (particularly the order scaling step) is not effective in such cases (\Sref{sss:combination}); the exposures would need to be spectrophotometrically flux calibrated to allow reliable combination.
\item ${\rm Spec.\ status} = 4$: More than one object with similar magnitude aligned in the slit. This applies to two lensed quasars J110633$-$182124 and J145433$-$401232. Producing separate, successfully resolved spectra would require non-standard reduction steps not implemented here. The observations of another lensed quasar, J091127$+$055054, had a second, much fainter object aligned in the slit; this did not affect the data reduction steps so we provide a final spectrum for this object but urge caution in using it.
\item ${\rm Spec.\ status} = 5$: A lack of appropriate calibration exposures in the ESO UVES archive precluded the basic data reduction steps required. This applies to three quasars (J030449$-$000813, J223337$-$603329, J033032$-$270438).
\end{itemize}

\section{Data calibration and reduction}\label{s:reduction}

\subsection{Science and calibration data selection}\label{ss:selection}

UVES is a two-arm, grating cross-dispersed echelle spectrograph mounted on the Nasmyth platform of Unit Telescope 2 of the VLT \citep{Dekker:2000:534}. Combined, the two arms can cover a very broad wavelength range ($\sim$3050--10500\,\AA), albeit with gaps depending on the wavelength settings chosen. The blue arm camera contains a single CCD chip, while the red arm camera contains a two-chip mosaic. Most observations use both arms simultaneously, with the quasar light split into the two arms by a dichroic mirror, in two of nine standard wavelength settings named according to the central wavelength, in nm: 346, 390 and 437 for the blue arm; 520, 564, 580, 600, 760 and 860 for the red arm. However, some observations use a single arm only, and many different non-standard wavelength settings. The wavelength settings used for the DR1 quasar observations are specified in \Tref{t:cat}. Each setting is characterised by a different, nominal wavelength coverage.

We only consider exposures taken through an entrance slit; UVES has an image slicer option but we exclude such data from DR1. The slit width and on-chip binning determine the nominal resolving power, i.e.\ that expected for a fully illuminated slit, as is the case for ThAr exposures. However, the quasar exposure's resolving power will be somewhat larger than this, especially if the seeing FWHM is significantly less than the slit width. \Tref{t:cat} therefore provides the range of slit widths, binnings and seeing during the observations as a guide (\Sref{s:database} discusses the nominal resolving power reported in \Tref{t:cat} for the final spectra). We do not include quasar observations made through UVES's iodine absorption cell; these require additional calibration exposures and cannot be combined with non-absorption cell observations of the same quasars. Finally, we exclude exposures taken with the Fibre Large Array Multi Element Spectrograph (FLAMES) mode of UVES. Given the above spectrograph details, a range of calibration exposures are required to reduce each quasar exposure.

The default operations model for UVES is that all calibration exposures are taken in the morning after each night's observations. This means that some exposures are used to calibrate more than one quasar exposure, and that associating calibrations with exposures requires a matching algorithm. We requested all available UVES ``science'' exposures of the DR1 quasars, within 1 arcminute search boxes of their adopted coordinates (from SDSS or SuperCosmos), plus matching calibration exposures, from the ESO Science Archive. However, for many quasars the calibration matching algorithm was clearly imperfect so additional, manual requests for a large number of calibration exposures around the observation dates of many quasars were made as well. This resulted in a large database of potential calibration files. We used a custom-written code, \headsort\ \citep{Murphy:2016:UVESheadsort}, to ensure that the best-matching calibrations were selected within a specified ``calibration period'' before and after each quasar exposure. This generally meant selecting the calibration exposure(s) closest in time to the corresponding quasar exposure for five different calibration types:
\begin{itemize}
\item Wavelength calibration: A single thorium-argon (ThAr) exposure with the same spectrograph settings (i.e.\ wavelength setting, on-chip binning and slit-width), was generally selected. Given the UVES operations model, in most cases the ThAr exposure was taken at least several hours after the quasar exposure. Indeed, the median time difference for all 3088 processed DR1 exposures is 5.4\,h. However, preference was given to ``attached'' ThAr exposures, i.e.\ those taken immediately after quasar exposures without any grating angle changes. An attached ThAr exposure was identified as having the same grating encoder value as the corresponding quasar exposure. In a very small number of cases, particularly for exposures taken before 2001, a slightly different slit width was allowed for the matched ThAr exposure compared to the quasar exposure.
\item Order format and definition: ThAr and quartz lamp exposures taken through a short slit are used to identify the echelle orders and define a baseline trace across the CCD. A single exposure of each type with the same spectrograph settings (except for the much shorter slit), was selected in all cases.
\item Flat field: Five quartz lamp exposures with the same spectrograph settings were selected. In a small number of cases, especially for early UVES data (before 2003), some quasar exposures only had 3 or 4 matching flat field exposures; rarely, only a single flat field exposure could be found for quasar exposures taken before 2002.
\item Bias: The five bias (zero-duration) exposures taken on the same CCD as the quasar exposure were selected in all but rare cases from early UVES operations (before 2002).
\end{itemize}

\subsection{Reduction with \headsort\ and the ESO Common Pipeline Library}\label{ss:cpl}

After determining the best set of calibration exposures for a given quasar exposure, \headsort\ outputs a reduction script for use with ESO's Common Pipeline Library ({\sc cpl}, version 4.7.8\footnote{{\urlstyle{rm}\url{http://www.eso.org/observing/dfo/quality/UVES/pipeline/pipe_reduc.html}}}) of UVES data reduction routines, specifically via the ESO Recipe Execution Tool ({\sc esorex}) command-line interface. This provides a highly streamlined data reduction pipeline -- typically, a quasar exposure can be matched with calibrations and fully reduced within several minutes -- while allowing low-level access to the data reduction parameters for improving the reduction if required.

Most of the reduction steps are standard for UVES data and are explained in detail in the UVES {\sc cpl} pipeline manual\footnote{\urlstyle{rm}\url{https://www.eso.org/sci/software/pipelines/uves/uves-pipe-recipes.html}}. Briefly, these standard steps are:
\begin{enumerate}
\item ThAr lines are identified on the format definition frame and used to constrain a physical model of the UVES echellogram. This identifies the diffraction order numbers and spectral setup of the exposure which assists the order definition [step (ii)] and enables the automatic wavelength calibration in step (iv) below.
\item The order definition exposure is used to establish a baseline trace for object light along each echelle order. This acts as an initial guide for extracting the quasar flux.
\item The bias and flat-field exposures are combined to form masters which are used to correct the quasar exposure for bias and dark-current offsets and pixel-to-pixel sensitivity variations in the subsequent steps.
\item The ThAr flux is extracted along the default trace in the wavelength calibration exposure (corrected for the blaze function using the master flat) and the ThAr lines are automatically, iteratively matched with those in the list carefully selected for UVES in \citet{Murphy:2007:221}. This allows a polynomial (air) wavelength solution to be established for the entire CCD (i.e.\ air wavelength versus pixel position for each echelle order).
\item The quasar flux is optimally extracted, with weights determined by averaging the quasar flux along small spectral sections (normally 32 pixels) and either fitting a Gaussian function to this average profile or using it directly, depending on its \SN. The sky flux is extracted simultaneously in this process and is subtracted from the quasar flux in each extracted spectral pixel. The 1$\sigma$ flux uncertainty is also determined from the quasar flux, sky flux and CCD noise characteristics. The flux and uncertainty spectra are corrected for the blaze function using the master flat.
\end{enumerate}

Step (iv) was then repeated for the DR1 spectra to improve their wavelength calibration. The optimal extraction weights from step (v) were used to re-extract the ThAr spectra and perform a refined wavelength calibration process. This ensures that the same pixels, with the same statistical weights, are being used to establish the wavelength scale for the quasar spectrum (e.g.\ it naturally negates the effects of spatially tilted ThAr lines on the CCD). \headsort's reduction scripts also modify {\sc cpl}'s defaults for the wavelength polynomial degree, the number of ThAr lines to search for and select before performing the iterative polynomial fitting, and the tolerance allowed between the fitted and expected wavelength of ThAr lines. Typically, these new defaults simultaneously increase the number of lines used in the wavelength calibration, reduce the residuals around the final wavelength solution, and marginally improve the accuracy of the solution (due to increasing the polynomial degree). In some cases, particularly with the very blue wavelength settings (e.g.\ the standard 346 and 390-nm settings), these new defaults were modified manually to achieve a more robust wavelength solution (i.e.\ to increase the number of ThAr lines used).

\begin{figure}
\begin{center}
\includegraphics[width=0.90\columnwidth]{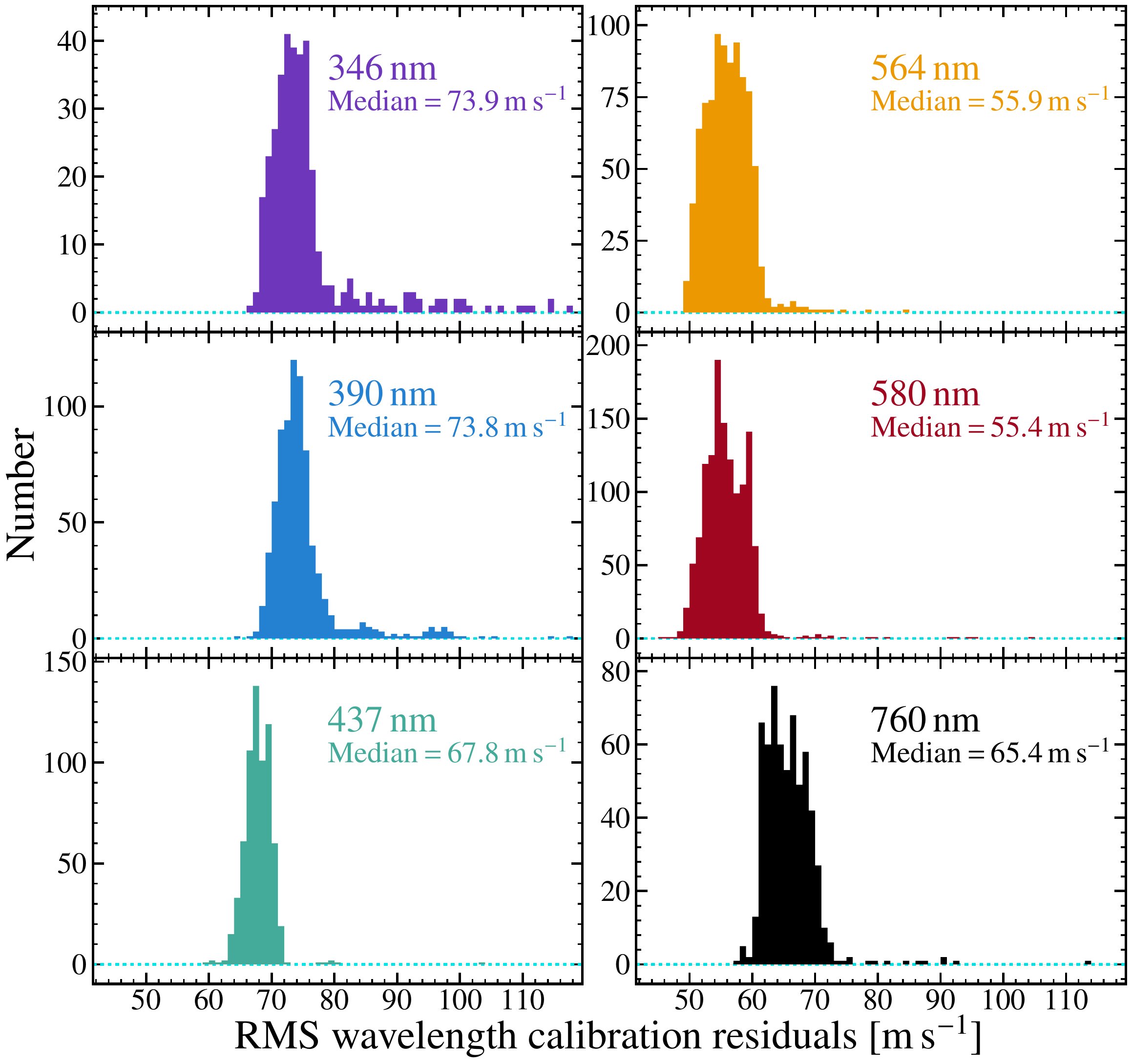}
\vspace{-1em}
\caption{Typical RMS wavelength calibration residuals. The distributions show the calibration results for each DR1 quasar exposure taken with a 1-arcsecond-wide slit and 2$\times$2 on-chip binning in the six most common wavelength settings (as labelled).}
\label{f:wavcal}
\end{center}
\end{figure}

\Fref{f:wavcal} shows the resulting root-mean-square deviation from the mean (RMS) of the wavelength calibration residuals for each CCD chip for all DR1 quasar exposures taken with a 1-arcsecond-wide slit and 2$\times$2 on-chip binning in the six most commonly-used wavelength settings. Together, these exposures comprise 53\% of all DR1 quasar exposures. In all but the bluest two settings, our approach to the wavelength calibration produced very similar residuals for almost all exposures. For the 390 and particularly the 346-nm settings, some exposures had substantially larger residuals. This is mainly due to the strong variation in UVES's total efficiency across the wavelength ranges covered by these settings. This causes a deficiency in the number of ThAr lines found above the intensity threshold set by the {\sc cpl} pipeline in the bluest orders. Although the {\sc cpl} pipeline addresses this problem in most cases, in some cases the tolerance for accepting calibration lines had to be increased so that enough lines could be found to provide a robust wavelength solution. This, in turn, causes the observed increase in the RMS of the wavelength calibration residuals in such cases.

After step (v) above, the {\sc cpl} pipeline redisperses the flux and uncertainty arrays onto a linear wavelength grid (i.e.\ all pixels have the same size in wavelength), merges the spectra from adjacent spectral orders, and corrects the spectral shape using an estimate of the instrument response curve. However, because the resolving power remains reasonably constant across the wavelength range of grating cross-dispersed echelle spectrographs, and UVES covers more than a factor of three in wavelength range ($\sim$3050--10500\,\AA), a constant dispersion in wavelength is inappropriate; it inevitably oversamples the resolution element in the bluest parts of the spectrum and/or undersamples it in the reddest parts. Also, merging adjacent orders should account for small instrument response and/or blaze correction imperfections and variations by scaling their relative flux before averaging, but the accuracy of this is severely limited in a single exposure due to lack of \SN. However, almost all quasars in DR1 were observed in multiple exposures, so there is an opportunity to improve the merging of adjacent orders by considering all exposures together. And, finally, if the spectra from multiple exposures are to be combined, they will have to be redispersed, again, onto a common wavelength grid after correction for heliocentric motions. For these reasons, we use the not-redispersed, extracted flux and uncertainty arrays of each order (not flux calibrated), from every exposure, to produce each quasar's final spectrum. This was performed using the custom-written code, \popler, described below. All relevant pipeline products are provided for every DR1 exposure in \citet{Murphy:2018:UVESSQUADDR1}.

\section{\popler: UVES Post-pipeline Echelle Reduction}\label{s:popler}

\popler\ \citep{Murphy:2016:UVESpopler} was designed specifically for combining the UVES data reduction pipeline products to produce a final, continuum-fitted spectrum (however, it can also use products from many other data reduction pipelines, including those often used for Keck/HIRES and Subaru/HDS high-resolution spectra). Below we summarise the overall approach of \popler\ (\Sref{ss:popler_summary}) and how it was applied to create the DR1 spectra (\Sref{ss:poplerDR1}).

\subsection{Summary of \popler\ operation}\label{ss:popler_summary}

\popler\ reads the extracted flux and uncertainty arrays for each echelle order of each quasar exposure and the wavelength calibration polynomials derived from their corresponding ThAr exposures. Operation then proceeds in two phases: the automatic and manual phases. It is important to note that both phases are entirely reproducible and transparent: all parameters of the automatic phase, and relevant details of all manual ``actions'' subsequently performed in the manual phase, are recorded in a \popler\ log (UPL) file; any user can understand how a spectrum has been formed and modified, and re-run both the entire process themselves. UPL files for all DR1 quasars are provided in \citet{Murphy:2018:UVESSQUADDR1}.

The automatic phase attempts to combine the spectra from all orders in all exposures and perform a basic continuum fit. Its main steps are:
\begin{enumerate}
\item Data validation: Reject pixels whose uncertainty indicates problems in the extraction (e.g.\ negative or extremely small uncertainties). This normally occurs near the order edges for UVES {\sc cpl}-reduced spectra.
\item Residual cosmic ray rejection: Reject pixels, and their immediate neighbours, whose flux is much larger than the mean flux for their neighbouring 34 pixels. This rejects ``cosmic rays'' and/or bad pixels not already rejected in the optimal extraction step of UVES {\sc cpl}-reduced spectra.
\item Vacuum and heliocentric corrections: The wavelength scales for the individual exposures are converted from air wavelengths to vacuum, and their correction for heliocentric motion is calculated and applied.
\item Redispersion: A common log-linear, vacuum--heliocentric wavelength scale is established, with a constant velocity dispersion specified by the user, that covers the remaining pixels in the contributing exposures. The flux and uncertainty spectra from all exposures are linearly redispersed onto this common grid.
\item Order scaling and combination: The spectra in all echelle orders are combined in an interactive process starting from the highest \SN\ order. It is combined with the next highest ``rank'' order: that with the highest combination of \SN\ and wavelength overlap. The next highest rank order is combined with the previous two, and so on until all orders are combined. The flux (and uncertainty) in each order is optimally scaled to match the combined spectrum from the previous iteration. For each spectral pixel, the combined flux is the weighted mean of the fluxes from the contributing spectra, which is determined through an iterative clipping process to remove discrepant values.
\item Continuum fitting: Each contiguous section of the combined spectrum is broken into ``chunks'', typically 20000\,\kms\ wide below the quasar \lya\ emission line and 2500\,\kms\ above it, which overlap half of the adjacent chunks. An iterative polynomial fit is performed to each chunk: at each iteration, pixels with flux significantly below (typically $>$1.4$\sigma$) or above (typically $>$3.0$\sigma$) the current fit are rejected for the next iteration. To form a smooth, final continuum, the final fits from adjacent chunks are averaged with a weight that decreases linearly from unity at the chunk's centre to zero at its edge.
\end{enumerate}

The automatic phase of \popler\ generally produces excellent ``quick-look'' spectra that are entirely adequate for many scientific goals, particularly those focussing on individual absorption systems whose transitions collectively occupy only a small fraction of the pixels. However, individual UVES exposures nearly always contain some artefacts that inhibit larger, statistical studies (and are often a nuisance to others as well) because, for example, they can mimic real absorption features in blind searches. The automatic continuum fits redwards of the \lya\ forest are generally very reliable, except in the vicinity of absorption features wider than the chunk size or across very narrow quasar emission lines. However, the automatic continuum in the \lya\ forest is not generally useful; reliable automatic continuum placement is a notorious problem in quasar spectroscopy that limits the speed with which high-resolution spectra can be analysed. Unfortunately, we have not solved that problem here. For these reasons, a manual phase of operation is required.

The manual phase of \popler\ allows interactive ``actions'' to be performed on the contributing echelle orders or combined spectrum to improve the quality of the latter and its continuum fit. These actions include:
\begin{itemize}
\item Clip (and unclip) pixels from contributing orders or the combined spectrum.
\item Manually fit or draw (spline) a new continuum to part of the combined spectrum.
\item Automatically fit the continuum for the entire spectrum again.
\item Manually fit or draw (spline) a continuum to (part of) a contributing order to reshape its flux (and uncertainty) array to that of the combined continuum.
\item Scale an order's flux and uncertainty array by a constant factor.
\item Rerun the automatic order scaling algorithm starting from the highest rank order not manually scaled by the user.
\end{itemize}
In general, a user will select portions of a spectrum to manually improve using the above actions based on their specific scientific goals. For example, for studying the intergalactic medium, it will be important to remove artefacts and re-fit the continuum in the \lya\ forest region. In \Sref{ss:poplerDR1} below we describe the approach to improving the DR1 quasar spectra for use towards as many different scientific goals as possible, particularly large statistical studies of DLAs, the intergalactic medium and metal absorption systems.

\subsection{Creation of UVES SQUAD DR1 spectra with \popler}\label{ss:poplerDR1}

For DR1, \popler\ (version 1.00) was used to create the final quasar spectra. We provide the complete record of parameters used for the automatic phase, and all subsequent manual actions for all DR1 quasars as UPL files in \citet{Murphy:2018:UVESSQUADDR1}. The detailed, specific treatment of each quasar is therefore transparent and any user may reproduce a quasar's final DR1 spectrum with a single execution of \popler\ (with the UPL as its argument). A key aspect of this approach is that users may further improve the DR1 spectrum by using \popler\ to add manual actions to the UPL file. Indeed, we welcome improved UPL files from the user community for inclusion in subsequent data releases.

It is important to note that we produce a single final spectrum for each quasar. That is, we combine all available {\sc cpl}-reduced exposures of a quasar regardless of variations in slit width and CCD binning. While a large range of slit widths are available for UVES (0.3--10\,arcsec), in practice the range used for a specific quasar is very narrow, presumably because achieving a threshold \SN\ is most often the immediate observational goal and, for faint (i.e.\ most) quasars, this is severely affected by the choice of slit width (UVES is a natural-seeing instrument). For example, 385 of the 467 final spectra comprise exposures with a single slit width\footnote{The two arms of UVES can have different slit widths for a dual-arm observation. However, for this example we have ignored cases where the different arms used consistent but different slit widths. That is, the number of spectra for which a single slit width was used at any given wavelength will be somewhat larger than 385.}, while only 16 combine exposures with slit widths differing by more than 0.3\,arcsec. Nevertheless, different slit widths and CCD binnings will produce individual exposures with different effective resolving powers, thereby affecting the resolution of the final spectrum. A nominal, mean resolving power is calculated for each final spectrum in \Sref{sss:Rnom} below. If separate combination of exposures of different resolutions is required, \popler\ can easily be used to construct such ``sub-spectra'' of a given quasar using its UPL file, as discussed in \Sref{ss:subspec}. For example, this technique has been employed in the analysis of J051707$-$441055 by \citet{Kotus:2017:3679}.

To make each DR1 quasar spectrum useful for as many scientific goals as possible, our approach was to ``clean'' it to at least a minimum standard in the manual phase of \popler. Clearly, this cleaning process is the most time-consuming stage, and all authors contributed to it, so ensuring a strictly uniform standard for all DR1 quasars was not practical. Nevertheless, the following cleaning steps were taken for each quasar in DR1 with a view to making the final spectrum as useful as possible.

\subsubsection{Artefact and bad data removal}\label{sss:artefacts}

\begin{figure}
\begin{center}
\includegraphics[width=0.90\columnwidth]{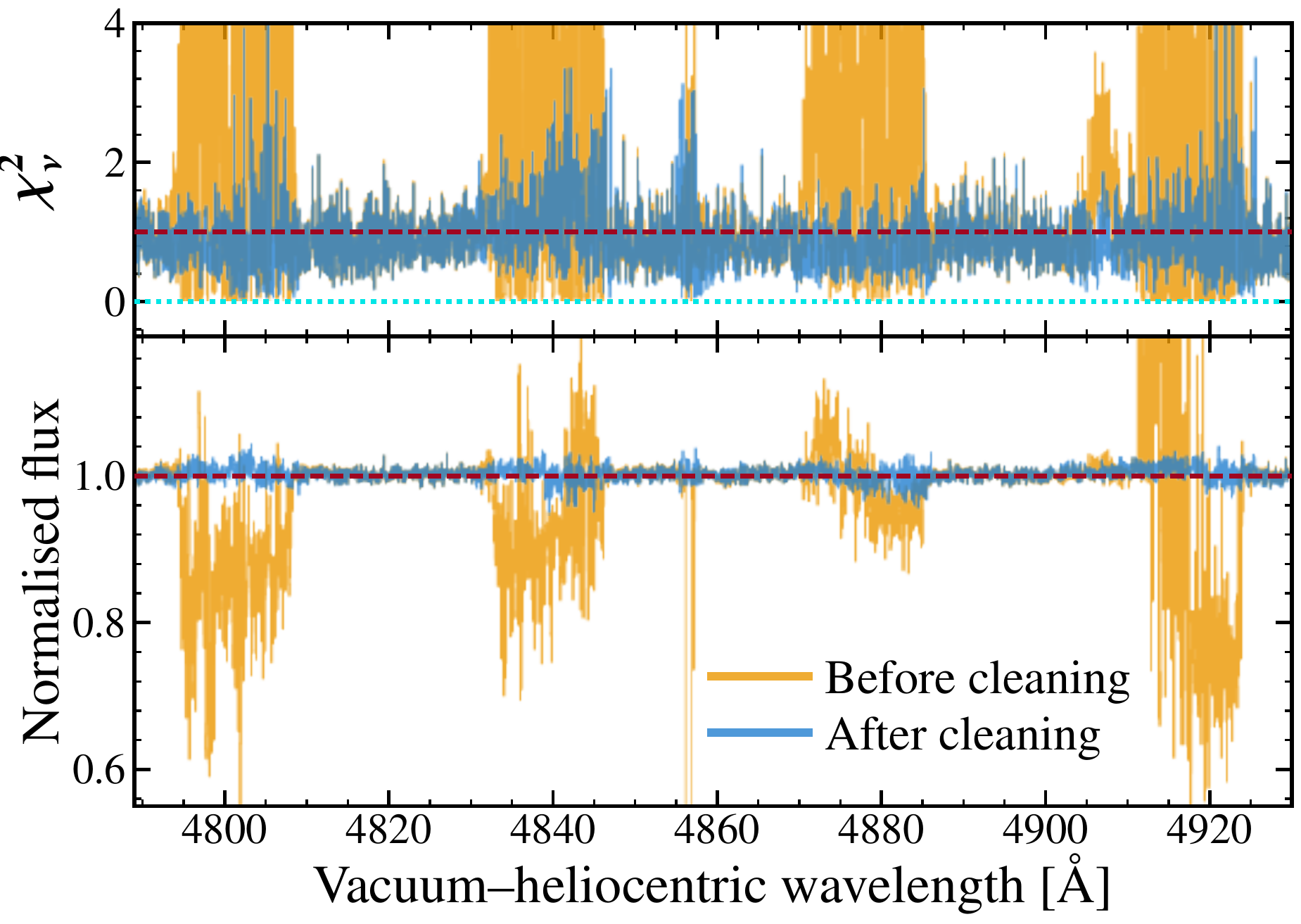}
\vspace{-1em}
\caption{Example of manual cleaning of obvious artefacts and bad data from the spectrum of J014333$-$391700. Lower panel: The orange (light, background) spectrum is the result of the automatic phase of \popler\ and clearly shows the effect of bad CCD rows in the first 4 orders of red arm spectra (580-nm wavelength setting; not all contributing exposures were affected). The blue (dark, foreground) spectrum shows the result after the manual artefact removal stage in \popler: the most obvious artefacts were removed, but there are still some low-level effects from other contributing exposures that were not manually removed (e.g.\ at 4840, 4880 and 4920\,\AA). Upper panel: The $\chi^2$ spectra before (orange, background) and after (blue, foreground) artefact removal. $\chi^2$ for a given pixel is that of the contributing pixels' flux values relative to their combined value (weighted mean). The $\chi^2$ spectra are used in \popler\ to help identify regions where artefacts may be present.}
\label{f:baddata}
\end{center}
\end{figure}

The {\sc cpl}-reduced UVES spectra often contain very obvious artefacts that are similar, though not identical, in different spectra. Thus, they are not removed by the iterative clipping process when the contributing order spectra are combined [step (v) in \Sref{ss:popler_summary}] and can corrupt the final spectrum. Manually removing them from the contributing spectra can often leave a relatively uncorrupted, contiguous region in the final, combined spectrum. A prominent and common example occurs in the bluest 4--5 orders of the red arm spectra due to several bad pixel rows in the corresponding CCD. An example of this problem is shown in \Fref{f:baddata}. For each quasar, we visually scanned the spectrum in \popler\ to identify such artefacts. Clearly, the flux spectrum is one important guide here, as can be seen in \Fref{f:baddata}, and we removed artefacts that obviously affected the flux spectrum. However, \popler\ also displays the $\chi_\nu^2$ spectrum: for each pixel, this is the $\chi_\nu^2$ of the contributing pixel fluxes around their weighted mean value. This assists in identifying regions where the contributing exposures do not match as closely as expected (given their uncertainties); it tends to help find artefacts that have a more subtle effect on the final flux spectrum. \Fref{f:baddata} contains an example at $\approx$4905--4910\,\AA: the significant increase in the $\chi_\nu^2$ spectrum here corresponds to only a small effect on the final flux spectrum. However, to reduce the time for cleaning all DR1 spectra, in many cases we did not remove some of these more subtle artefacts from contributing exposures if they did not affect an obvious absorption feature.

Another, less common, artefact in {\sc cpl}-reduced UVES spectra is that of ``bends'': echelle order spectra that have different shapes where they overlap. This can occur for several reasons, e.g.\ time-evolution in the flat-field lamp spectral shape, or poor extractions of the quasar flux near order edges, perhaps due to poorly constrained object traces. When very severe, these affected the final flux spectrum, and so were corrected. More subtle cases were still evident in the $\chi_\nu^2$ spectra and were corrected if they affected an obvious absorption line. Bends in contributing orders were corrected either by removing the bent section or by fitting a continuum to the order (or part thereof) and re-normalising it to match the combined spectrum's continuum shape.

\subsubsection{Order rescaling and combination}\label{sss:combination}

In spectral regions with very low \SN, or in echelle orders affected by severe artefacts, the relative scaling between an echelle order's spectrum and the combined spectrum [step (v) in \Sref{ss:popler_summary}] can be very poorly or spuriously determined. This occurs frequently in the bluest orders of the 346 and 390-nm settings. It also occurs if the broad trough of a DLA straddles two echelle orders, and below the Lyman limit in the rest frame of DLAs and Lyman limit systems. In these latter examples, there is simply no flux to allow a relative scaling between adjacent orders; this is certainly a disadvantage of the order-scaling algorithm in \popler. To address this in DR1 spectra we manually adjusted the scaling of the highest-ranked order with an obvious scaling problem and re-ran the automatic scaling algorithm starting at that order. This process was repeated for lower-ranked orders to achieve a final spectrum that, visually, appears properly scaled. For the extreme blue orders, where the \SN\ degrades significantly, the best manual scaling factor to choose is often quite unclear, so there may be significant scaling differences between orders in regions of final spectra with $\SN\la5$ per pixel.

\subsubsection{Continuum fitting}\label{sss:continuum}

As discussed in \Sref{ss:popler_summary}, the continuum fit in \popler's automatic phase is generally not useful in the \lya\ forest, near wide absorption features or over narrow emission lines. For the wide absorption and narrow emission features, we manually fit a new continuum only around the problematic region. This was relatively straight-forward, except for broad absorption-line quasars (BALs), because there are many pixels that are clearly not absorbed so the true continuum is easily discerned by the human eye. However, for the \lya\ forest, our approach was to manually fit the continuum in the entire region below the \lya\ emission line of all DR1 spectra. The well-known problem is that few \lya\ forest pixels are unabsorbed  (except perhaps at redshifts $z\la2$), so the true continuum level is usually not at all clear. Our fitting approach is to manually select seemingly unabsorbed ``peaks'' in the \lya\ forest and interpolate between them with a low-order polynomial. This is done in chunks of spectrum ranging from $\sim$2000 to $\sim$50000\,\kms\ wide, depending on how variable the true continuum appears to be. The continuum fits to neighbouring chunks are blended together in a user-defined overlap region to ensure a smoothly-varying final continuum. In some chunks it is not possible to perform a polynomial fit in this way; for example, if BALs or DLAs fall near an emission line (most often the \lya\ emission line) the human eye can discern an approximate shape for a continuum fit but there clearly may be no pixels without substantial absorption to enable a fit. In these chunks, a continuum was simply drawn using a cubic spline function. Our \lya\ continuum fits are, therefore, necessarily subjective and uncertain; however, we expect that they are likely more accurate, and more predictably biased, than algorithmic approaches (certainly the ones currently available in \popler).

\begin{figure}
\begin{center}
\includegraphics[width=0.90\columnwidth]{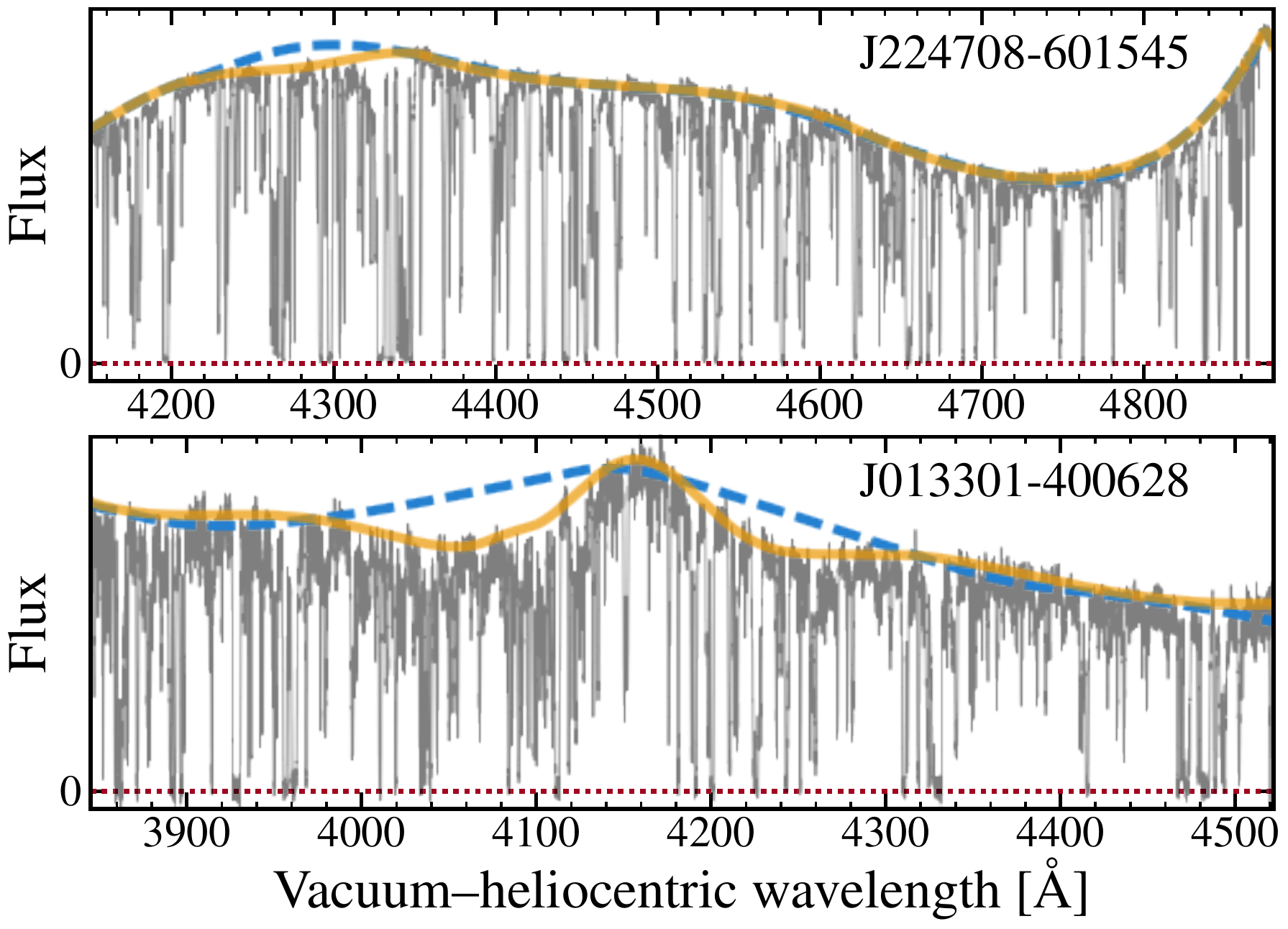}
\vspace{-1em}
\caption{Example of automatic (blue dashed lines) and manual (orange solid lines) continuum fits in the \lya\ forest of two DR1 quasar spectra. For J224708$-$601545 (upper panel), the automatic and manual fits are very similar; in this case the parameters of the automatic continuum fitting algorithm of \popler\ were highly tuned to produce a reasonable fit, and this is aided by the very high \SN\ of the spectrum. For the lower-\SN\ spectrum of J013301$-$400628 (lower panel), no combination of fitting parameters could achieve a reasonable fit to the region shown here; the automatic fit shown reflects \popler's default \lya\ forest fitting parameters: individual chunks of $\approx$250\,\AA\ are fit with a fourth-order polynomial. However, six separate manual continuum fits, in overlapping chunks, contributed to the final manual fit for the section of spectrum shown, with polynomial orders varying between 4 and 15; higher orders were needed to fit a reasonable continuum over the emission line at 4160\,\AA.}
\label{f:Lyacont}
\end{center}
\end{figure}

\Fref{f:Lyacont} compares the automatic and manual continuum fits in part of the \lya\ forest in two $\zem\approx3$ quasars. While the automatic continuum fit to the very high-\SN\ spectrum of J224708$-$601545 appears reasonable, on close inspection it is clearly too low in most regions and obviously too high around 4300\,\AA. However, for the lower-\SN\ spectrum of J013301$-$400628, the automatic continuum is completely inadequate. This is caused mainly by the emission line at 4160\,\AA. The manual fits shown in \Fref{f:Lyacont} are clearly more accurate and useful for statistical studies of the \lya\ forest, and even for more detailed studies of these individual lines of sight. However, even by eye, one can identify potential problems with our manual fits. For example, the manual continuum redwards of $\sim$4300\,\AA\ in \Fref{f:Lyacont} for J013301$-$400628 may by too high in general, perhaps by as much as $\sim$2\%. We discuss the biases in our continuum fits in \Sref{sss:art_cont}.

\subsubsection{Quality control}\label{sss:quality}

All the authors, and several others (see Acknowledgements), contributed to the manual cleaning and continuum fitting steps outlined above. Of course, this may lead to varying quality and homogeneity among the final spectra. To reduce this, one author (MTM) reviewed all DR1 spectra and modified or added manual actions to improve and homogenise them, where necessary. While the purpose of the general cleaning steps above is to ensure a minimum quality and usefulness for all DR1 spectra, some spectra -- or, most often, certain aspects of some spectra -- have received much more extensive attention, including manual changes to the spectrum not described above. These are generally spectra that have already been published elsewhere. One example is the very detailed study of J051707$-$441055 to constrain cosmological variations in the fine-structure constant by \citet{Kotus:2017:3679}. Beyond the basic cleaning steps outlined above, this study focussed on correcting the individual exposures for known, long-range distortions of the UVES wavelength scale \citep[e.g.][]{Rahmani:2013:861,Whitmore:2015:446} and velocity shifts between exposures caused by varying alignment of the quasar within the UVES slits. Such improvements are included in the DR1 versions of the spectra when available.

\section{Database of final spectra}\label{s:database}

The final DR1 spectral database is available in \citet{Murphy:2018:UVESSQUADDR1}. Each quasar's final spectrum is provided in standard FITS format \citep{Wells:1981:363}, with several FITS headers containing extensive information about the spectrum itself, the exposures that contributed to it and information about their extraction and calibration. We therefore expect that, for almost all scientific uses, only these final spectrum FITS files will be needed. However, for each quasar the database also contains the \popler\ log (UPL) file and {\sc cpl} pipeline products from each contributing exposure. This allows any user to reproduce or modify the final spectrum using \popler. Furthermore, we also provide all the reduction scripts and lists of raw input science and calibration data; this allows the entire reduction procedure to be reproduced or modified if desired.

\subsection{Basic spectral properties}\label{ss:properties}

\Tref{t:cat} summarises information about the final spectra relevant for most scientific uses. These properties were determined as described below. We emphasise that the full observational information for every exposure is contained within the FITS header for each quasar \citep[see documentation in][]{Murphy:2018:UVESSQUADDR1}.

\subsubsection{Dispersion}\label{sss:dispersion}

The log-linear dispersion per pixel, expressed as a velocity in \kms, was chosen according to the on-chip binning used for the contributing exposures. The native pixel scale of UVES is $\approx$1.3\,\kms\,pix$^{-1}$ so this was the dispersion set for spectra for which most or all contributing exposures were unbinned in the spectral direction (i.e.\ 1$\times$1 binning). However, most quasars had all, or almost all, 2$\times$2 or 2$\times$1-binned (spectral\,$\times$\,spatial) contributing exposures, so we set a 2.5\,\kms\,pix$^{-1}$ dispersion for these spectra. In more mixed cases we set intermediate dispersion values.

\subsubsection{Wavelength coverage}\label{sss:coverage}

Each quasar spectrum is accompanied by a pixel status spectrum whose integer value encodes whether each pixel is valid or the reason it is invalid. On a pixel-by-pixel basis, this array defines the detailed wavelength coverage map of the spectrum. However, for absorption-line searches, a more useful definition ignores single invalid pixels within larger, contiguous valid regions. Each DR1 FITS header therefore includes a wavelength coverage map in which valid chunks must be at least 100\,\kms\ wide and contain gaps (runs of invalid pixels) no wider than 10\,\kms. \Tref{t:cat} shows an abridged version of this wavelength coverage map.

\begin{figure}
\begin{center}
\includegraphics[width=0.90\columnwidth]{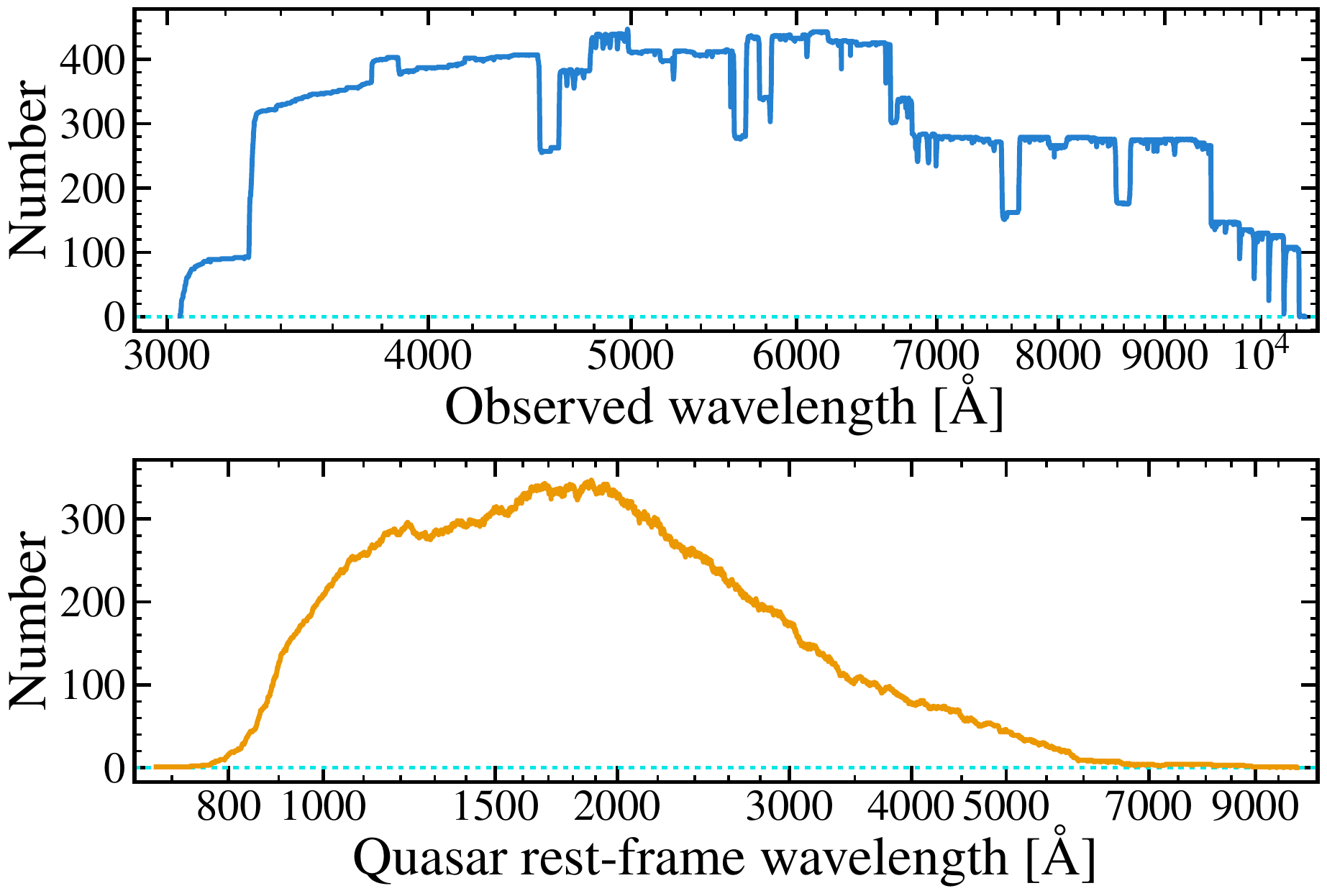}
\vspace{-1em}
\caption{Total wavelength coverage of the 467 final DR1 quasar spectra in the observed frame (upper panel) and common quasar rest frame (lower panel). The wavelength coverage of an individual spectrum is determined by requiring that valid chunks (contiguous runs of valid pixels) must be at least 100\,\kms\ wide and contain gaps (runs of invalid pixels) no wider than 10\,\kms.}
\label{f:wlcov}
\end{center}
\end{figure}

The upper panel of \Fref{f:wlcov} illustrates the total wavelength coverage of all 467 DR1 quasars with final spectra. The many detailed features in this map generally reflect the different wavelength settings used in the UVES observations. For example, the broad bump at $\sim$3800\,\AA\ is where the 390 and 437\,nm settings overlap, while the dip at $\sim$4500\,\AA\ is where the 390-nm wavelength coverage ends and where that of the 564-nm setting begins. The series of narrow dips redwards of $\sim$9500\,\AA\ are due to gaps in wavelength coverage between neighbouring echelle orders (i.e.\ where the free spectral range exceeds the CCD width). The lower panel of \Fref{f:wlcov} shows the total wavelength coverage in the common quasar rest frame. Here the focus on rest wavelengths $\la$2800\,\AA\ is evident, which is driven by the relative lack of strong absorption lines redwards of the Mg{\sc \,ii} doublet ($\lambda\lambda$2796/2803).

\subsubsection{Continuum-to-noise ratio (\CN)}\label{sss:CNR}

\begin{figure}
\begin{center}
\includegraphics[width=0.90\columnwidth]{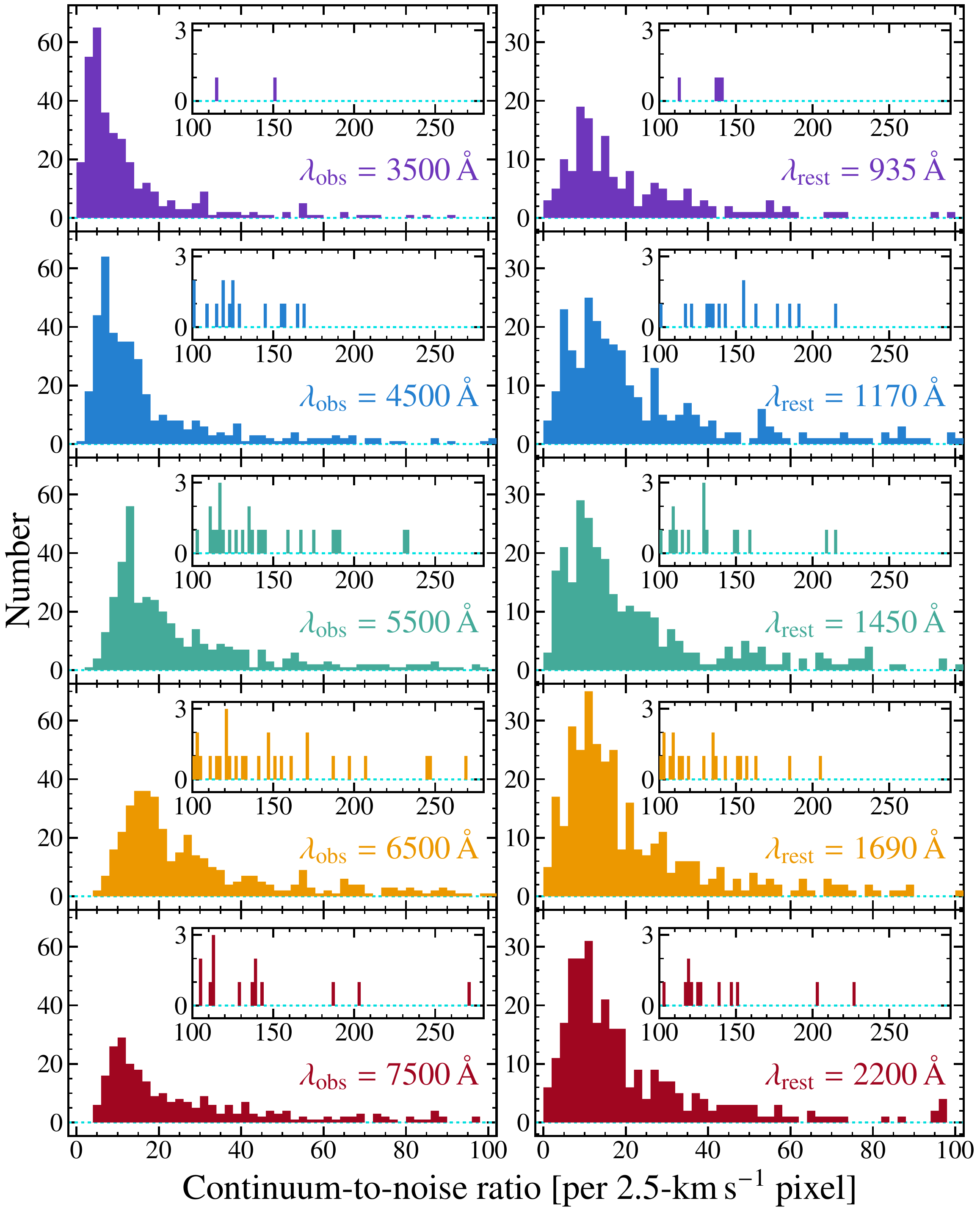}
\vspace{-1em}
\caption{Distribution of continuum-to-noise ratio (\CN) per 2.5-\kms\ pixel for the 467 DR1 quasars with final spectra. The left panel shows the \CN\ distributions at a series of representative observed wavelengths. The right panel shows the \CN\ distributions in the common quasar rest frame at two representative \lya\ forest wavelengths (935 and 1170\,\AA) and three relatively emission-line free wavelengths redwards of \lya. The \CN\ is measured as the median value within a 1000-\kms\ window around the nominated wavelength. The inset in each panel shows the high-\CN\ tail of each distribution. Note that a small number of spectra have even higher \CN\ at some wavelengths than the insets cover. For example, J051707$-$441055 has $\CN=342$ per 2.5-\kms\ pixel at 5500\,\AA, the highest in the DR1 sample.}
\label{f:cnr}
\end{center}
\end{figure}

Each DR1 FITS header provides the median \CN\ of the spectrum in bins of 1000\,\kms. \Tref{t:cat} presents these \CN\ values for the bins with wavelength centres closest to 3500, 4500, 5500, 6500 and 7500\,\AA. The left panel of \Fref{f:cnr} shows the \CN\ distribution for the DR1 quasars at these wavelengths. Here, for uniform treatment of quasars with different dispersions, the \CN\ in has been converted a per-2.5-\kms-pixel value for all quasars. Most DR1 quasar spectra have \CN\ in the range 5--60. However, a substantial number of spectra (26 at 5500\,\AA\ and 28 at 6500\,\AA) have $\CN>100$\,per 2.5-\kms\ pixel. Of course, the \CN\ in the most sensitive region of the spectrograph ($\sim$5500--6500) has a non-zero minimum, but the minimum extends to essentially zero for the bluer regions, particularly at wavelengths $\sim$3500\,\AA. The right panel of \Fref{f:cnr} shows the \CN\ distributions at five wavelengths in the common quasar rest-frame which characterise the data quality in the \lya\ forest near the Lyman limit (935\,\AA) and \lya\ emission line (1170\,\AA), and in three regions redwards of \lya\ -- 1450, 1690, 2200\,\AA\ -- which are known to be relatively free of quasar emission lines \citep[e.g.][]{vandenBerk:2001:549,Murphy:2016:1043}

\subsubsection{Nominal resolving power ($R_{\rm nom}$)}\label{sss:Rnom}

$R_{\rm nom}$ is the mean resolving power of the contributing exposures, in 1000\,\kms-wide bins, determined from their slit widths, assuming the slit is uniformly illuminated. For quasar exposures where the seeing was similar to, or smaller than, the slit width, the real resolving power will be somewhat higher than the nominal value; for example, \citet{Kotus:2017:3679} found that an $\approx$10\% increase in resolving power is typically expected. Each DR1 FITS header provides $R_{\rm nom}$ for all bins (within the wavelength coverage of the spectrum), while \Tref{t:cat} includes values only for the bins with wavelength centres closest to 3500, 4500, 5500, 6500 and 7500\,\AA. $R_{\rm nom}$ was modelled as a second-order polynomial of slit width, $d_{\rm slit}$, in arcseconds: $R_{\rm nom}=a_0+a_1d_{\rm slit}+a_2d_{\rm slit}^2$. The polynomial coefficients, $a_i$, shown in \Tref{t:nom_resol} were derived by fitting the resolving power against slit width of the ThAr exposures in ESO's UVES quality control database\footnote{See \urlstyle{rm}\url{http://archive.eso.org/bin/qc1_cgi}} for years 2010--2016. Seperate sets of coefficients were derived for the blue arm (from the 390-nm setting) and red arm (580-nm setting), and for unbinned and 2$\times$2-binned ThAr exposures (with 0.4--1.2 and 0.8--1.4\,arcsecond slit widths, respectively). The two CCD chips in the red arm were found to have very similar resolving powers in all cases, so they have been treated together and assigned the same coefficients. Some or all of the contributing exposures to 11 DR1 quasars were binned 3$\times$2; however, their nominal resolving powers were assumed to be the same as for 2$\times$2-binned exposures.

\begin{table}
\caption{Polynomial coefficients used to determine the nominal resolving power (i.e.\ for a uniformly illuminated slit), as a function of slit width: $R_{\rm nom}=a_0+a_1d_{\rm slit}+a_2d_{\rm slit}^2$ for $d_{\rm slit}$ in arcseconds. See text in \Sref{s:database} for details.}
\begin{center}
\label{t:nom_resol}
\begin{tabular}{lcccc}
\hline
 \multicolumn{1}{c}{Arm} & Binning    & $a_0$  & $a_1$ & $a_2$   \\
\hline
 Blue                    & None       & 10033  & 63237 & $-$24986  \\
 Blue                    & 2$\times$2 & 22011  & 50563 & $-$22803  \\
 Red                     & None       & 8533.3 & 52709 & $-$16005  \\
 Red                     & 2$\times$2 & 28846  & 28505 & $-$9533.3 \\
\hline
\end{tabular}
\end{center}
\end{table}

\subsection{Remaining artefacts, systematic effects and limitations}\label{ss:artefacts}

While we invested significant effort to ensure that at least a minimum standard of cleaning and quality control was applied to each spectrum (see \Sref{ss:poplerDR1}), there are numerous remaining artefacts in all DR1 spectra. We discuss the most prominent of these below. It is important that users of the DR1 spectra consider the effect these remaining artefacts may have on their analyses; we expect that most statistical analyses of the spectra will be somewhat sensitive to one or more of these artefacts. Users are encouraged to discuss these and other potential (or more subtle) systematic effects with the corresponding author (MTM).

\subsubsection{Continuum errors}\label{sss:art_cont}

\begin{figure}
\begin{center}
\includegraphics[width=0.90\columnwidth]{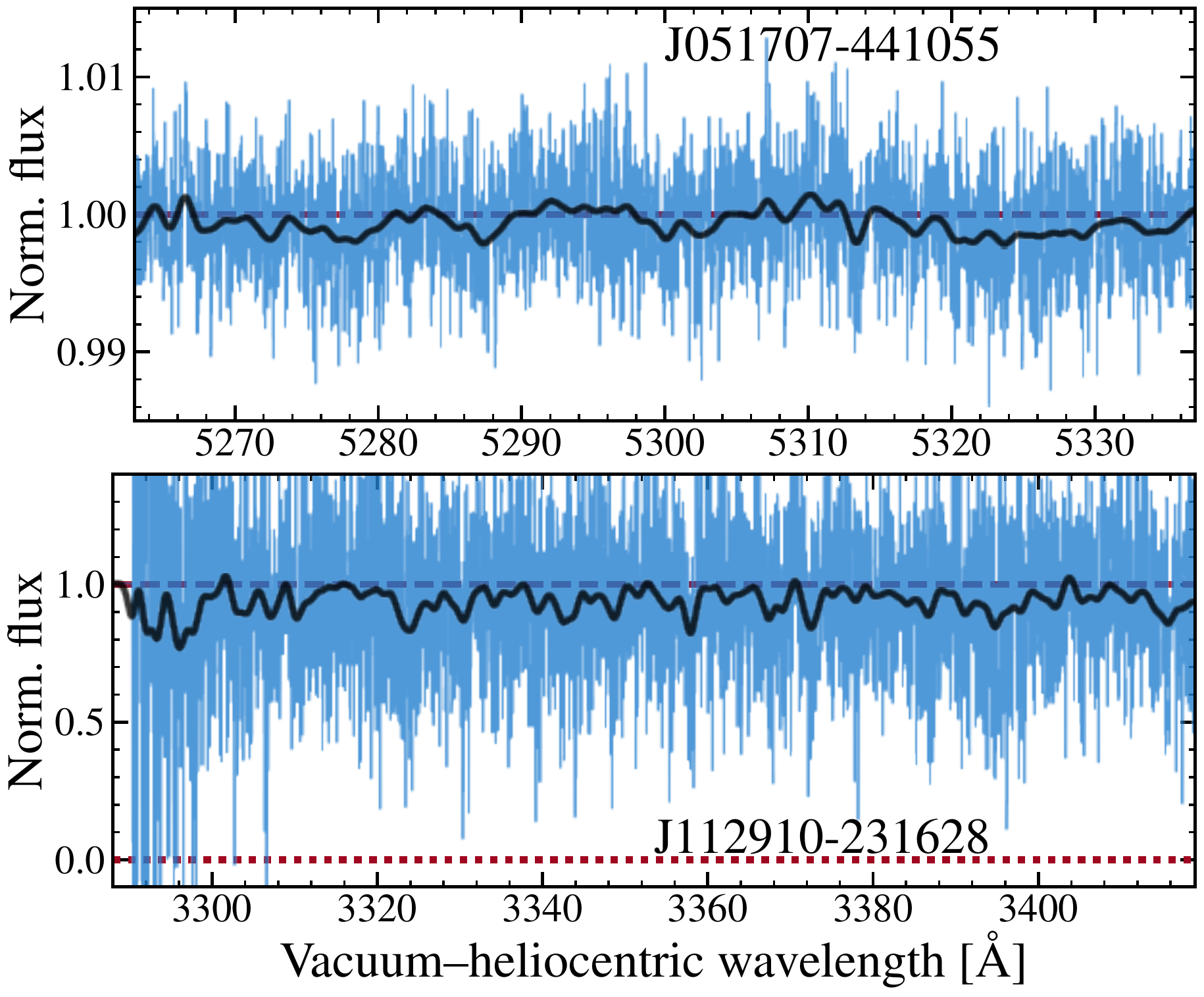}
\vspace{-1em}
\caption{Examples of overestimated automatic continuum level (redwards of the \lya\ emission line). The top panel shows a very high \SN\ region while the bottom panel shows a very low \SN\ region, i.e.\ the extreme blue end of the 390-nm standard wavelength setting. The spectra (blue/grey) have been smoothed (black) by a Gaussian filter (with a 20-pixel kernel) to aid the eye in discerning the bias in the continuum level. In both cases, the continuum is generally overestimated by up to $\sim$0.25 times the flux RMS. That is, the magnitude of the bias scales with the \SN; analyses of lower-\SN\ DR1 spectra are more likely to be affected by this bias.}
\label{f:cont_overest}
\end{center}
\end{figure}

As discussed in \Sref{ss:popler_summary} and \Sref{sss:continuum}, the continuum in most regions redwards of the \lya\ emission line is automatically fit, while it is manually fit in the \lya\ forest. In addition to the inherent inaccuracy in defining a \lya\ forest continuum (see \Sref{sss:continuum}), these two approaches cause two significant systematic continuum errors in the DR1 spectra:
\begin{enumerate}
\item Overestimated automatic continuum fit: Even in (by-eye) completely unabsorbed regions redwards of the \lya\ emission line, the automatic continuum level is generally slightly overestimated. The mean continuum-normalised flux in such regions lies 0.05--0.25$\bar{\sigma}$ below unity, for $\bar{\sigma}$ the mean normalised flux uncertainty. This continuum overestimate is caused by the asymmetric prejudice built in to our iterative continuum fitting algorithm (see \Sref{ss:popler_summary}): absorption features are far more common than (spurious) emission features (e.g.\ cosmic rays), so the algorithm rejects pixels with flux (default values) $>$3.0$\sigma$ above and $>$1.4$\sigma$ below the current fit for the next iteration. For high \SN\ DR1 spectra, this effect is clearly a small fraction of the continuum level and therefore unlikely to cause problems for most analyses. The upper panel of \Fref{f:cont_overest} shows an example. However, the overestimate will be a larger proportion of the continuum level for lower-\SN\ regions of all DR1 spectra; for example, the lower \SN\ at the blue extreme of many spectra causes a noticeable overestimate, as illustrated in the lower panel of \Fref{f:cont_overest}.
\item Redshift-dependent bias in manual continuum fits of the \lya\ forest: As described in \Sref{sss:continuum}, our approach to manually fitting a continuum in the \lya\ forest involved visually identifying ``seemingly unabsorbed peaks'' and interpolating between them with the same iterative polynomial fitting algorithm used redwards of the \lya\ emission line. Upon very close inspection of the DR1 spectra, there appear to be convincingly unabsorbed continuum regions of the \lya\ forest at $z\la3$; i.e.\ the lower number density of forest lines leaves more truly unabsorbed regions to fit. Therefore, we expect at least the same overestimation of the continuum in the $z\la3$ \lya\ forest as described in point (i) above. However, the increasing \lya\ forest line density means that few (or no) truly unabsorbed regions exist at higher redshifts, so our fits will underestimate the true continuum. While it is likely to increase with redshift, it is difficult to estimate the magnitude of this bias. However, we refer users of DR1 spectra to the study by \citet{Becker:2007:72} as a guide: using a similar continuum fitting approach, fitting theoretical models of the \lya\ flux probability distribution required upward adjustments to their continuum levels by $\la$3\% at $z\sim3$, and by $\sim$7--17\% at $z\sim4$--5.
\end{enumerate}

Finally, the incidence of Lyman-limit systems (where there is significant remaining flux bluewards of the limit) and broad absorption line (BAL) features makes the definition of the continuum dependent on the scientific question being addressed: for example, what constitutes the continuum is very different when studying a Lyman limit or the \lya\ forest bluewards of it. Generally, in such cases, we attempted to fit or interpolate a continuum that would be useful for most users. However, we urge those studying Lyman-limit systems and BALs in DR1 spectra to redefine the continuum placement accordingly.

\subsubsection{Telluric features}\label{sss:telluric}

\begin{figure}
\begin{center}
\includegraphics[width=0.90\columnwidth]{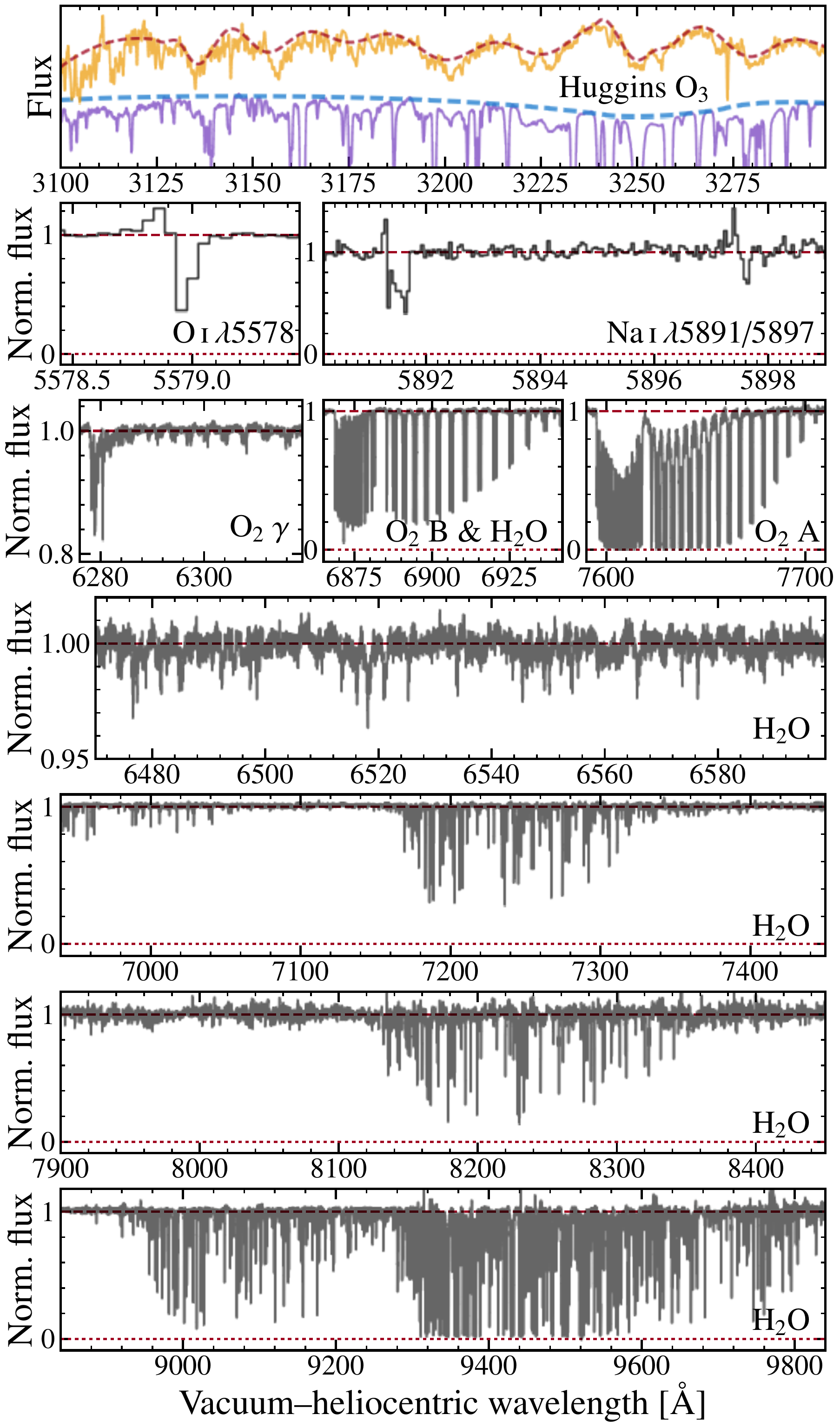}
\vspace{-1em}
\caption{Common, strong telluric features in the DR1 quasar spectra. The top panel shows two spectra (J051707$-$441055: lower, purple; J083052$+$241059: upper, orange; both are Gaussian-smoothed with a 6-pixel kernel for clarity) affected by the broad Huggins O$_3$ bands at $\la$3300\,\AA. This is easily seen in the upper spectrum which does not cover the \lya\ forest: our undulating continuum fit (dashed line) largely matches the band shapes (though not completely; e.g.\ the broad dips below the continuum fit near $\approx$3156, 3202 and 3225\,\AA). These bands are not fitted in the lower spectrum because it covers the \lya\ forest. The second row of panels show poorly subtracted telluric emission lines: \ion{O}{i}\,$\lambda$5578 in J124913$-$055919 and \ion{Na}{i}\,$\lambda$5891/5897 in J001306$+$000431). The other panels show strong molecular absorption bands in J051707$-$441055.}
\label{f:telluric}
\end{center}
\end{figure}

\Fref{f:telluric} illustrates the strongest common telluric features in the DR1 spectra. In general, no attempt was made to remove telluric features in individual spectra. While, in principle, variation of the heliocentric velocity among a large number of exposures, plus our iterative removal of outlying data when combining exposures, can remove many telluric features from the final spectrum, these criteria are rarely met in practice. This means that most DR1 spectra contain many telluric absorption lines -- particularly the O$_2$ $\gamma$, B and A bands ($\sim$6300, 6880 and 7620\,\AA, respectively) and the H$_2$O bands at $\approx$6470--6600, 6830--7450, 7820--8620 and 8780--10000\,\AA\ -- and residuals from imperfectly subtracted sky emission lines, particularly at $\ga$7000\,\AA, the strong \ion{O}{i}\,$\lambda$5578 sky emission line and the Na{\sc \,i}\,$\lambda$5891/5897 doublet. In the few spectra with high \SN\ at $\la$3400\,\AA, the broad Huggins ozone bands are visible. If these bands occurred redwards of the \lya\ forest (i.e.\ for quasars at $\zem\la1.5$), they were reasonably well fit by our continuum-fitting process, but our fits are not likely to reflect the complex shapes of individual bands. However, for higher redshift quasars, our continuum fitting approach generally ignored the Huggins bands; users of such spectra should be cautious of the continuum fit and corresponding normalised flux spectrum below $\approx$3300\,\AA.

\subsubsection{Cosmic rays and bad pixels}\label{sss:cosmic}

\begin{figure}
\begin{center}
\includegraphics[width=0.90\columnwidth]{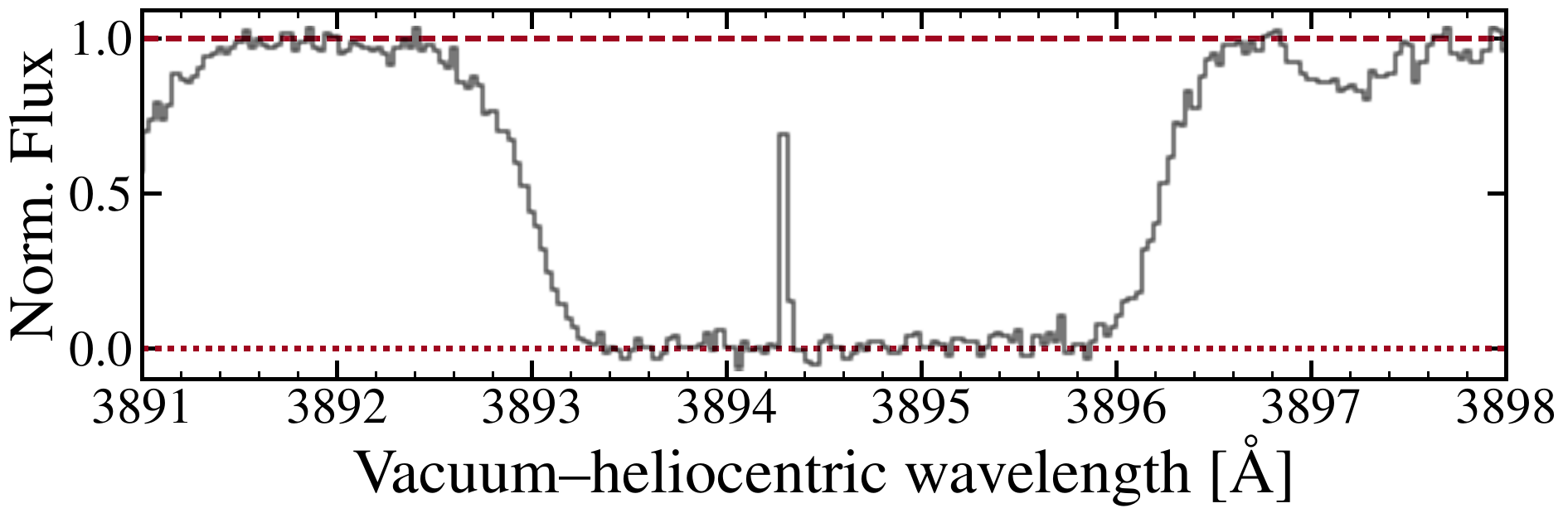}
\vspace{-1em}
\caption{Example of a `cosmic ray' (at 3894.2\,\AA) in the saturated core of \lya\ forest line in the DR1 spectrum of J033108$-$252443. \popler\ does not remove such artefacts because similar, real spectral features are often seen in saturated metal lines, and our manual cleaning process did not attempt to identify and remove all such artefacts.}
\label{f:cosmic}
\end{center}
\end{figure}

The {\sc cpl} data reduction suite, and the post-reduction processing and combination of exposures in \popler, both attempt to identify and mask `cosmic rays' and bad pixels. However, many remain unidentified and, at least, imperfectly removed from individual exposures. DR1 spectra with fewer contributing exposures therefore contain many, generally narrow ($\la$5 pixels wide) remaining cosmic rays and bad pixel residuals. These are much less common in spectra with more than $\sim$5 contributing exposures. Nevertheless, even in such cases, users should be cautious of residual cosmic ray and bad pixel artefacts in deep absorption lines: as \Fref{f:cosmic} illustrates \popler\ does not remove sharp, positive flux spikes in regions of low local relative flux because these can be real velocity structure in metal-line absorption systems. These were generally not removed in our manual cleaning process.

\subsubsection{Unidentified absorption artefacts}\label{sss:unidentified}

During the manual cleaning process, several features were noticed in many spectra that, in some cases, were clearly not due to real absorption systems: they had slightly different positions and shapes in different exposures of the same quasar, similar to the artefacts from bad rows of CCD pixels discussed in \Sref{sss:artefacts}, but much narrower and weaker in general. When these features were clearly spurious they were manually removed. However, in many spectra, the author cleaning the spectrum either did not notice these artefacts -- they vary in strength considerably from quasar to quasar, and can be weak (or apparently absent) and not obvious to visual inspection -- or there was not enough evidence to confirm they were not real absorption lines (e.g.\ there was not a significant difference between the features in different exposures). Upon completion of the cleaning of all DR1 spectra, it was clear that similar features were often found at similar wavelengths in different quasar spectra, confirming their spurious origin. However, their origin is not currently clear. Simple checks for bad pixel runs, ThAr remnants and flat-field features did not reveal a clear cause.

To more systematically reveal these unidentified features, and other common remaining artefacts, we combined the final spectra of the 131 DR1 quasars at $z<1.5$. The redshift criterion ensures that the composite is not contaminated by the \lya\ forest. The spectra were redispersed onto a common vacuum--heliocentric wavelength scale with 2.5-\kms\,per pixel dispersion and combined using a clipped mean for each pixel. A contributing pixel with flux more than 3$\sigma$ below, or 4$\sigma$ above, the mean was removed ($\sigma$ is its flux uncertainty) to avoid real absorption lines or sky-line emission residuals and reveal features common to many spectra.

\begin{figure}
\begin{center}
\includegraphics[width=0.90\columnwidth]{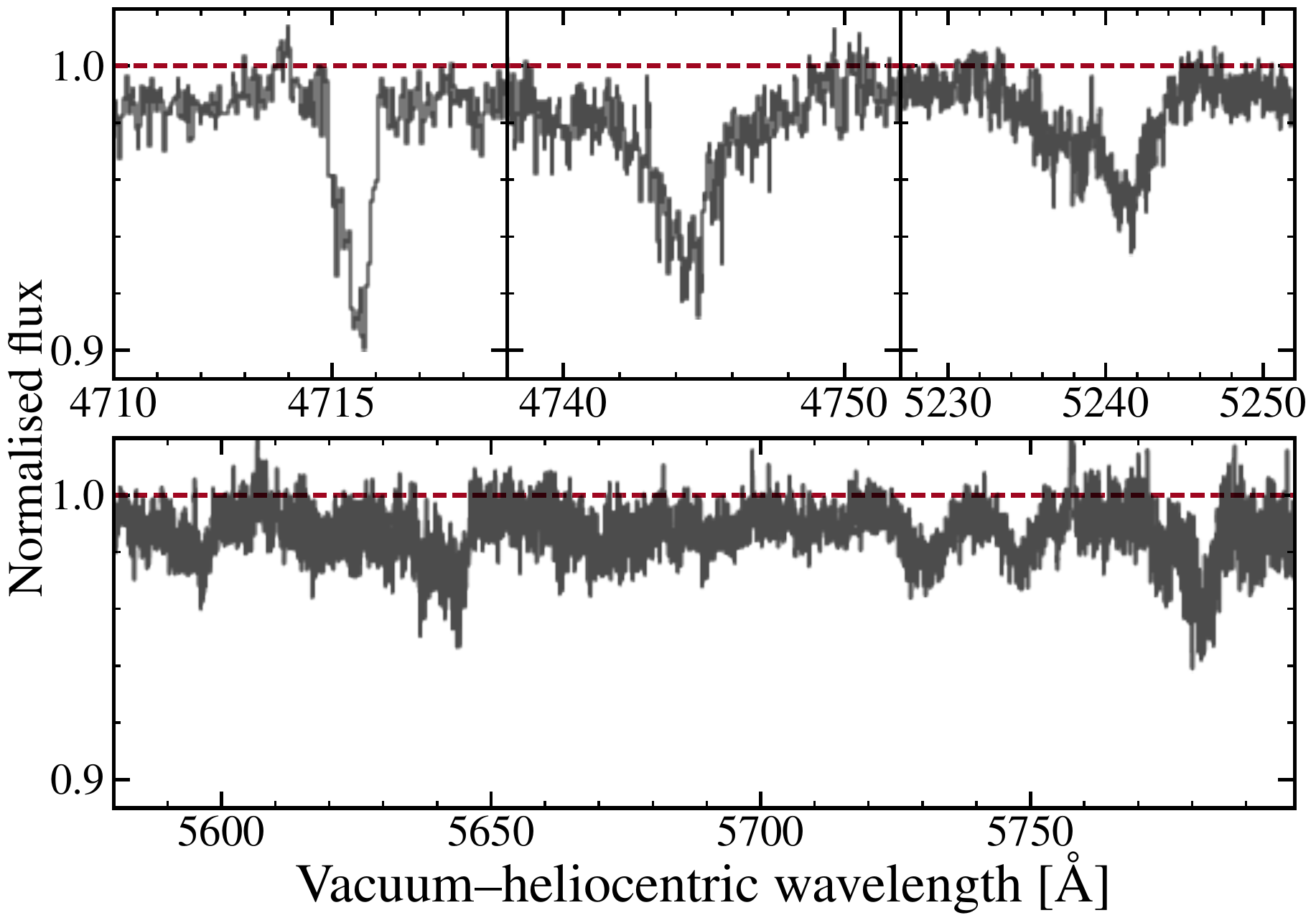}
\vspace{-1em}
\caption{Main unidentified artefacts revealed by the clipped mean composite spectrum of $\zem<1.5$ DR1 quasars. These features are detectable in many, but certainly not all, individual DR1 spectra. In some cases they were removed in the manual cleaning process, but not in most cases, as the composite spectrum shows.}
\label{f:unidentified}
\end{center}
\end{figure}

\Fref{f:unidentified} shows the main unidentified artefacts revealed by the composite spectrum at 4716, 4744, 5240, and 5580--5800\,\AA. The width and shape of the composite features reflects those found in individual spectra. However, they do seem to appear at slightly different (vacuum--heliocentric) wavelengths in individual spectra, so the composite features may be somewhat broadened. The composite spectrum also reveals many weaker features. We provide the clipped mean composite DR1 spectrum in \citet{Murphy:2018:UVESSQUADDR1} so that users can utilise it directly to identify and mask spurious spectral features that may affect their absorption line surveys.

\subsubsection{Underestimated uncertainties at low flux levels}\label{sss:low_flux}

\begin{figure}
\begin{center}
\includegraphics[width=0.90\columnwidth]{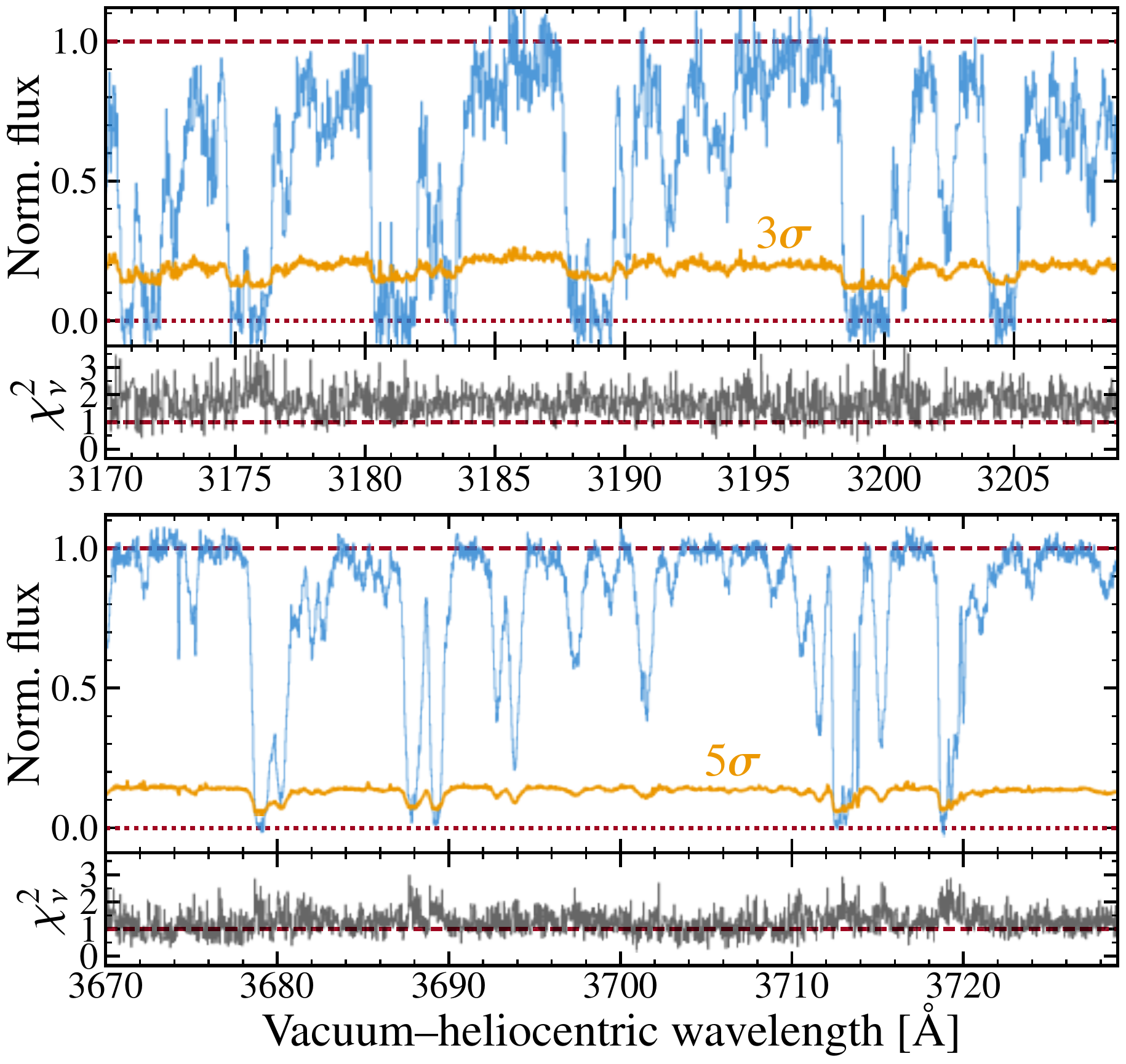}
\vspace{-1em}
\caption{Example of the underestimated uncertainty spectrum in regions of low flux. Top two panels: Final DR1 spectrum (blue, dark) of J033108$-$252443 in a region where all 14 contributing exposures have low \SN\ ($\la$5\,per pixel) and, therefore, where the uncertainty spectrum (orange, light; shown as 3$\sigma$ for clarity) is dominated by sky photon noise; this is also indicated by its low contrast between areas of unabsorbed continuum and saturated lines. The corresponding $\chi^2_\nu$ spectrum ($\chi^2$ per degree of freedom of the flux in the contributing exposures around the weighted mean flux) is generally higher than expected -- i.e.\ $\approx$2 (c.f.\ expectation of $\approx$1) -- demonstrating that the uncertainty spectrum is underestimated. Lower two panels: Same as upper two panels but for a higher \SN\ region. Here, the quasar photon noise dominates and $\chi^2_\nu$ is $\approx$1 in unabsorbed regions. However, in strong or saturated lines, where the sky photon noise dominates, $\chi^2_\nu$ increases to $\approx$2. This indicates that the sky noise contribution is underestimated.}
\label{f:unc_underestimate}
\end{center}
\end{figure}

Common to all DR1 spectra is that the flux uncertainty arrays of individual {\sc cpl}-reduced exposures are underestimated when the quasar flux is low. This is easily noticed in \popler\ as peaks in the $\chi^2_\nu$ spectrum (see \Sref{sss:artefacts}) in strong and, especially, saturated absorption lines, or where the \SN\ of individual exposures is $\la$5\,per pixel. It is therefore most noticeable in the \lya\ forest. \Fref{f:unc_underestimate} illustrates two examples in the \lya\ forest of one DR1 quasar spectrum. $\chi^2_\nu$ typically reaches $\sim$2 in such regions, indicating that the uncertainty array is underestimated by a factor of $\sim$1.4. However, this factor depends on the \SN\ of individual exposures: it tends to be larger for lower \SN\ exposures. We suspect that the {\sc cpl} reduction pipeline underestimates the noise contribution from the sky flux during the optimal extraction. This problem also existed in the previous, {\sc eso-midas} data reduction code for UVES. To compensate some of its effects on absorption line studies, some authors have increased the flux uncertainty estimate in the cores of deep/saturated absorption lines in UVES spectra \citep[e.g.][]{King:2012:3370}.

\subsubsection{Bad data in individual exposures}\label{sss:bad_data}

As noted in \Sref{sss:artefacts}, our approach for removing bad data from contributing exposures was to do so when they affected an obvious absorption feature. Artefacts from remaining bad data may still affect, or even mask, very weak absorption features that were not noticed by eye in the manual cleaning process. Users aiming to detect weak absorption features in the individual DR1 spectra are advised to inspect the flux, uncertainty and $\chi^2_\nu$ spectra -- of both the combined spectra and their contributing exposures -- in detail. Indeed, \popler\ was specifically designed to display these details to allow such specific quality control steps.

\subsubsection{Blaze function variations and remnants}\label{sss:blaze}

\begin{figure}
\begin{center}
\includegraphics[width=0.90\columnwidth]{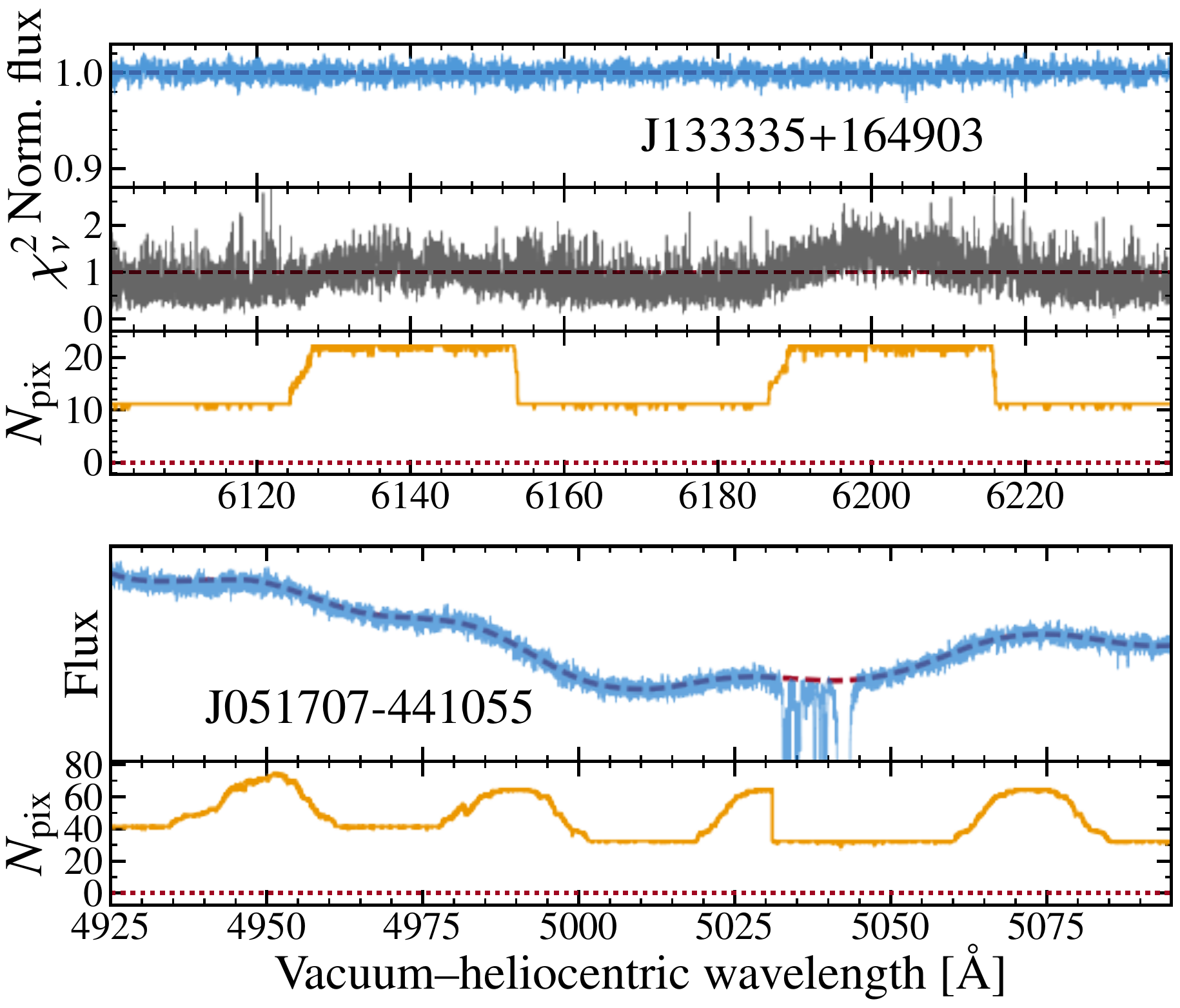}
\vspace{-1em}
\caption{Examples of artefacts remaining from the blaze function removal. Top panels: Spectra from overlapping echelle orders may have different shapes in the overlap region -- ``bends'' -- causing higher $\chi^2_\nu$ values and, potentially, spurious effects on weak absorption features in the combined spectrum. The $N_{\rm pix}$ spectrum shows the number of contributing echelle orders. The increase from $N_{\rm pix}\approx11$ to $\approx$22 indicates the overlap region of neighbouring echelle orders; the increase in $\chi^2_\nu$ in these overlap regions is due to bends. In this example, no adverse effects on the normalised flux spectrum are apparent. Lower panels: Imperfect blaze function removal leaves broad undulations in the flux spectrum that coincide with the overlap region between echelle orders (indicated by the $N_{\rm pix}$ spectrum). In this example, note the sharp drop in the $N_{\rm pix}$ spectrum at $\approx$5031\,\AA: the low-\SN\ red ends of the echelle orders covering $\approx$4990--5040\,\AA\ have been manually removed to ensure they did not affect the strong absorption line system at $\approx$5031--5045\,\AA.}
\label{f:bends}
\end{center}
\end{figure}

The blaze function for each echelle order of an exposure is approximated using the master flat field by the {\sc cpl} reduction software. This, in principle, can change from exposure to exposure as the alignment of optical elements change slightly with time or wavelength setting, or with changes in the flat field lamp spectrum (e.g.\ as it ages) or illumination pattern on the CCDs. As explained in \Sref{sss:artefacts}, these effects (and possibly others) cause ``bends'' between the spectral shapes of spectra from overlapping orders, the most obvious of which we attempted to identify and correct in the manual cleaning process. Our priority was to address this problem when it affected an obvious absorption line. However, for very weak absorption features (especially broader ones) that may not have been noticed by eye, weak bends may not have been removed. And, where obvious absorption lines were not found, noticeable bends will still be present in the DR1 spectra, particularly in high \SN\ cases. An example of remaining bends in the spectrum of the bright quasar J133335$+$164903 is shown in the upper panels of \Fref{f:bends}.
 
The correction for the blaze function also appears to be imperfect in systematic ways. Many DR1 spectra therefore contain remnants of the blaze function that appear as ripples or undulations in the flux spectrum over echelle-order scales. An example is shown in the lower panels of \Fref{f:bends}. These undulations have amplitudes $\la$5\% of the continuum level, and are usually substantially smaller. In non-\lya\ forest regions, our continuum fitting approach will largely correct for these blaze remnants, as is evident in \Fref{f:bends}. However, in \lya\ forest regions, where individual continuum fits cover a larger range of wavelengths, and where the forest obscures such broad, shallow undulations, these blaze remnants may still significantly affect the final normalised flux spectrum.

\subsubsection{Zero level errors}\label{sss:zero}

\begin{figure}
\begin{center}
\includegraphics[width=0.90\columnwidth]{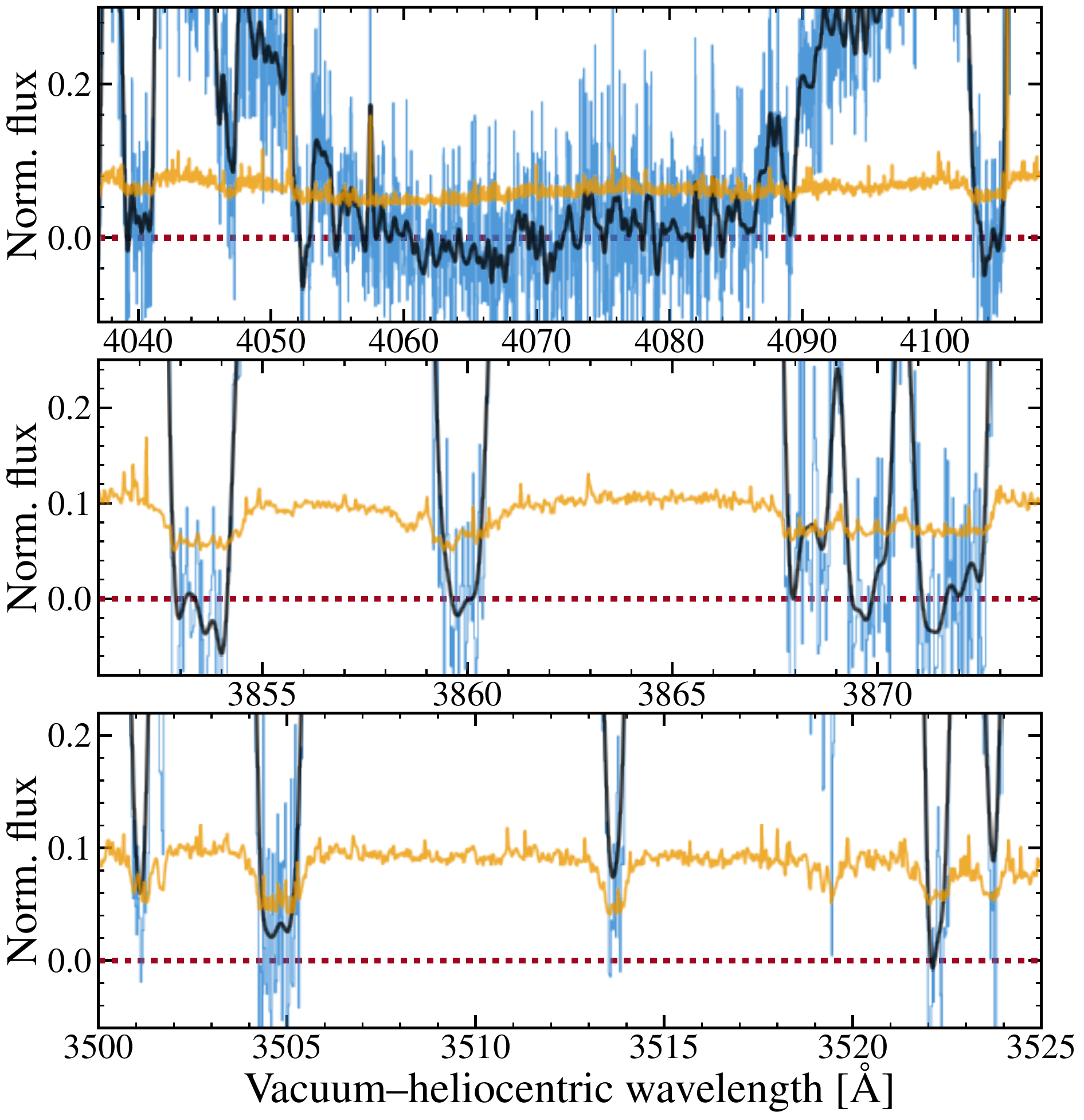}
\vspace{-1em}
\caption{Examples of zero-level errors in the DR1 spectra. The spectra (blue histograms) have been smoothed (black line) to show a running average flux, compared with the 1$\sigma$ uncertainty array (light, orange line). Top panel: A DLA trough in the spectrum of J020900$-$455026 where the zero level varies from $\approx$3\% too high ($\approx$4061--4068\,\AA) to $\approx$2\% too low ($\approx$4072--4084\,\AA) over a relatively short wavelength range. Middle panel: Saturated \lya\ forest line cores where the zero level is overestimated by $\approx$1--3\% (same spectrum as top panel). Bottom panel: Saturated \lya\ forest lines in the spectrum of J231359$-$370446 where the zero level is underestimated by $\approx$2--3\%.}
\label{f:zero}
\end{center}
\end{figure}

Inspection of {\sc cpl}-extracted UVES exposures with relatively low \SN\ often reveals imperfect zero levels: the average flux in saturated line cores is significantly different to zero. This indicates that the sky flux level is inaccurately measured in the optimal extraction process. \Fref{f:zero} shows some typical examples of this problem in DLA absorption troughs (upper panel) and saturated \lya\ forest line cores (lower two panels). As the figure illustrates, the zero-level error in some final spectra can be of order $\sim$2--4\%; in extreme cases -- exposures with very low \SN, where the trace of the quasar is not well determined -- the error can be up to $\sim$10\%. \Fref{f:zero} also shows that the zero level can be overestimated or underestimated (compare the lower two panels) and that, in some cases, it can vary from too low to too high over relatively short wavelength ranges (upper panel); however, in most cases the zero level error appears to have the same sign and not vary substantially in magnitude over much larger wavelength ranges (typically $\ga$300\,\AA). We have not attempted to correct for these zero-level errors in the DR1 spectra so users should account for them when, for example, modelling strong or saturated absorption lines (e.g. DLAs).

\subsubsection{Wavelength scale shifts and distortions}\label{sss:wave}

The wavelength calibration accuracy of UVES has been the specific focus of many quasar absorption studies, particularly those seeking to constrain possible variations in the electromagnetic fine-structure constant and proton-to-electron mass ratio (using metal-line and H$_2$ absorption, respectively). The wavelength scale is set by comparison with a ThAr lamp exposure, and several effects shift and/or distort the true quasar wavelength scale with respect to this:
\begin{enumerate}
\item Mechanical drifts and changes in the refractive index of air were designed to be compensated for by resetting the grating angles \citep{Dekker:2000:534} which, in practice, is limited to $\sim$0.1--0.2\,\kms\ accuracy;
\item Differences in alignment of the quasar in a slit between exposures, and/or between the two slits (i.e.\ between the spectrograph arms), produces (approximately) velocity-space shifts of, typically, up to $\sim$0.4\,\kms\ \citep[e.g.][]{Molaro:2013:A68,Rahmani:2013:861,Evans:2014:128,Kotus:2017:3679} and up to $\sim$2\,\kms\ in extreme cases;
\item Intra-order distortions (i.e.\ wavelength-dependent shifts on echelle order scales), that have a largely (but not entirely) repeated shape and amplitude from order to order, have amplitudes of typically $\sim$0.1\,\kms\ \citep[up to 0.4\,\kms\ in extreme cases; e.g.][]{Whitmore:2010:89,Whitmore:2015:446};
\item Long-range distortions between the science object and ThAr wavelength scales have been identified that have magnitudes of typically $\sim$0.1\,\kms\ per 1000\,\AA\ and up to $\sim$4 times larger in extreme cases \citep[e.g.][]{Rahmani:2013:861,Bagdonaite:2014:10,Whitmore:2015:446}.
\end{enumerate}
The magnitude, sign and shape of the latter two distortion effects is quite variable, and can change substantially over $\sim$1--3 day periods.

In general, the individual exposures and final DR1 spectra are not corrected for the above effects. However, they are relatively small -- typically $\la$20\,\% of a (unbinned) pixel -- so are not likely to significantly affect most applications. A small number of DR1 spectra have been corrected using asteroid or iodine-cell stellar observations \citep[e.g.][]{Evans:2014:128}, solar twin stars \citep[e.g.][]{Dapra:2015:489,Dapra:2016:192} or spectra of the same object on better-calibrated spectrographs \citep[e.g.][]{Kotus:2017:3679}. The UPL files include these corrections in such cases.

\section{Scientific uses}\label{s:use}

As described in \Sref{s:intro}, numerous scientific questions can be addressed with the DR1 quasar spectra. In this section we seek to highlight and assist the large-scale statistical studies of quasar absorption systems that are possible with such a large sample of high-resolution spectra. Specifically, we illustrate how the DR1 spectra will be useful for detailed DLA studies, absorption line surveys, and studies of time-variable absorption lines.

\subsection{Damped \lya\ system studies}\label{ss:dlas}

DLAs contain most of the neutral hydrogen in the universe at all epochs currently probed, from $z\sim5$ down to 0 \citep[see review by][]{Wolfe:2005:861}. Their high \ion{H}{i} column densities -- $\NHI\ge2\times10^{20}$\,\pcmsq, by definition -- shield the gas from ionising radiation, allowing it to remain highly neutral, presumably making DLA gas available for later star formation. DLAs also contain a large proportion of the universe's metals; studying their chemical abundances and metallicities -- and how these evolve with redshift -- are therefore important elements in understanding galaxy formation and evolution. DLA metal abundances and metallicities can be very accurately measured, owing to the simple relationship between optical depth and column density, and the neutrality of the DLA gas (i.e.\ no corrections for ionised hydrogen or metals are generally required). The most accurate and precise DLA metal-line measurements are possible in high-resolution spectra because the metal line velocity structures can be resolved.

For these reasons, the DR1 quasar spectra offer an excellent opportunity for detailed studies of a large sample of DLAs. To assist such work, we have identified 155 DLAs towards the 467 DR1 quasars with final spectra and catalogued them in \Tref{t:dla}. While 137 of the DLAs in \Tref{t:dla} have previously been reported in the literature, the other 18 are reported here for the first time (to our knowledge).

\setlength{\tabcolsep}{0.29em}
\begin{table}
  \caption{The DR1 sample of 155 DLAs. 132 DR1 quasar spectra have previously reported DLAs (i.e.\ $\NHI \ge 2\times10^{20}$\,\pcmsq; references 1--22 in the final column) where the \lya\ transition is detected in the DR1 spectrum. 18 new DLAs are reported below (reference 23 in the final column) for which we have measured \NHI\ from the DR1 spectra.}
\label{t:dla}
\begin{center}
{\footnotesize 
\begin{tabular}{lccccr}
\hline
DR1 Name & $z_{\rm em,Adopt}$ & $\zab$ & $\log(N_\textsc{h\scriptsize{\,i}}/\textrm{cm}^{-2})$ & $\sigma(\log N_\textsc{h\scriptsize{\,i}})$ & Ref.$^{\rm a}$ \\
\hline
 J000149$-$015939 &      2.815 &                    2.095 &                   20.65 &               0.10 &                20 \\
 J000149$-$015939 &      2.815 &                    2.154 &                   20.3  &               0.1  &                20 \\
 J000443$-$555044 &      2.100 &                    1.886 &                    20.4 &                0.1 &                23 \\
 J000651$-$620803 &      4.455 &                    2.970 &                   20.7  &               0.2  &                20 \\
 J000651$-$620803 &      4.455 &                    3.202 &                   20.8  &               0.1  &                20 \\
 J000815$-$095854 &      1.955 &                    1.768 &                   20.85 &               0.15 &                20 \\
 J001306$+$000431 &      2.164 &                    2.025 &                   20.95 &               0.10 &                20 \\
 J001602$-$001225 &      2.085 &                    1.973 &                   20.83 &               0.05 &                20 \\
 J003023$-$181956 &      2.550 &                    2.402 &                   21.75 &               0.10 &                20 \\
 J004054$-$091526 &      4.976 &                    4.740 &                   20.39 &               0.11 &                20 \\
 J004131$-$493611 &      3.240 &                    2.248 &                   20.46 &               0.13 &                20 \\
 J004216$-$333754 &      2.480 &                    2.224 &                    20.6 &                0.1 &                20 \\
 J004435$-$262259 &      2.980 &                    2.549 &                    21.5 &                0.2 &                23 \\
 J004508$-$291432 &      2.388 &                    1.809 &                   20.4  &               0.1  &                20 \\
 J004508$-$291432 &      2.388 &                    1.936 &                   20.5  &               0.1  &                20 \\
 J005127$-$280433 &      2.256 &                    2.071 &                   20.45 &               0.10 &                12 \\
 J010104$-$285801 &      3.070 &                    2.671 &                    21.1 &                0.1 &                20 \\
 J010311$+$131617 &      2.705 &                    2.309 &                   21.35 &               0.08 &                20 \\
 J010516$-$184642 &      3.025 &                    2.370 &                   21.00 &               0.08 &                20 \\
 J011453$+$031457 &      2.810 &                    2.423 &                   20.90 &               0.10 &                10 \\
 J011504$-$302514 &      2.985 &                    2.418 &                   20.50 &               0.08 &                20 \\
 J011504$-$302514 &      2.985 &                    2.702 &                   20.3  &               0.1  &                20 \\
 J012517$-$001828 &      2.274 &                    1.761 &                   20.78 &               0.07 &                20 \\
 J012550$-$535225 &      3.180 &                    2.837 &                    21.2 &                0.1 &                23 \\
 J013340$+$040059 &      4.172 &                    3.692 &                   20.68 &               0.15 &                20 \\
 J013340$+$040059 &      4.172 &                    3.773 &                   20.42 &               0.10 &                20 \\
 J013754$-$270736 &      3.210 &                    2.107 &                   20.30 &               0.15 &                10 \\
 J013754$-$270736 &      3.210 &                    2.800 &                   21.00 &               0.10 &                10 \\
 J013901$-$082444 &      3.013 &                    2.677 &                   20.70 &               0.15 &                20 \\
 J014049$-$083942 &      3.713 &                    3.696 &                   20.75 &               0.15 &                17 \\
 J014214$+$002324 &      3.370 &                    3.348 &                   20.38 &               0.05 &                17 \\
 J020900$-$455026 &      2.520 &                    2.349 &                    21.0 &                0.1 &                23 \\
 J020944$+$051713 &      4.184 &                    3.666 &                   20.47 &               0.10 &                20 \\
 J020944$+$051713 &      4.184 &                    3.863 &                   20.43 &               0.15 &                20 \\
 J021741$-$370059 &      2.910 &                    2.429 &                   20.62 &               0.08 &                20 \\
 J021741$-$370059 &      2.910 &                    2.514 &                   20.46 &               0.09 &                20 \\
 J021857$+$081727 &      2.991 &                    2.293 &                   20.45 &               0.16 &                20 \\
 J024449$-$290449 &      3.230 &                    2.560 &                    20.8 &                0.2 &                23 \\
 J025240$-$553832 &      2.370 &                    2.340 &                    20.6 &                0.1 &                23 \\
 J025634$-$401300 &      2.290 &                    2.046 &                   20.45 &               0.08 &                20 \\
 J030211$-$314030 &      2.370 &                    2.179 &                    20.8 &                0.1 &                12 \\
 J030722$-$494548 &      4.728 &                    4.466 &                   20.67 &               0.09 &                 4 \\
 J032412$-$320259 &      3.302 &                    2.243 &                    20.5 &                0.1 &                23 \\
 J033025$-$495403 &      2.230 &                    1.893 &                    21.2 &                0.2 &                23 \\
 J033413$-$161205 &      4.363 &                    3.557 &                   21.12 &               0.15 &                 8 \\
 J033854$-$000521 &      3.050 &                    2.230 &                   21.05 &               0.25 &                20 \\
 J033900$-$013317 &      3.197 &                    3.062 &                   21.20 &               0.09 &                20 \\
 J034943$-$381030 &      3.205 &                    3.025 &                   20.73 &               0.05 &                20 \\
 J040718$-$441013 &      3.000 &                    1.913 &                   20.8  &               0.1  &                20 \\
 J040718$-$441013 &      3.000 &                    2.551 &                   21.15 &               0.15 &                20 \\
 J040718$-$441013 &      3.000 &                    2.595 &                   21.05 &               0.10 &                20 \\
 J040718$-$441013 &      3.000 &                    2.621 &                   20.45 &               0.10 &                20 \\
 J041656$-$284340 &      2.090 &                    1.719 &                    21.2 &                0.2 &                23 \\
 J042353$-$261801 &      2.277 &                    2.157 &                   20.65 &               0.10 &                12 \\
 J042644$-$520819 &      2.250 &                    2.224 &                    20.3 &                0.1 &                12 \\
 J043255$-$355030 &      2.280 &                    1.961 &                    20.7 &                0.1 &                23 \\
 J043403$-$435547 &      2.649 &                    2.302 &                   20.78 &               0.10 &                20 \\
 J044017$-$433308 &      2.863 &                    2.347 &                   20.78 &               0.12 &                20 \\
 J044534$-$354704 &      2.610 &                    2.408 &                    20.3 &                0.1 &                23 \\
 J045313$-$130555 &      2.300 &                    2.067 &                   20.50 &               0.07 &                20 \\
\end{tabular}
}
\end{center}
\end{table}

\begin{table}
  \contcaption{$\!\!.$ The DR1 sample of 155 DLAs.}
\begin{center}
{\footnotesize 
\begin{tabular}{lccccr}
\hline
DR1 Name & $z_{\rm em,Adopt}$ & $\zab$ & $\log(N_\textsc{h\scriptsize{\,i}}/\textrm{cm}^{-2})$ & $\sigma(\log N_\textsc{h\scriptsize{\,i}})$ & Ref. \\
\hline
 J050112$-$015914 &      2.286 &                    2.040 &                    21.7 &                0.1 &                20 \\
 J053007$-$250329 &      2.813 &                    2.141 &                   20.95 &               0.05 &                20 \\
 J053007$-$250329 &      2.813 &                    2.811 &                   21.35 &               0.07 &                20 \\
 J055246$-$363727 &      2.317 &                    1.962 &                   20.70 &               0.08 &                20 \\
 J060008$-$504036 &      3.130 &                    2.149 &                   20.40 &               0.12 &                20 \\
 J064326$-$504112 &      3.090 &                    2.659 &                   20.95 &               0.08 &                20 \\
 J080916$+$053941 &      2.555 &                    2.319 &                   20.39 &               0.22 &                15 \\
 J081634$+$144612 &      3.846 &                    3.287 &                    22.0 &                0.1 &                19 \\
 J082003$+$155932 &      1.954 &                    1.926 &                    21.0 &                0.2 &                23 \\
 J083932$+$111206 &      2.671 &                    2.465 &                   20.58 &               0.10 &                 5 \\
 J084424$+$124546 &      2.496 &                    1.864 &                   21.0  &               0.1  &                20 \\
 J084424$+$124546 &      2.496 &                    2.375 &                   21.05 &               0.10 &                20 \\
 J084424$+$124546 &      2.496 &                    2.476 &                   20.8  &               0.1  &                20 \\
 J091613$+$070224 &      2.786 &                    2.618 &                   20.35 &               0.10 &                20 \\
 J093509$-$333237 &      2.906 &                    2.682 &                    20.5 &                0.1 &                20 \\
 J094008$+$023209 &      3.218 &                    2.565 &                   20.63 &               0.05 &                 9 \\
 J094438$+$194111 &      3.186 &                    2.655 &                   20.56 &               0.03 &                15 \\
 J095355$-$050418 &      4.369 &                    3.858 &                   20.6  &               0.1  &                20 \\
 J095355$-$050418 &      4.369 &                    4.203 &                   20.55 &               0.10 &                20 \\
 J095500$-$013006 &      4.426 &                    4.024 &                   20.55 &               0.10 &                20 \\
 J103842$-$272912 &      3.090 &                    2.792 &                   20.65 &               0.13 &                20 \\
 J103909$-$231326 &      3.130 &                    2.777 &                   20.93 &               0.05 &                20 \\
 J104252$+$011736 &      2.440 &                    2.267 &                   20.75 &               0.15 &                 9 \\
 J105744$+$062914 &      3.147 &                    2.500 &                   20.55 &               0.05 &                 9 \\
 J105800$-$302455 &      2.523 &                    1.904 &                   21.54 &               0.10 &                20 \\
 J110855$+$120953 &      3.672 &                    3.396 &                   20.65 &               0.06 &                18 \\
 J111109$+$144238 &      3.100 &                    2.600 &                   21.35 &               0.15 &                16 \\
 J111113$-$080401 &      3.922 &                    3.608 &                   20.37 &               0.07 &                20 \\
 J111119$+$133603 &      3.475 &                    3.201 &                   21.20 &               0.15 &                16 \\
 J111350$-$153333 &      3.370 &                    3.265 &                   21.30 &               0.05 &                20 \\
 J112010$-$134625 &      3.958 &                    3.350 &                   20.95 &               0.10 &                10 \\
 J115122$+$020426 &      2.397 &                    1.969 &                   20.84 &               0.14 &                22 \\
 J115411$+$063427 &      2.755 &                    1.775 &                   21.30 &               0.08 &                20 \\
 J115538$+$053050 &      3.463 &                    2.608 &                   20.37 &               0.11 &                20 \\
 J115538$+$053050 &      3.463 &                    3.327 &                   21.0  &               0.1  &                20 \\
 J115944$+$011206 &      2.002 &                    1.944 &                    21.8 &                0.1 &                 1 \\
 J120523$-$074232 &      4.695 &                    4.383 &                   20.60 &               0.14 &                20 \\
 J120550$+$020131 &      2.132 &                    1.747 &                    20.4 &                0.1 &                20 \\
 J121134$+$090220 &      3.287 &                    2.584 &                    21.4 &                0.1 &                20 \\
 J121303$+$171423 &      2.569 &                    1.892 &                   20.70 &               0.08 &                20 \\
 J122040$+$092326 &      3.140 &                    3.133 &                   20.75 &               0.20 &                13 \\
 J122607$+$173649 &      2.942 &                    2.466 &                    21.4 &                0.1 &                20 \\
 J122848$-$010414 &      2.647 &                    2.263 &                   20.40 &               0.15 &                 9 \\
 J123313$-$102518 &      1.931 &                    1.931 &                   20.48 &               0.10 &                20 \\
 J123437$+$075843 &      2.578 &                    2.338 &                   20.90 &               0.08 &                20 \\
 J124020$+$145535 &      3.105 &                    3.024 &                   20.45 &               0.05 &                14 \\
 J124524$-$000938 &      2.092 &                    1.824 &                   20.45 &               0.10 &                20 \\
 J124924$-$023339 &      2.120 &                    1.781 &                   21.45 &               0.15 &                20 \\
 J125316$+$114720 &      3.285 &                    2.944 &                   20.35 &               0.15 &                16 \\
 J133335$+$164903 &      2.089 &                    1.776 &                   21.15 &               0.07 &                20 \\
 J133941$+$054822 &      2.982 &                    2.585 &                   20.45 &               0.15 &                 9 \\
 J134002$+$110630 &      2.919 &                    2.796 &                   20.95 &               0.10 &                11 \\
 J135334$-$031022 &      3.005 &                    2.560 &                   20.35 &               0.15 &                 9 \\
 J135646$-$110128 &      3.000 &                    2.501 &                   20.44 &               0.05 &                20 \\
 J135646$-$110128 &      3.000 &                    2.967 &                   20.8  &               0.1  &                20 \\
 J141217$+$091624 &      2.849 &                    2.019 &                   20.65 &               0.10 &                20 \\
 J141217$+$091624 &      2.849 &                    2.456 &                   20.53 &               0.08 &                20 \\
 J144331$+$272436 &      4.430 &                    4.224 &                   20.95 &               0.08 &                20 \\
 J145418$+$121053 &      3.252 &                    2.255 &                   20.30 &               0.10 &                23 \\
 J145418$+$121053 &      3.252 &                    2.469 &                   20.39 &               0.10 &                3  \\
 J172323$+$224357 &      4.520 &                    3.697 &                   20.35 &               0.10 &                20 \\
 J210244$-$355307 &      3.090 &                    3.083 &                   20.98 &               0.08 &                10 \\
 J213605$-$430818 &      2.420 &                    1.916 &                   20.74 &               0.09 &                20 \\
 J214159$-$441325 &      3.170 &                    2.383 &                   20.60 &               0.05 &                20 \\
 J214159$-$441325 &      3.170 &                    2.852 &                   20.98 &               0.05 &                20 \\
\end{tabular}
}
\end{center}
\end{table}

\begin{table}
  \contcaption{$\!\!.$ The DR1 sample of 155 DLAs.}
\begin{center}
{\footnotesize 
\begin{tabular}{lccccr}
\hline
DR1 Name & $z_{\rm em,Adopt}$ & $\zab$ & $\log(N_\textsc{h\scriptsize{\,i}}/\textrm{cm}^{-2})$ & $\sigma(\log N_\textsc{h\scriptsize{\,i}})$ & Ref. \\
\hline
 J215502$+$135825 &      4.256 &                    3.316 &                   20.55 &               0.15 &                 6 \\
 J220852$-$194359 &      2.558 &                    1.921 &                   20.67 &               0.05 &                20 \\
 J220852$-$194359 &      2.558 &                    2.076 &                   20.44 &               0.05 &                20 \\
 J222540$-$392436 &      2.180 &                    2.154 &                   20.85 &               0.10 &                20 \\
 J222826$-$400957 &      2.020 &                    1.965 &                   20.65 &               0.10 &                20 \\
 J223235$+$024755 &      2.150 &                    1.864 &                    20.9 &                0.1 &                20 \\
 J223408$+$000001 &      3.025 &                    2.066 &                   20.56 &               0.10 &                 2 \\
 J223953$-$055220 &      4.558 &                    4.079 &                   20.55 &               0.10 &                20 \\
 J224708$-$601545 &      3.005 &                    2.331 &                   20.65 &               0.05 &                10 \\
 J224931$-$441730 &      2.180 &                    1.893 &                    20.4 &                0.1 &                23 \\
 J225310$-$365815 &      3.200 &                    2.741 &                    20.5 &                0.1 &                23 \\
 J225719$-$100104 &      2.080 &                    1.836 &                   20.38 &               0.07 &                21 \\
 J230001$-$341319 &      2.200 &                    2.129 &                    20.5 &                0.1 &                23 \\
 J231359$-$370446 &      2.476 &                    2.182 &                   20.48 &               0.13 &                20 \\
 J231646$-$404120 &      2.448 &                    1.857 &                    20.9 &                0.1 &                 7 \\
 J231857$-$095551 &      2.665 &                    1.930 &                    21.2 &                0.2 &                23 \\
 J232115$+$142131 &      2.571 &                    2.573 &                    20.6 &                0.2 &                 9 \\
 J232128$-$105122 &      2.960 &                    1.629 &                   20.52 &               0.14 &                20 \\
 J232128$-$105122 &      2.960 &                    1.989 &                   20.68 &               0.05 &                20 \\
 J233156$-$090802 &      2.661 &                    2.143 &                   21.15 &               0.15 &                21 \\
 J233446$-$090812 &      3.317 &                    3.057 &                   20.50 &               0.07 &                20 \\
 J234403$+$034226 &      4.239 &                    3.219 &                   21.35 &               0.07 &                 7 \\
 J234451$+$343348 &      3.053 &                    2.909 &                    21.1 &                0.1 &                20 \\
 J234512$-$433814 &      3.190 &                    2.612 &                    20.6 &                0.1 &                23 \\
 J234625$+$124743 &      2.568 &                    2.574 &                   20.98 &               0.04 &                20 \\
 J234628$+$124858 &      2.515 &                    2.431 &                   20.40 &               0.07 &                20 \\
 J235057$-$005209 &      3.021 &                    2.426 &                   20.5  &               0.1  &                20 \\
 J235057$-$005209 &      3.021 &                    2.615 &                   21.30 &               0.08 &                20 \\
 J235129$-$142756 &      2.940 &                    2.279 &                   20.56 &               0.08 &                20 \\
 J235702$-$004824 &      2.998 &                    2.479 &                   20.41 &               0.08 &                20 \\
\hline
\end{tabular}
}
\end{center}
\begin{minipage}{\columnwidth}
{\footnotesize
$^{\rm a}$References:
1: \citet{Wolfe:1981:460};
2: \citet{Prochaska:1999:369};
3: \citet{Petitjean:2000:L26};
4: \citet{Dessauges-Zavadsky:2001:426};
5: \citet{Prochaska:2001:21};
6: \citet{Prochaska:2003:227};
7: \citet{Prochaska:2003:L9};
8: \citet{Peroux:2005:479};
9: \citet{Prochaska:2005:123};
10: \citet{Ledoux:2006:71};
11: \citet{Prochaska:2007:29};
12: \citet{Noterdaeme:2008:327};
13: \citet{Prochaska:2008:1002};
14: \citet{Wolfe:2008:881};
15: \citet{Noterdaeme:2009:1087};
16: \citet{Prochaska:2009:1543};
17: \citet{Ellison:2010:1435};
18: \citet{Penprase:2010:1};
19: \citet{Guimaraes:2012:147};
20: \citet{Zafar:2013:A141};
21: \citet{Ledoux:2015:A8};
22: \citet{Parks:2018:1151};
23: This work.
}
\end{minipage}
\end{table}

\setlength{\tabcolsep}{\oldtabcolsep}

All DLA candidates were identified by visually checking the final DR1 spectrum; the very broad, damped \lya\ line-shape is easily identified in such high-resolution spectra. Note that we only report DLAs where the \lya\ transition is covered by the DR1 spectrum, even when a DLA has been previously reported\footnote{For example, the following DLAs were identified in the UVES DLA compilation of \citet{Zafar:2013:A141} from previous literature, but are not covered by the DR1 spectra: $\zab=3.776$ towards J000651$-$620803; $\zab=3.385$ towards J020346$+$113445; $\zab=3.178$ towards J033755$-$120404; and $\zab=3.448$ towards J142107$-$064356.}, and we do not report systems with $\NHI<2\times10^{20}$\,\pcmsq. The literature was then searched for previous reports of each DLA (primarily using the NASA Extragalactic Database and NASA Astrophysics Data System) and \NHI\ measurements. The \NHI\ measurement from the highest resolution, highest quality spectrum is provided in \Tref{t:dla} for these 137 literature DLAs. This sample builds on the 97 DLAs (with \lya\ coverage) identified in the ESO UVES ``advanced data products'' sample of 250 quasar spectra by \citet{Zafar:2013:A141} (which contains the original DLA references). Note that we identified two absorbers \citep[included in][]{Zafar:2013:A141}, originally discovered in low-resolution spectra, that the DR1 spectra clearly show are not DLAs; they have much lower column densities than the DLA threshold: $\zab=4.060$ towards J033829$+$002156 \citep{Peroux:2001:1799} and $\zab=1.947$ towards J034957$-$210247 \citep{Ellison:2001:393}. These are excluded from the DR1 DLA catalogue. We also found the previously reported redshift to be inaccurate (by more than $\sim$200\,\kms) for three DLAs and have corrected these in \Tref{t:dla}: $\zab=2.302$ towards J043403$-$435547 \citep{Ellison:2001:393}; $\zab=2.338$ towards J123437$+$075843 \citep{Zafar:2013:A141}; and $\zab=2.574$ towards J234625$+$124743 \citep{Zafar:2013:A141}.

\begin{figure*}
\begin{center}
\includegraphics[width=0.75\textwidth]{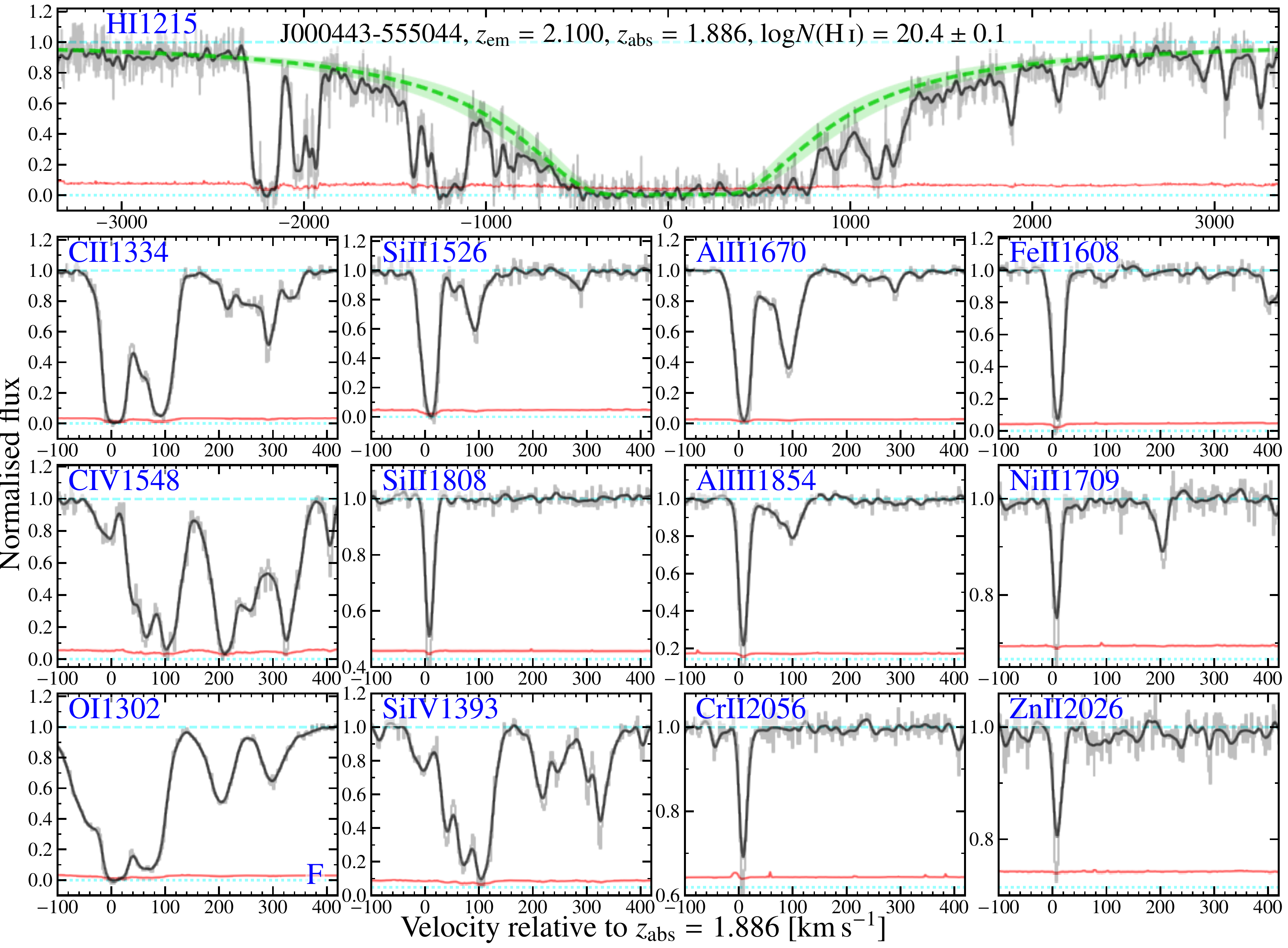}
\vspace{-0.5em}
\caption{Example of a DLA in the DR1 spectrum of J000443$-$555044 that has not been previously reported (to our knowledge). Each panel shows the portion of spectrum (light grey histogram) covering the labelled transition in velocity space around a common redshift, $\zab=1.886$. To aid visual inspection, the spectrum has been Gaussian-smoothed (shown in the black line). A single-component Voigt profile with an \ion{H}{i} column density that best matches the \lya\ profile, $\NHI=10^{20.4\pm0.1}$\,\pcmsq, is shown in the upper panel (green dashed line; shading indicates the \NHI\ uncertainty). The ``F'' in the \tran{O}{i}{1302} panel indicates that is falls within the \lya\ forest.}
\label{f:egDLA}
\end{center}
\end{figure*}

For the 18 new DLAs, \Tref{t:dla} provides the \NHI\ and representative uncertainty measured directly from the final DR1 spectra. To determine \NHI\ we overlaid a single-component Voigt profile over the spectrum with {\sc vpguess} \citep{Liske:2014:VPGUESS}. The redshift was fixed at the approximate optical-depth weighted mean centroid of unsaturated metal lines associated with the DLA, and the Doppler $b$ parameter was fixed to 20\,\kms. \NHI\ was then adjusted to best match the damped \lya\ profile shape. \Fref{f:egDLA} shows the \lya\ and strongest associated metal line transitions for one new DLA ($\zab=1.886$ towards J000443$-$555044). The Voigt profile with the best matching \NHI\ value is shown for comparison with the \lya\ transition. The Supporting Information online provides similar plots for all 18 new DLAs.

\begin{figure}
\begin{center}
\includegraphics[width=0.90\columnwidth]{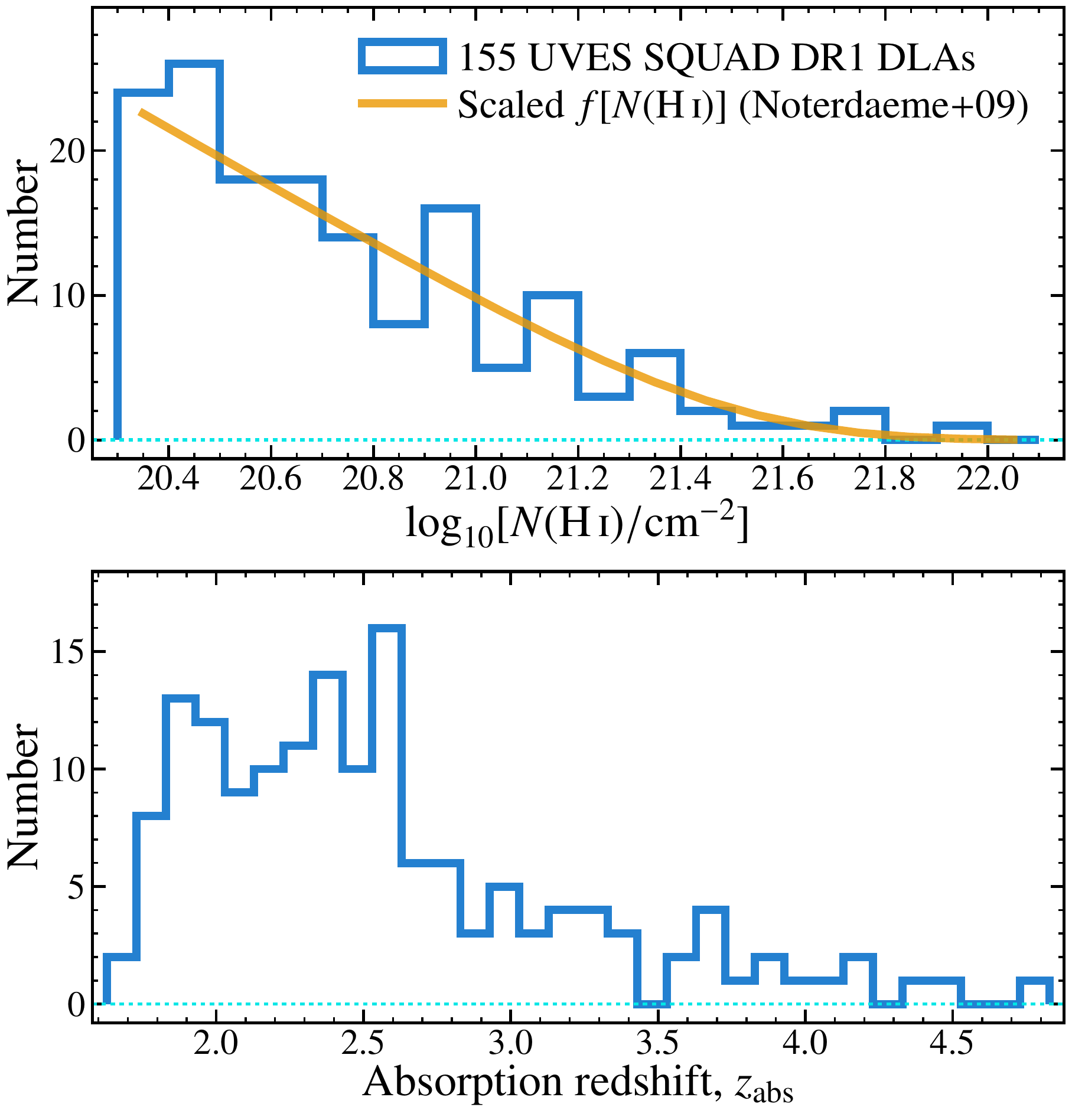}
\vspace{-1em}
\caption{\ion{H}{i} column density and absorption redshift distributions for the 155 DLAs identified in the final DR1 spectra. Because the original UVES observations often targeted known DLAs, they will be significantly over-represented in the DR1 spectra. The upper panel compares the \lNHI\ distribution (histogram) shape with the Gamma function fit of the frequency distribution, $f(\NHI)$, derived from the SDSS survey by \citet{Noterdaeme:2009:1087}, scaled so that its integral for $\lNHI\ge20.3$ is 155 (orange line). The absorption redshift (\zab) distribution shown in the lower panel should be interpreted in conjunction with the emission redshift distribution in \Fref{f:sky+zem} and the \lya\ detection sensitivity functions in Figs.\ \ref{f:trancov} and \ref{f:gz}.}
\label{f:DLAdists}
\end{center}
\end{figure}

The DR1 sample of 155 DLAs is the largest sample from a single high-resolution spectrograph to date. In this context it offers unprecedented opportunities for studying the range of chemical properties of DLAs, where the high resolution allows the detection and detailed modelling of the associated metal lines. However, it is important to note that a substantial proportion of UVES quasar observations specifically target quasars with known DLAs: \citet{Zafar:2013:A141} searched the proposal titles and abstracts for references to DLAs for their sample of 250 UVES quasar spectra, finding that $\sim$45 were targeted due to specific, known DLAs. While this ensures a large DLA sample from the DR1 spectra, it also ensures the sample is biased: DLAs will be over-represented in the DR1 spectra and their column density, redshift and metallicity distributions should be interpreted cautiously as a result. \Fref{f:DLAdists} illustrates the distribution of \ion{H}{i} column densities and absorption redshifts in the DR1 DLA sample. The \zab\ distribution in the lower panel is clearly a strong function of the emission redshift distribution (\Fref{f:sky+zem}), but also the wavelength coverage and detectability of broad \lya\ lines; the following section (\Sref{ss:surveys}) details these aspects. The upper panel of \Fref{f:DLAdists} compares the shape of the DR1 \lNHI\ distribution with the Gamma function fit to that of the SDSS DLA survey by \citet{Noterdaeme:2009:1087}. The fit has been scaled to yield a cumulative number of DLAs equal to the number in DR1. Overall, the relative representation of low and high \NHI\ DLAs in DR1 approximately reflects the expectation from a blind survey. However, very high \NHI\ DLAs are somewhat over-represented: there are 4 systems with $\lNHI\ge21.7$ in DR1, while 0.8 are expected from the scaled fit from SDSS in \Fref{f:DLAdists}.

\subsection{Absorption line surveys}\label{ss:surveys}

\begin{figure}
\begin{center}
\includegraphics[width=0.90\columnwidth]{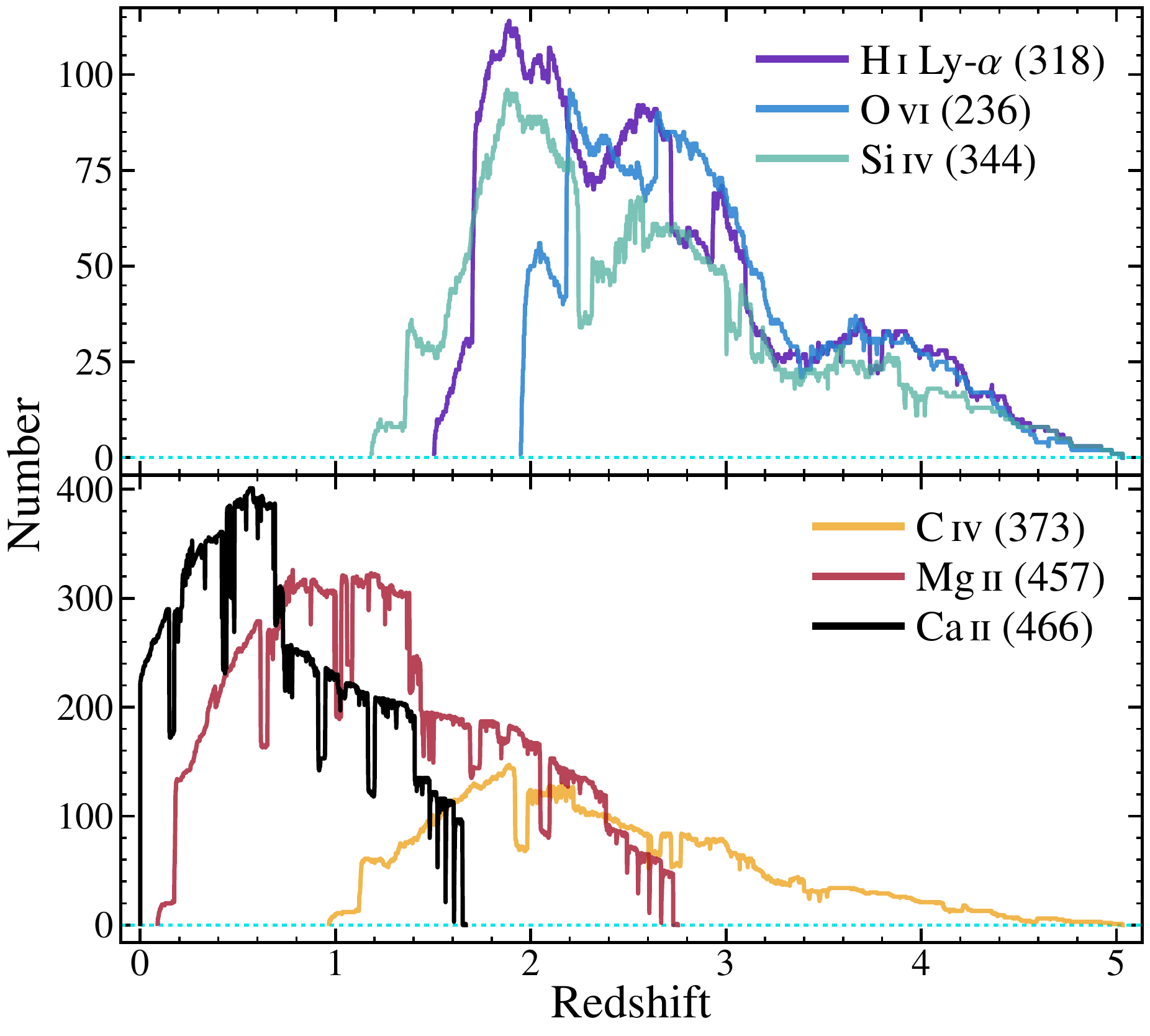}
\vspace{-1em}
\caption{Number of DR1 quasar spectra covering the most commonly surveyed transitions as a function of redshift: \ion{H}{i} \lya, \tran{O}{vi}{1031}, \tran{C}{iv}{1548}, \tran{Mg}{ii}{2796}, \tran{Si}{iv}{1393} and \tran{Ca}{ii}{3934}. The total number of spectra covering each transition, integrated over all redshifts, is given in parentheses after the species label. Note the different vertical axis ranges for the top and bottom panels.}
\label{f:trancov}
\end{center}
\end{figure}

\Fref{f:trancov} illustrates the total DR1 spectral coverage of strong absorption lines that are commonly surveyed in cosmological studies: \ion{H}{i} \lya, \tran{O}{vi}{1031}, \tran{C}{iv}{1548}, \tran{Mg}{ii}{2796}, \tran{Si}{iv}{1393} and \tran{Ca}{ii}{3934}. The metal lines occur in doublets and so are easily identified, especially in high-resolution spectra because their (identical) velocity structures are resolved. In \Fref{f:trancov} only the strongest member of the doublets, as listed above, is considered. For \ion{H}{i} \lya, \Fref{f:trancov} includes only the \lya\ forest, i.e.\ the redshift path between the \lya\ and \lyb\ quasar emission lines, and for \ion{O}{vi}, which occurs only in the \lya\ forest, all redshifts up to the emission redshift are included. For \ion{C}{iv}, \ion{Mg}{ii}, {Si}{iv} and \ion{Ca}{ii}, the redshift path extends down to the \lya\ emission line (or zero redshift, for \ion{Ca}{ii}). \Fref{f:trancov} also shows the total number of DR1 spectra that cover each transition at any redshift (in parentheses in the legends). For example, 68\% of the spectra cover some part of the \lya\ forest, while almost all DR1 spectra cover \ion{Mg}{ii} (457 of 467) and \ion{Ca}{ii} (all but one of 467). Even with these simple considerations, it is clear from \Fref{f:trancov} that the DR1 spectra provide an extensive dataset in which to survey absorption lines at high resolution.

\begin{figure}
\begin{center}
\includegraphics[width=0.90\columnwidth]{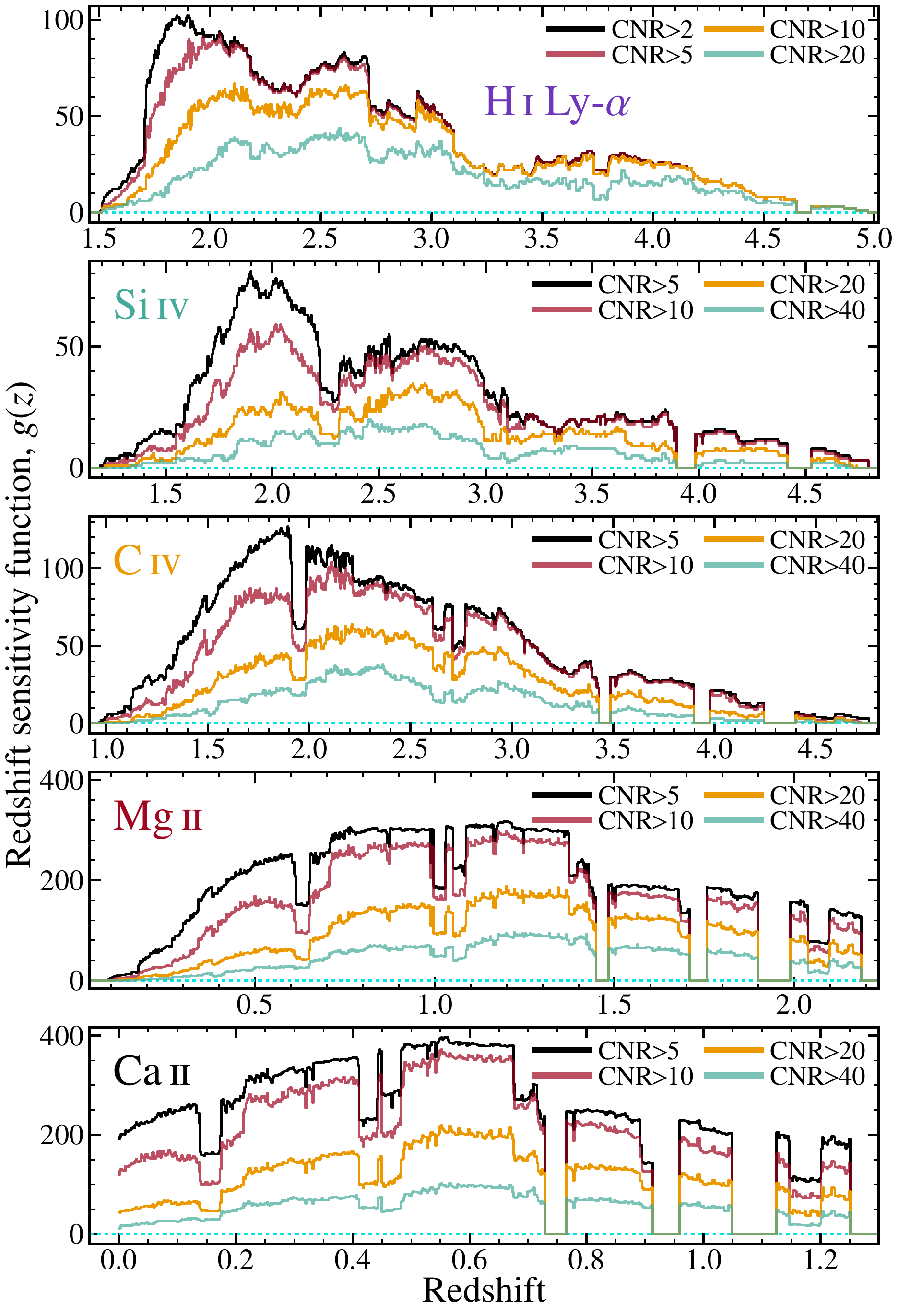}
\vspace{-1em}
\caption{Total redshift sensitivity function, $g(z)$, of the DR1 spectra for \ion{H}{i} \lya\ and the most commonly surveyed metal-line doublets outside the \lya\ forest: \ion{C}{iv}\,$\lambda\lambda$1548/1550, \ion{Mg}{ii}\,$\lambda\lambda$2796/2803, \ion{Si}{iv}\,$\lambda\lambda$1393/1402 and \ion{Ca}{ii}\,$\lambda\lambda$3934/3969. The strongest telluric absorption features (\Fref{f:telluric}) and the proximity zones near the relevant emission lines are excluded from $g(z)$, causing its shape to differ from the corresponding wavelength coverage function in \Fref{f:trancov}. $g(z)$ is calculated for four representative continuum-to-noise ratio (\CN) values for each species. This illustrates the large number of spectra in which relatively weak absorption lines can be detected.}
\label{f:gz}
\end{center}
\end{figure}

However, there are usually many other considerations required for statistical absorption line surveys in order to account for selection effects and biases. Typically, the redshift sensitivity function, $g(z)$, is a key metric: this is the total number of spectra in which the spectral feature being considered can, in principle, be detected once known selection effects are accounted for. That is, $g(z)$ is the spectral coverage of a given transition, or doublet, once additional detection effects are included, such as the \SN, proximity to quasar emission lines, telluric features and BALs that may preclude detection etc. \Fref{f:gz} presents $g(z)$ for the \lya\ forest region (regardless of the presence of DLAs and BALs), and the most commonly surveyed strong metal-line doublets outside the forest, i.e.\ \ion{C}{iv}\,$\lambda\lambda$1548/1550, \ion{Mg}{ii}\,$\lambda\lambda$2796/2803, \ion{Si}{iv}\,$\lambda\lambda$1393/1402 and \ion{Ca}{ii}\,$\lambda\lambda$3934/3969. Here, $g(z)$ masks out the strongest regions of telluric absorption (the O$_2$ and H$_2$O bands with $<$80\% transmitted flux in \Fref{f:telluric}). For all species we excluded the region 3000\,\kms\ bluewards of the quasar emission redshift to avoid proximity effects from the quasar. For the metal doublets, $g(z)$ extends down to 3000\,\kms\ redwards of the \lya\ emission line, while for \lya\ it extends down to 3000\,\kms\ redwards of the \lyb\ emission line to avoid including \lyb\ absorption.

The $g(z)$ functions in \Fref{f:gz} have broadly similar shapes and magnitudes to their raw wavelength coverage counterparts in \Fref{f:trancov}. However, avoiding the telluric and proximity regions reduces the number of sight-lines at all redshifts and reduces the sensitivity to zero for significant path lengths in the metal doublets, particularly \ion{Mg}{ii} and \ion{Ca}{ii}. To demonstrate how absorption line surveys will depend on the data quality, $g(z)$ is calculated for four representative \CN\ values for each species. This utilised the 1000-\kms-binned \CN\ maps of the quasar spectra described in \Sref{sss:CNR}. Note that the $g(z)$ functions in \Fref{f:gz} do not avoid other selection effects, such as broad absorption-line quasars, that some users should take into account for specific surveys. Finally, as is the case for DLAs (\Sref{ss:dlas} above), many DR1 quasars were targeted for observations with UVES due to the presence of metal-line absorbers. We therefore urge caution in determining the incidence rates of such systems with the DR1 sample. However, the proposal names and abstracts for the original observations do not indicate targeting of specific, weak absorbers. That is, there is likely little or no such bias for absorbers with equivalent widths below the detection threshold for lower-resolution surveys (e.g.\ $\la$0.2\,\AA\ for SDSS spectra). The $g(z)$ functions in \Fref{f:gz} indicate that very large redshift path lengths can be surveyed with the DR1 spectra, even for weak systems where higher \CN\ spectra are required (i.e.\ $\CN\ge20$).

\subsection{Time-variability studies using sub-spectra}\label{ss:subspec}

The DR1 spectra were taken over a 16-year period, so they are potentially useful in constraining the time-variability of absorption systems. This is particularly relevant for both narrow and broad metal absorption lines associated with outflows from the quasar central engines and/or host galaxies, as these lines are often observed to vary on time-scales shorter than $\sim$10 years \citep[e.g][]{Hamann:2011:1957,RodriguezHidalgo:2013:14}. Amongst the 467 quasars with final DR1 spectra, 92 comprise exposures taken more than a year apart, and 11 were observed over more than a decade interval. Significantly time-variable absorption lines can be easily identified in the final spectra using the same methods used to identify artefacts in \Sref{sss:artefacts} (see \Fref{f:baddata}), particularly by using \popler\ to simultaneously view the flux and $\chi^2_\nu$ spectra.

\begin{figure}
\begin{center}
\includegraphics[width=0.90\columnwidth]{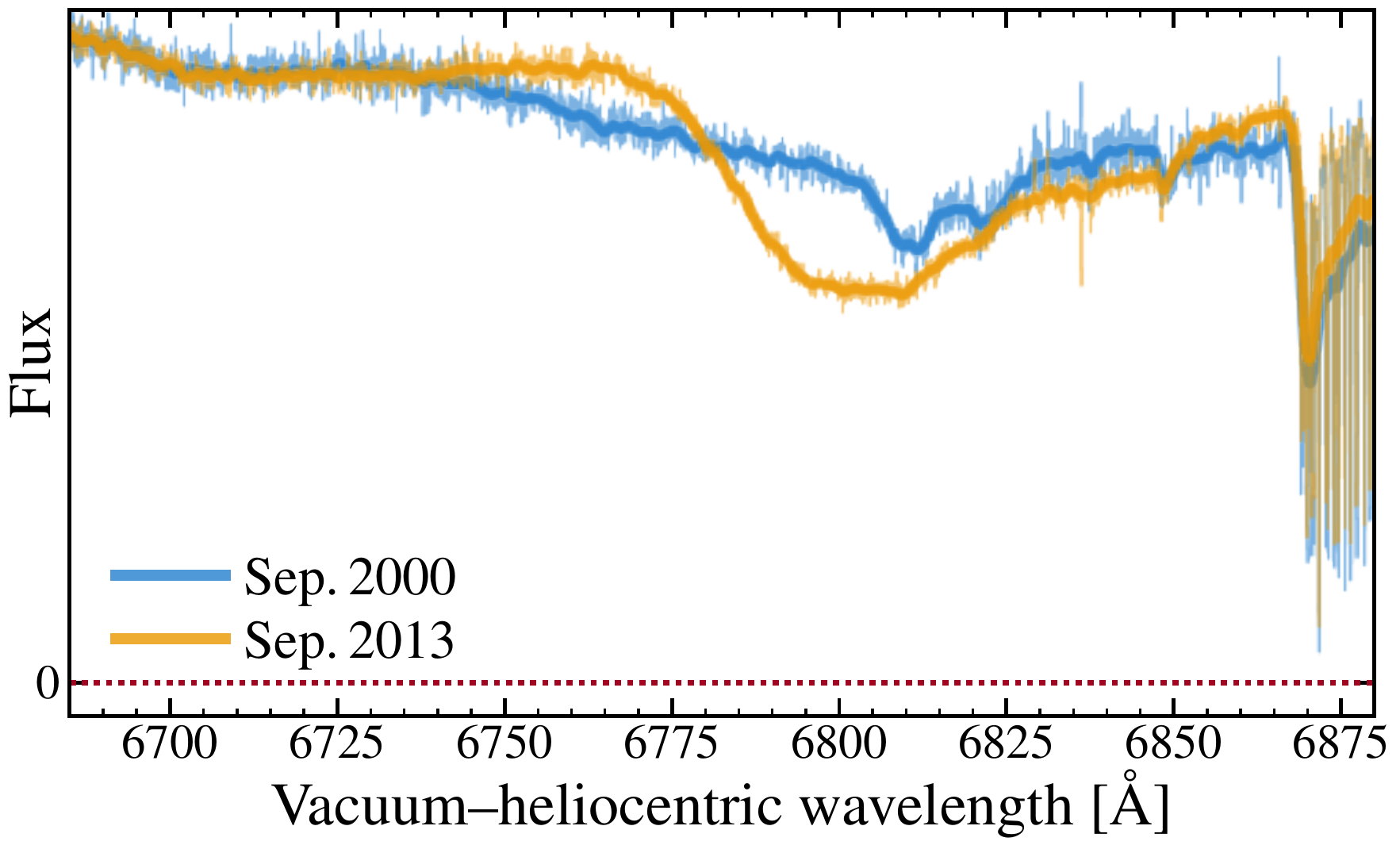}
\vspace{-1em}
\caption{Example of a time-variable \ion{C}{iv} broad absorption line feature $\sim$21000\,\kms\ bluewards of the emission redshift of J005758$-$264314. Two ``sub-spectra'' are shown (see text), comprising 11 and 13 exposures from Sep.\ 2000 and 2013, respectively. All exposures were combined to form the DR1 final spectrum of this quasar, but its $\chi^2_\nu$ spectrum (not shown) increases dramatically in the spectral region shown, sign-posting potential time-variable absorption. Each sub-spectrum (histograms) has been smoothed with a 10-pixel-wide Gaussian kernel for clarity (lines). Note that the absorption redwards of 6865\,\AA\ is the telluric O$_2$ B band.}
\label{f:broadCIV}
\end{center}
\end{figure}

Once time-variable absorption has been identified -- or if it is suspected but higher \SN\ is required for detection -- \popler\ includes a facility to create ``sub-spectra'' for comparative analyses: subsets of the original exposures can be selected for combination after the manual phase. This allows, for example, the best possible combination of exposures taken in one year to be compared with that from another year, and to identify or study time-variable absorption with the highest possible \SN\ in each case. The advantage of this approach is that sub-spectra are only split into separate spectra after being processed, cleaned, and continuum-fitted together; their treatment has been identical and observed differences between them are therefore more easily studied and interpreted. \popler\ sub-spectra have previously been employed in isolating instrumental systematic effects in UVES and HIRES spectra for studies of possible variations in the fine-structure constant and proton-to-electron mass ratio \citep[e.g.][]{Dapra:2015:489,Murphy:2016:2461,Murphy:2017:4930}. However, \Fref{f:broadCIV} illustrates their utility in studying time-variability in a broad absorption-line system that is highly blueshifted with respect to the quasar, presumably due to a high-velocity outflow. A systematic search for, and study of, time-variable absorption lines in, e.g., BAL quasars could therefore be undertaken using the full DR1 data products in \citet[][i.e.\ both the final spectrum and the extracted, contributing exposures]{Murphy:2018:UVESSQUADDR1}.

\section{Conclusions}\label{s:conclusion}

We have presented the first data release of the UVES SQUAD: 475 quasars, 467 of which have final, combined spectra (see \Tref{t:cat}). This is the largest database of high-resolution quasar spectra. The DR1 spectra are fully reproducible -- from the raw archival UVES exposures to final, combined and continuum-fitted spectra -- with a few commands using open-source software. All reduced exposures and files required to produce the final spectra and figures in this paper are provided in \citet{Murphy:2018:UVESSQUADDR1}. We also documented software tools for preparing and executing the UVES data reduction pipeline \citep[\headsort;][]{Murphy:2016:UVESheadsort}, and for combining the extracted exposures of a quasar, and cleaning and continuum fitting the result \citep[\popler;][]{Murphy:2016:UVESpopler}. The latter tool enables users of the DR1 spectra to improve the spectra, or modify them for their particular projects; we encourage and welcome improvements in this way and can make them available in future data releases.

A primary motivation for constructing DR1 was to enable statistical analyses of large quasar and absorber samples, particularly those requiring high resolution spectra (e.g.\ weak metal lines). We highlighted three broad areas where the DR1 spectra may be especially useful: DLA studies, absorption-line surveys, and studies of variable absorption, particularly outflows from quasars. To assist DLA studies, we presented a catalogue of 155 DLAs whose \lya\ lines are recorded in the DR1 spectra, 18 of which have not been reported previously (see \Tref{t:dla}). The \ion{H}{i} column densities from the literature are provided, plus values for the 18 new DLAs measured directly from the DR1 spectra. For absorption line surveys, the redshift sensitivity functions of the DR1 sample are presented in \Fref{f:gz} for the most commonly surveyed ionic species (\ion{H}{i}, \ion{Si}{iv}, \ion{C}{iv}, \ion{Mg}{ii} and \ion{Ca}{ii}). Indeed, several absorption-line surveys have already been conducted using earlier, preliminary versions of the UVES SQUAD spectra \citep[e.g.][]{Richter:2011:A12,BenBekhti:2012:A110,Nielsen:2013:114,Mas-Ribas:2018:50,Mathes:2018}. To assist studies of variable absorption lines, the \popler\ software can be used to construct ``sub-spectra'': sub-sets of exposures combined in the same way, with the same continuum, as the final spectrum. The highest \SN\ sub-spectra from different epochs can then be compared as accurately as possible. Finally, while we have invested considerable effort in cleaning all DR1 spectra to a minimum standard, many artefacts remain and are important to consider for such applications -- see \Sref{ss:artefacts}.

The second data release of UVES SQUAD is currently in preparation. It is anticipated to include spectra of another $\sim$180 quasars whose first exposure was taken between 30 June 2008 (the end date for first exposures in DR1) and early 2018. DR1 quasars with new exposures taken after 30 June 2008 will also be updated in DR2. While we have attempted to identify every quasar observed by UVES in the DR1 acquisition period (see \Sref{s:selection}), it remains possible that a small number may have escaped our selection criteria. We welcome knowledge of such quasars from the community and will include spectra of these in DR2 where possible.

\section*{Acknowledgements}

We thank Julija Bagdonaite, Tyler Evans, Sr\dj an Kotu\v{s}, Adrian Malec, Max Spolaor and Jonathan Whitmore for assistance in preliminary reduction and cleaning of several spectra, and Lachlan Thomson and Luka Giles for initial assistance with the DLA catalogue. We thank John Webb for work and discussions on the earlier database of {\sc midas}-reduced UVES spectra that contributed to \citet{King:2012:3370}. We thank the referee, Tae-Sun Kim, for comments that improved the paper. MTM thanks the Australian Research Council for \textsl{Discovery Projects} grants DP0877998, DP110100866 and DP130100568, and GGK and MTM thank it for \textsl{Discovery Projects} grant DP170103470, which supported this work. The UVES SQUAD spectra are based on observations collected at the European Organisation for Astronomical Research in the Southern Hemisphere under ESO programmes listed in \Tref{t:cat}. This research made use of: the NASA/IPAC Extragalactic Database (NED), which is operated by the Jet Propulsion Laboratory, California Institute of Technology, under contract with the National Aeronautics and Space Administration; NASA's Astrophysics Data System; {\sc astropy} \citep{Astropy:2013:A33}; and {\sc matplotlib} \citep{Hunter:2007:90}.

%%%%%%%%%%%%%%%%%%%%%%%%%%%%%%%%%%%%%%%%%%%%%%%%%%



%\bibliographystyle{mnras}
%\bibliography{ref}

\begin{thebibliography}{}
\makeatletter
\relax
\def\mn@urlcharsother{\let\do\@makeother \do\$\do\&\do\#\do\^\do\_\do\%\do\~}
\def\mn@doi{\begingroup\mn@urlcharsother \@ifnextchar [ {\mn@doi@}
  {\mn@doi@[]}}
\def\mn@doi@[#1]#2{\def\@tempa{#1}\ifx\@tempa\@empty \href
  {http://dx.doi.org/#2} {doi:#2}\else \href {http://dx.doi.org/#2} {#1}\fi
  \endgroup}
\def\mn@eprint#1#2{\mn@eprint@#1:#2::\@nil}
\def\mn@eprint@arXiv#1{\href {http://arxiv.org/abs/#1} {{\tt arXiv:#1}}}
\def\mn@eprint@dblp#1{\href {http://dblp.uni-trier.de/rec/bibtex/#1.xml}
  {dblp:#1}}
\def\mn@eprint@#1:#2:#3:#4\@nil{\def\@tempa {#1}\def\@tempb {#2}\def\@tempc
  {#3}\ifx \@tempc \@empty \let \@tempc \@tempb \let \@tempb \@tempa \fi \ifx
  \@tempb \@empty \def\@tempb {arXiv}\fi \@ifundefined
  {mn@eprint@\@tempb}{\@tempb:\@tempc}{\expandafter \expandafter \csname
  mn@eprint@\@tempb\endcsname \expandafter{\@tempc}}}

\bibitem[\protect\citeauthoryear{Abolfathi et~al.,}{Abolfathi
  et~al.}{2018}]{Abolfathi:2018:42}
Abolfathi B. et~al., 2018, \mn@doi [ApJS] {10.3847/1538-4365/aa9e8a}, \href
  {http://adsabs.harvard.edu/abs/2018ApJS..235...42A} {235, 42}

\bibitem[\protect\citeauthoryear{{Astropy Collaboration},}{{Astropy
  Collaboration}}{2013}]{Astropy:2013:A33}
{Astropy Collaboration}, 2013, \mn@doi [A\&A]
  {10.1051/0004-6361/201322068}, \href
  {http://adsabs.harvard.edu/abs/2013A%26A...558A..33A} {558, A33}

\bibitem[\protect\citeauthoryear{Bagdonaite, Ubachs, Murphy  \&
  Whitmore}{Bagdonaite et~al.}{2014}]{Bagdonaite:2014:10}
Bagdonaite J.,  Ubachs W.,  Murphy M.~T.,   Whitmore J.~B.,  2014, \mn@doi
  [ApJ] {10.1088/0004-637X/782/1/10}, \href
  {http://adsabs.harvard.edu/abs/2014ApJ...782...10B} {782, 10}

\bibitem[\protect\citeauthoryear{Becker, Rauch  \& Sargent}{Becker
  et~al.}{2007}]{Becker:2007:72}
Becker G.~D.,  Rauch M.,   Sargent W. L.~W.,  2007, \mn@doi [ApJ]
  {10.1086/517866}, \href {http://adsabs.harvard.edu/abs/2007ApJ...662...72B}
  {662, 72}

\bibitem[\protect\citeauthoryear{Ben~Bekhti, Winkel, Richter, Kerp, Klein  \&
  Murphy}{Ben~Bekhti et~al.}{2012}]{BenBekhti:2012:A110}
Ben~Bekhti N.,  Winkel B.,  Richter P.,  Kerp J.,  Klein U.,   Murphy M.~T.,
  2012, \mn@doi [A\&A] {10.1051/0004-6361/201118673}, \href
  {http://adsabs.harvard.edu/abs/2012A%26A...542A.110B} {542, A110}

\bibitem[\protect\citeauthoryear{Boera, Murphy, Becker  \& Bolton}{Boera
  et~al.}{2014}]{Boera:2014:1916}
Boera E.,  Murphy M.~T.,  Becker G.~D.,   Bolton J.~S.,  2014, \mn@doi [MNRAS]
  {10.1093/mnras/stu660}, \href
  {http://adsabs.harvard.edu/abs/2014MNRAS.441.1916B} {441, 1916}

\bibitem[\protect\citeauthoryear{Dapr{\`a}, Bagdonaite, Murphy  \&
  Ubachs}{Dapr{\`a} et~al.}{2015}]{Dapra:2015:489}
Dapr{\`a} M.,  Bagdonaite J.,  Murphy M.~T.,   Ubachs W.,  2015, \mn@doi
  [MNRAS] {10.1093/mnras/stv1998}, \href
  {http://adsabs.harvard.edu/abs/2015MNRAS.454..489D} {454, 489}

\bibitem[\protect\citeauthoryear{Dapr{\`a}, Niu, Salumbides, Murphy  \&
  Ubachs}{Dapr{\`a} et~al.}{2016}]{Dapra:2016:192}
Dapr{\`a} M.,  Niu M.~L.,  Salumbides E.~J.,  Murphy M.~T.,   Ubachs W.,  2016,
  \mn@doi [ApJ] {10.3847/0004-637X/826/2/192}, \href
  {http://adsabs.harvard.edu/abs/2016ApJ...826..192D} {826, 192}

\bibitem[\protect\citeauthoryear{Dekker, D'Odorico, Kaufer, Delabre  \&
  Kotzlowski}{Dekker et~al.}{2000}]{Dekker:2000:534}
Dekker H.,  D'Odorico S.,  Kaufer A.,  Delabre B.,   Kotzlowski H.,  2000, in
  Iye M.,  Moorwood A. F.~M.,  eds, Proc. SPIE Vol. 4008. SPIE, pp 534--545,
  \mn@doi{10.1117/12.395512}
  

\bibitem[\protect\citeauthoryear{Dessauges-Zavadsky, D'Odorico, McMahon,
  Molaro, Ledoux, P{\'e}roux  \& Storrie-Lombardi}{Dessauges-Zavadsky
  et~al.}{2001}]{Dessauges-Zavadsky:2001:426}
Dessauges-Zavadsky M.,  D'Odorico S.,  McMahon R.~G.,  Molaro P.,  Ledoux C.,
  P{\'e}roux C.,   Storrie-Lombardi L.~J.,  2001, \mn@doi [A\&A]
  {10.1051/0004-6361:20010217}, \href
  {http://adsabs.harvard.edu/abs/2001A%26A...370..426D} {370, 426}

\bibitem[\protect\citeauthoryear{Dessauges-Zavadsky, P{\'e}roux, Kim, D'Odorico
   \& McMahon}{Dessauges-Zavadsky et~al.}{2003}]{Dessauges-Zavadsky:2003:447}
Dessauges-Zavadsky M.,  P{\'e}roux C.,  Kim T.-S.,  D'Odorico S.,   McMahon
  R.~G.,  2003, \mn@doi [MNRAS] {10.1046/j.1365-8711.2003.06949.x}, 345, 447

\bibitem[\protect\citeauthoryear{Ellison, Yan, Hook, Pettini, Wall  \&
  Shaver}{Ellison et~al.}{2001}]{Ellison:2001:393}
Ellison S.~L.,  Yan L.,  Hook I.~M.,  Pettini M.,  Wall J.~V.,   Shaver P.,
  2001, \mn@doi [A\&A] {10.1051/0004-6361:20011281}, \href
  {http://adsabs.harvard.edu/abs/2001A%26A...379..393E} {379, 393}

\bibitem[\protect\citeauthoryear{Ellison, Prochaska, Hennawi, Lopez, Usher,
  Wolfe, Russell  \& Benn}{Ellison et~al.}{2010}]{Ellison:2010:1435}
Ellison S.~L.,  Prochaska J.~X.,  Hennawi J.,  Lopez S.,  Usher C.,  Wolfe
  A.~M.,  Russell D.~M.,   Benn C.~R.,  2010, \mn@doi [MNRAS]
  {10.1111/j.1365-2966.2010.16780.x}, \href
  {http://adsabs.harvard.edu/abs/2010MNRAS.406.1435E} {406, 1435}

\bibitem[\protect\citeauthoryear{Evans et~al.,}{Evans
  et~al.}{2014}]{Evans:2014:128}
Evans T.~M. et~al., 2014, \mn@doi [MNRAS] {10.1093/mnras/stu1754}, \href
  {http://adsabs.harvard.edu/abs/2014MNRAS.445..128E} {445, 128}

\bibitem[\protect\citeauthoryear{Flesch}{Flesch}{2015}]{Flesch:2015:e010}
Flesch E.~W.,  2015, \mn@doi [PASA] {10.1017/pasa.2015.10}, \href
  {http://adsabs.harvard.edu/abs/2015PASA...32...10F} {32, e010}

\bibitem[\protect\citeauthoryear{Guimar{\~a}es, Noterdaeme, Petitjean, Ledoux,
  Srianand, Lopez  \& Rahmani}{Guimar{\~a}es et~al.}{2012}]{Guimaraes:2012:147}
Guimar{\~a}es R.,  Noterdaeme P.,  Petitjean P.,  Ledoux C.,  Srianand R.,
  Lopez S.,   Rahmani H.,  2012, \mn@doi [AJ] {10.1088/0004-6256/143/6/147},
  \href {http://adsabs.harvard.edu/abs/2012AJ....143..147G} {143, 147}

\bibitem[\protect\citeauthoryear{Hamann, Kanekar, Prochaska, Murphy, Ellison,
  Malec, Milutinovic  \& Ubachs}{Hamann et~al.}{2011}]{Hamann:2011:1957}
Hamann F.,  Kanekar N.,  Prochaska J.~X.,  Murphy M.~T.,  Ellison S.,  Malec
  A.~L.,  Milutinovic N.,   Ubachs W.,  2011, \mn@doi [MNRAS]
  {10.1111/j.1365-2966.2010.17575.x}, \href
  {http://adsabs.harvard.edu/abs/2011MNRAS.410.1957H} {410, 1957}

\bibitem[\protect\citeauthoryear{Hambly et~al.,}{Hambly
  et~al.}{2001}]{Hambly:2001:1279}
Hambly N.~C. et~al., 2001, \mn@doi [MNRAS]
  {10.1111/j.1365-2966.2001.04660.x}, \href
  {http://adsabs.harvard.edu/abs/2001MNRAS.326.1279H} {326, 1279}

\bibitem[\protect\citeauthoryear{Hunter}{Hunter}{2007}]{Hunter:2007:90}
Hunter J.~D.,  2007, \mn@doi [Computing in Science and Engineering]
  {10.1109/MCSE.2007.55}, \href
  {http://adsabs.harvard.edu/abs/2007CSE.....9...90H} {9, 90}

\bibitem[\protect\citeauthoryear{Kim, Viel, Haehnelt, Carswell  \&
  Cristiani}{Kim et~al.}{2004}]{Kim:2004:355}
Kim T.-S.,  Viel M.,  Haehnelt M.~G.,  Carswell R.~F.,   Cristiani S.,  2004,
  \mn@doi [MNRAS] {10.1111/j.1365-2966.2004.07221.x}, \href
  {http://adsabs.harvard.edu/abs/2004MNRAS.347..355K} {347, 355}

\bibitem[\protect\citeauthoryear{King, Webb, Murphy  \& Carswell}{King
  et~al.}{2008}]{King:2008:251304}
King J.~A.,  Webb J.~K.,  Murphy M.~T.,   Carswell R.~F.,  2008, \mn@doi [Phys.
  Rev. Lett.] {10.1103/PhysRevLett.101.251304}, \href
  {http://adsabs.harvard.edu/abs/2008PhRvL.101y1304K} {101, 251304}

\bibitem[\protect\citeauthoryear{King, Webb, Murphy, Flambaum, Carswell,
  Bainbridge, Wilczynska  \& Koch}{King et~al.}{2012}]{King:2012:3370}
King J.~A.,  Webb J.~K.,  Murphy M.~T.,  Flambaum V.~V.,  Carswell R.~F.,
  Bainbridge M.~B.,  Wilczynska M.~R.,   Koch F.~E.,  2012, \mn@doi [MNRAS]
  {10.1111/j.1365-2966.2012.20852.x}, \href
  {http://adsabs.harvard.edu/abs/2012MNRAS.422.3370K} {422, 3370}

\bibitem[\protect\citeauthoryear{Kotu{\v s}, Murphy  \& Carswell}{Kotu{\v s}
  et~al.}{2017}]{Kotus:2017:3679}
Kotu{\v s} S.~M.,  Murphy M.~T.,   Carswell R.~F.,  2017, \mn@doi [MNRAS]
  {10.1093/mnras/stw2543}, \href
  {http://adsabs.harvard.edu/abs/2017MNRAS.464.3679K} {464, 3679}

\bibitem[\protect\citeauthoryear{Ledoux, Petitjean  \& Srianand}{Ledoux
  et~al.}{2003}]{Ledoux:2003:209}
Ledoux C.,  Petitjean P.,   Srianand R.,  2003, \mn@doi [MNRAS]
  {10.1046/j.1365-2966.2003.07082.x}, \href
  {http://adsabs.harvard.edu/abs/2003MNRAS.346..209L} {346, 209}

\bibitem[\protect\citeauthoryear{Ledoux, Petitjean, Fynbo, M{\o}ller  \&
  Srianand}{Ledoux et~al.}{2006}]{Ledoux:2006:71}
Ledoux C.,  Petitjean P.,  Fynbo J. P.~U.,  M{\o}ller P.,   Srianand R.,  2006,
  \mn@doi [A\&A] {10.1051/0004-6361:20054242}, \href
  {http://adsabs.harvard.edu/abs/2006A%26A...457...71L} {457, 71}

\bibitem[\protect\citeauthoryear{Ledoux, Noterdaeme, Petitjean  \&
  Srianand}{Ledoux et~al.}{2015}]{Ledoux:2015:A8}
Ledoux C.,  Noterdaeme P.,  Petitjean P.,   Srianand R.,  2015, \mn@doi [A\&A]
  {10.1051/0004-6361/201424122}, \href
  {http://adsabs.harvard.edu/abs/2015A%26A...580A...8L} {580, A8}

\bibitem[\protect\citeauthoryear{{Liske}}{{Liske}}{2014}]{Liske:2014:VPGUESS}
{Liske} J.,  2014, {vpguess: Fitting multiple Voigt profiles to spectroscopic
  data}, Astrophysics Source Code Library (\mn@eprint {ascl} {1408.016})

\bibitem[\protect\citeauthoryear{Mas-Ribas, Riemer-S{\o}rensen, Hennawi,
  Miralda-Escud{\'e}, O'Meara, P{\'e}rez-R{\`a}fols, Murphy  \& Webb}{Mas-Ribas
  et~al.}{2018}]{Mas-Ribas:2018:50}
Mas-Ribas L.,  Riemer-S{\o}rensen S.,  Hennawi J.~F.,  Miralda-Escud{\'e} J.,
  O'Meara J.~M.,  P{\'e}rez-R{\`a}fols I.,  Murphy M.~T.,   Webb J.~K.,  2018,
  \mn@doi [ApJ] {10.3847/1538-4357/aac81a}, \href
  {http://adsabs.harvard.edu/abs/2018ApJ...862...50M} {862, 50}

\bibitem[\protect\citeauthoryear{{Mathes}, {Churchill}  \& {Murphy}}{{Mathes}
  et~al.}{2018}]{Mathes:2018}
{Mathes} N.~L.,  {Churchill} C.~W.,   {Murphy} M.~T.,  2018, \apj, submitted,
  arXiv:1701.05624, \href {http://adsabs.harvard.edu/abs/2017arXiv170105624M}
  {}

\bibitem[\protect\citeauthoryear{Molaro, Bonifacio, Centuri{\'o}n, D'Odorico,
  Vladilo, Santin  \& Di~Marcantonio}{Molaro et~al.}{2000}]{Molaro:2000:54}
Molaro P.,  Bonifacio P.,  Centuri{\'o}n M.,  D'Odorico S.,  Vladilo G.,
  Santin P.,   Di~Marcantonio P.,  2000, \mn@doi [ApJ] {10.1086/309439}, 541,
  54

\bibitem[\protect\citeauthoryear{Molaro et~al.,}{Molaro
  et~al.}{2013}]{Molaro:2013:A68}
Molaro P. et~al., 2013, \mn@doi [A\&A] {10.1051/0004-6361/201321351}, \href
  {http://adsabs.harvard.edu/abs/2013A%26A...555A..68M} {555, A68}

\bibitem[\protect\citeauthoryear{Murphy}{Murphy}{2016a}]{Murphy:2016:UVESheadsort}
Murphy M.~T.,  2016a, {UVES\_headsort: VLT/UVES pipeline preparation},
  \mn@doi{10.5281/zenodo.44766}

\bibitem[\protect\citeauthoryear{Murphy}{Murphy}{2016b}]{Murphy:2016:UVESpopler}
Murphy M.~T.,  2016b, {UVES\_popler: POst PipeLine Echelle Reduction software},
  \mn@doi{10.5281/zenodo.44765}

\bibitem[\protect\citeauthoryear{Murphy \& Bernet}{Murphy \&
  Bernet}{2016}]{Murphy:2016:1043}
Murphy M.~T.,  Bernet M.~L.,  2016, \mn@doi [MNRAS] {10.1093/mnras/stv2420},
  \href {http://adsabs.harvard.edu/abs/2016MNRAS.455.1043M} {455, 1043}

\bibitem[\protect\citeauthoryear{Murphy \& Cooksey}{Murphy \&
  Cooksey}{2017}]{Murphy:2017:4930}
Murphy M.~T.,  Cooksey K.~L.,  2017, \mn@doi [MNRAS] {10.1093/mnras/stx1949},
  \href {http://adsabs.harvard.edu/abs/2017MNRAS.471.4930M} {471, 4930}

\bibitem[\protect\citeauthoryear{Murphy, Tzanavaris, Webb  \& Lovis}{Murphy
  et~al.}{2007}]{Murphy:2007:221}
Murphy M.~T.,  Tzanavaris P.,  Webb J.~K.,   Lovis C.,  2007, \mn@doi [MNRAS]
  {10.1111/j.1365-2966.2007.11768.x}, \href
  {http://adsabs.harvard.edu/abs/2007MNRAS.378..221M} {378, 221}

\bibitem[\protect\citeauthoryear{Murphy, Malec  \& Prochaska}{Murphy
  et~al.}{2016}]{Murphy:2016:2461}
Murphy M.~T.,  Malec A.~L.,   Prochaska J.~X.,  2016, \mn@doi [MNRAS]
  {10.1093/mnras/stw1482}, \href
  {http://adsabs.harvard.edu/abs/2016MNRAS.461.2461M} {461, 2461}

\bibitem[\protect\citeauthoryear{Murphy, Kacprzak, Savorgnan  \&
  Carswell}{Murphy et~al.}{2018}]{Murphy:2018:UVESSQUADDR1}
Murphy M.~T.,  Kacprzak G.~G.,  Savorgnan G. A.~D.,   Carswell R.~F.,  2018,
  {UVES SQUAD Data Release 1}, \mn@doi{10.5281/zenodo.1345974}
  {\urlstyle{rm}\url{https://github.com/MTMurphy77/UVES\_SQUAD\_DR1}}

\bibitem[\protect\citeauthoryear{Nielsen, Churchill, Kacprzak  \&
  Murphy}{Nielsen et~al.}{2013}]{Nielsen:2013:114}
Nielsen N.~M.,  Churchill C.~W.,  Kacprzak G.~G.,   Murphy M.~T.,  2013,
  \mn@doi [ApJ] {10.1088/0004-637X/776/2/114}, \href
  {http://adsabs.harvard.edu/abs/2013ApJ...776..114N} {776, 114}

\bibitem[\protect\citeauthoryear{Noterdaeme, Ledoux, Petitjean  \&
  Srianand}{Noterdaeme et~al.}{2008}]{Noterdaeme:2008:327}
Noterdaeme P.,  Ledoux C.,  Petitjean P.,   Srianand R.,  2008, \mn@doi [A\&A]
  {10.1051/0004-6361:20078780}, \href
  {http://adsabs.harvard.edu/abs/2008A%26A...481..327N} {481, 327}

\bibitem[\protect\citeauthoryear{Noterdaeme, Petitjean, Ledoux  \&
  Srianand}{Noterdaeme et~al.}{2009}]{Noterdaeme:2009:1087}
Noterdaeme P.,  Petitjean P.,  Ledoux C.,   Srianand R.,  2009, \mn@doi [A\&A]
  {10.1051/0004-6361/200912768}, \href
  {http://adsabs.harvard.edu/abs/2009A%26A...505.1087N} {505, 1087}

\bibitem[\protect\citeauthoryear{Noterdaeme, Petitjean, Srianand, Ledoux  \&
  L{\'o}pez}{Noterdaeme et~al.}{2011}]{Noterdaeme:2011:L7}
Noterdaeme P.,  Petitjean P.,  Srianand R.,  Ledoux C.,   L{\'o}pez S.,  2011,
  \mn@doi [A\&A] {10.1051/0004-6361/201016140}, \href
  {http://adsabs.harvard.edu/abs/2011A%26A...526L...7N} {526, L7}

\bibitem[\protect\citeauthoryear{Noterdaeme, Srianand, Rahmani, Petitjean,
  P{\^a}ris, Ledoux, Gupta  \& L{\'o}pez}{Noterdaeme
  et~al.}{2015}]{Noterdaeme:2015:A24}
Noterdaeme P.,  Srianand R.,  Rahmani H.,  Petitjean P.,  P{\^a}ris I.,  Ledoux
  C.,  Gupta N.,   L{\'o}pez S.,  2015, \mn@doi [A\&A]
  {10.1051/0004-6361/201425376}, \href
  {http://adsabs.harvard.edu/abs/2015A%26A...577A..24N} {577, A24}

\bibitem[\protect\citeauthoryear{Noterdaeme et~al.,}{Noterdaeme
  et~al.}{2017}]{Noterdaeme:2017:A82}
Noterdaeme P. et~al., 2017, \mn@doi [A\&A] {10.1051/0004-6361/201629173},
  \href {http://adsabs.harvard.edu/abs/2017A%26A...597A..82N} {597, A82}

\bibitem[\protect\citeauthoryear{O'Meara et~al.,}{O'Meara
  et~al.}{2015}]{OMeara:2015:111}
O'Meara J.~M. et~al., 2015, \mn@doi [AJ] {10.1088/0004-6256/150/4/111}, \href
  {http://adsabs.harvard.edu/abs/2015AJ....150..111O} {150, 111}

\bibitem[\protect\citeauthoryear{O'Meara, Lehner, Howk, Prochaska, Fox,
  Peeples, Tumlinson  \& O'Shea}{O'Meara et~al.}{2017}]{OMeara:2017:114}
O'Meara J.~M.,  Lehner N.,  Howk J.~C.,  Prochaska J.~X.,  Fox A.~J.,  Peeples
  M.~S.,  Tumlinson J.,   O'Shea B.~W.,  2017, \mn@doi [AJ]
  {10.3847/1538-3881/aa82b8}, \href
  {http://adsabs.harvard.edu/abs/2017AJ....154..114O} {154, 114}

\bibitem[\protect\citeauthoryear{P{\^a}ris et~al.,}{P{\^a}ris
  et~al.}{2018}]{Paris:2018:A51}
P{\^a}ris I. et~al., 2018, \mn@doi [A\&A] {10.1051/0004-6361/201732445},
  \href {http://adsabs.harvard.edu/abs/2018A%26A...613A..51P} {613, A51}

\bibitem[\protect\citeauthoryear{Parks, Prochaska, Dong  \& Cai}{Parks
  et~al.}{2018}]{Parks:2018:1151}
Parks D.,  Prochaska J.~X.,  Dong S.,   Cai Z.,  2018, \mn@doi [MNRAS]
  {10.1093/mnras/sty196}, \href
  {http://adsabs.harvard.edu/abs/2018MNRAS.476.1151P} {476, 1151}

\bibitem[\protect\citeauthoryear{Penprase, Prochaska, Sargent, Toro-Martinez
  \& Beeler}{Penprase et~al.}{2010}]{Penprase:2010:1}
Penprase B.~E.,  Prochaska J.~X.,  Sargent W. L.~W.,  Toro-Martinez I.,
  Beeler D.~J.,  2010, \mn@doi [ApJ] {10.1088/0004-637X/721/1/1}, \href
  {http://adsabs.harvard.edu/abs/2010ApJ...721....1P} {721, 1}

\bibitem[\protect\citeauthoryear{P{\'e}roux, Storrie-Lombardi, McMahon, Irwin
  \& Hook}{P{\'e}roux et~al.}{2001}]{Peroux:2001:1799}
P{\'e}roux C.,  Storrie-Lombardi L.~J.,  McMahon R.~G.,  Irwin M.,   Hook
  I.~M.,  2001, \mn@doi [AJ] {10.1086/319967}, \href
  {http://adsabs.harvard.edu/abs/2001AJ....121.1799P} {121, 1799}

\bibitem[\protect\citeauthoryear{P{\'e}roux, Dessauges-Zavadsky, D'Odorico,
  Sun~Kim  \& McMahon}{P{\'e}roux et~al.}{2005}]{Peroux:2005:479}
P{\'e}roux C.,  Dessauges-Zavadsky M.,  D'Odorico S.,  Sun~Kim T.,   McMahon
  R.~G.,  2005, \mn@doi [MNRAS] {10.1111/j.1365-2966.2005.09432.x}, \href
  {http://adsabs.harvard.edu/abs/2005MNRAS.363..479P} {363, 479}

\bibitem[\protect\citeauthoryear{Petitjean, Srianand  \& Ledoux}{Petitjean
  et~al.}{2000}]{Petitjean:2000:L26}
Petitjean P.,  Srianand R.,   Ledoux C.,  2000, A\&A, 364, L26

\bibitem[\protect\citeauthoryear{Pettini \& Cooke}{Pettini \&
  Cooke}{2012}]{Pettini:2012:2477}
Pettini M.,  Cooke R.,  2012, \mn@doi [MNRAS]
  {10.1111/j.1365-2966.2012.21665.x}, \href
  {http://adsabs.harvard.edu/abs/2012MNRAS.425.2477P} {425, 2477}

\bibitem[\protect\citeauthoryear{Pettini, Ellison, Bergeron  \&
  Petitjean}{Pettini et~al.}{2002}]{Pettini:2002:21}
Pettini M.,  Ellison S.~L.,  Bergeron J.,   Petitjean P.,  2002, A\&A, 391, 21

\bibitem[\protect\citeauthoryear{Pettini, Zych, Steidel  \& Chaffee}{Pettini
  et~al.}{2008a}]{Pettini:2008:2011}
Pettini M.,  Zych B.~J.,  Steidel C.~C.,   Chaffee F.~H.,  2008a, \mn@doi
  [MNRAS] {10.1111/j.1365-2966.2008.12951.x}, \href
  {http://adsabs.harvard.edu/abs/2008MNRAS.385.2011P} {385, 2011}

\bibitem[\protect\citeauthoryear{Pettini, Zych, Murphy, Lewis  \&
  Steidel}{Pettini et~al.}{2008b}]{Pettini:2008:1499}
Pettini M.,  Zych B.~J.,  Murphy M.~T.,  Lewis A.,   Steidel C.~C.,  2008b,
  \mn@doi [MNRAS] {10.1111/j.1365-2966.2008.13921.x}, \href
  {http://adsabs.harvard.edu/abs/2008MNRAS.391.1499P} {391, 1499}

\bibitem[\protect\citeauthoryear{Prochaska \& Wolfe}{Prochaska \&
  Wolfe}{1999}]{Prochaska:1999:369}
Prochaska J.~X.,  Wolfe A.~M.,  1999, \mn@doi [ApJS] {10.1086/313200}, 121, 369

\bibitem[\protect\citeauthoryear{Prochaska \& Wolfe}{Prochaska \&
  Wolfe}{2009}]{Prochaska:2009:1543}
Prochaska J.~X.,  Wolfe A.~M.,  2009, \mn@doi [ApJ]
  {10.1088/0004-637X/696/2/1543}, \href
  {http://adsabs.harvard.edu/abs/2009ApJ...696.1543P} {696, 1543}

\bibitem[\protect\citeauthoryear{Prochaska et~al.,}{Prochaska
  et~al.}{2001}]{Prochaska:2001:21}
Prochaska J.~X. et~al., 2001, \mn@doi [ApJS] {10.1086/322542}, 137, 21

\bibitem[\protect\citeauthoryear{Prochaska, Gawiser, Wolfe, Cooke  \&
  Gelino}{Prochaska et~al.}{2003a}]{Prochaska:2003:227}
Prochaska J.~X.,  Gawiser E.,  Wolfe A.~M.,  Cooke J.,   Gelino D.,  2003a,
  \mn@doi [ApJS] {10.1086/375839}, \href
  {http://adsabs.harvard.edu/abs/2003ApJS..147..227P} {147, 227}

\bibitem[\protect\citeauthoryear{Prochaska, Gawiser, Wolfe, Castro  \&
  Djorgovski}{Prochaska et~al.}{2003b}]{Prochaska:2003:L9}
Prochaska J.~X.,  Gawiser E.,  Wolfe A.~M.,  Castro S.,   Djorgovski S.~G.,
  2003b, \mn@doi [ApJ] {10.1086/378945}, \href
  {http://adsabs.harvard.edu/abs/2003ApJ...595L...9P} {595, L9}

\bibitem[\protect\citeauthoryear{Prochaska, Herbert-Fort  \& Wolfe}{Prochaska
  et~al.}{2005}]{Prochaska:2005:123}
Prochaska J.~X.,  Herbert-Fort S.,   Wolfe A.~M.,  2005, \mn@doi [ApJ]
  {10.1086/497287}, \href {http://adsabs.harvard.edu/abs/2005ApJ...635..123P}
  {635, 123}

\bibitem[\protect\citeauthoryear{Prochaska, Wolfe, Howk, Gawiser, Burles  \&
  Cooke}{Prochaska et~al.}{2007}]{Prochaska:2007:29}
Prochaska J.~X.,  Wolfe A.~M.,  Howk J.~C.,  Gawiser E.,  Burles S.~M.,   Cooke
  J.,  2007, \mn@doi [ApJS] {10.1086/513714}, \href
  {http://adsabs.harvard.edu/abs/2007ApJS..171...29P} {171, 29}

\bibitem[\protect\citeauthoryear{Prochaska, Hennawi  \& Herbert-Fort}{Prochaska
  et~al.}{2008}]{Prochaska:2008:1002}
Prochaska J.~X.,  Hennawi J.~F.,   Herbert-Fort S.,  2008, \mn@doi [ApJ]
  {10.1086/526508}, \href {http://adsabs.harvard.edu/abs/2008ApJ...675.1002P}
  {675, 1002}

\bibitem[\protect\citeauthoryear{Quast, Reimers  \& Levshakov}{Quast
  et~al.}{2004}]{Quast:2004:L7}
Quast R.,  Reimers D.,   Levshakov S.~A.,  2004, \mn@doi [A\&A]
  {10.1051/0004-6361:20040013}, \href
  {http://adsabs.harvard.edu/abs/2004A%26A...415L...7Q} {415, L7}

\bibitem[\protect\citeauthoryear{Rahmani et~al.,}{Rahmani
  et~al.}{2013}]{Rahmani:2013:861}
Rahmani H. et~al., 2013, \mn@doi [MNRAS] {10.1093/mnras/stt1356}, \href
  {http://adsabs.harvard.edu/abs/2013MNRAS.435..861R} {435, 861}

\bibitem[\protect\citeauthoryear{Richter, Krause, Fechner, Charlton  \&
  Murphy}{Richter et~al.}{2011}]{Richter:2011:A12}
Richter P.,  Krause F.,  Fechner C.,  Charlton J.~C.,   Murphy M.~T.,  2011,
  \mn@doi [A\&A] {10.1051/0004-6361/201015566}, \href
  {http://adsabs.harvard.edu/abs/2011A%26A...528A..12R} {528, A12}

\bibitem[\protect\citeauthoryear{Riemer-S{\o}rensen, Kotu{\v s}, Webb, Ali,
  Dumont, Murphy  \& Carswell}{Riemer-S{\o}rensen
  et~al.}{2017}]{Riemer-Sorensen:2017:3239}
Riemer-S{\o}rensen S.,  Kotu{\v s} S.,  Webb J.~K.,  Ali K.,  Dumont V.,
  Murphy M.~T.,   Carswell R.~F.,  2017, \mn@doi [MNRAS]
  {10.1093/mnras/stx681}, \href
  {http://adsabs.harvard.edu/abs/2017MNRAS.468.3239R} {468, 3239}

\bibitem[\protect\citeauthoryear{Rodr{\'\i}guez~Hidalgo, Eracleous, Charlton,
  Hamann, Murphy  \& Nestor}{Rodr{\'\i}guez~Hidalgo
  et~al.}{2013}]{RodriguezHidalgo:2013:14}
Rodr{\'\i}guez~Hidalgo P.,  Eracleous M.,  Charlton J.,  Hamann F.,  Murphy
  M.~T.,   Nestor D.,  2013, \mn@doi [ApJ] {10.1088/0004-637X/775/1/14}, \href
  {http://adsabs.harvard.edu/abs/2013ApJ...775...14R} {775, 14}

\bibitem[\protect\citeauthoryear{Schaye, Aguirre, Kim, Theuns, Rauch  \&
  Sargent}{Schaye et~al.}{2003}]{Schaye:2003:768}
Schaye J.,  Aguirre A.,  Kim T.-S.,  Theuns T.,  Rauch M.,   Sargent W. L.~W.,
  2003, \mn@doi [ApJ] {10.1086/378044}, 596, 768

\bibitem[\protect\citeauthoryear{Srianand, Noterdaeme, Ledoux  \&
  Petitjean}{Srianand et~al.}{2008}]{Srianand:2008:L39}
Srianand R.,  Noterdaeme P.,  Ledoux C.,   Petitjean P.,  2008, \mn@doi [A\&A]
  {10.1051/0004-6361:200809727}, \href
  {http://adsabs.harvard.edu/abs/2008A%26A...482L..39S} {482, L39}

\bibitem[\protect\citeauthoryear{Vanden~Berk et~al.,}{Vanden~Berk
  et~al.}{2001}]{vandenBerk:2001:549}
Vanden~Berk D.~E. et~al., 2001, \mn@doi [AJ] {10.1086/321167}, \href
  {http://adsabs.harvard.edu/abs/2001AJ....122..549V} {122, 549}

\bibitem[\protect\citeauthoryear{Vogt et~al.,}{Vogt
  et~al.}{1994}]{Vogt:1994:362}
Vogt S.~S. et~al., 1994, Proc. SPIE Instrumentation in Astronomy VIII, \href
  {http://adsabs.harvard.edu/abs/1994SPIE.2198..362V} {2198, 362}

\bibitem[\protect\citeauthoryear{Wells, Greisen  \& Harten}{Wells
  et~al.}{1981}]{Wells:1981:363}
Wells D.~C.,  Greisen E.~W.,   Harten R.~H.,  1981, Astronomy and Astrophysics
  Supplement, 44, 363

\bibitem[\protect\citeauthoryear{Whitmore \& Murphy}{Whitmore \&
  Murphy}{2015}]{Whitmore:2015:446}
Whitmore J.~B.,  Murphy M.~T.,  2015, \mn@doi [MNRAS] {10.1093/mnras/stu2420},
  \href {http://adsabs.harvard.edu/abs/2015MNRAS.447..446W} {447, 446}

\bibitem[\protect\citeauthoryear{Whitmore, Murphy  \& Griest}{Whitmore
  et~al.}{2010}]{Whitmore:2010:89}
Whitmore J.~B.,  Murphy M.~T.,   Griest K.,  2010, \mn@doi [ApJ]
  {10.1088/0004-637X/723/1/89}, \href
  {http://adsabs.harvard.edu/abs/2010ApJ...723...89W} {723, 89}

\bibitem[\protect\citeauthoryear{Wolfe, Briggs  \& Jauncey}{Wolfe
  et~al.}{1981}]{Wolfe:1981:460}
Wolfe A.~M.,  Briggs F.~H.,   Jauncey D.~L.,  1981, \mn@doi [ApJ]
  {10.1086/159170}, 248, 460

\bibitem[\protect\citeauthoryear{Wolfe, Gawiser  \& Prochaska}{Wolfe
  et~al.}{2005}]{Wolfe:2005:861}
Wolfe A.~M.,  Gawiser E.,   Prochaska J.~X.,  2005, \mn@doi [Annual Review of
  Astronomy and Astrophysics] {10.1146/annurev.astro.42.053102.133950}, \href
  {http://adsabs.harvard.edu/abs/2005ARA%26A..43..861W} {43, 861}

\bibitem[\protect\citeauthoryear{Wolfe, Prochaska, Jorgenson  \&
  Rafelski}{Wolfe et~al.}{2008}]{Wolfe:2008:881}
Wolfe A.~M.,  Prochaska J.~X.,  Jorgenson R.~A.,   Rafelski M.,  2008, \mn@doi
  [ApJ] {10.1086/588090}, \href
  {http://adsabs.harvard.edu/abs/2008ApJ...681..881W} {681, 881}

\bibitem[\protect\citeauthoryear{Zafar, Popping  \& P{\'e}roux}{Zafar
  et~al.}{2013a}]{Zafar:2013:A140}
Zafar T.,  Popping A.,   P{\'e}roux C.,  2013a, \mn@doi [A\&A]
  {10.1051/0004-6361/201321153}, \href
  {http://adsabs.harvard.edu/abs/2013A%26A...556A.140Z} {556, A140}

\bibitem[\protect\citeauthoryear{Zafar, P{\'e}roux, Popping, Milliard,
  Deharveng  \& Frank}{Zafar et~al.}{2013b}]{Zafar:2013:A141}
Zafar T.,  P{\'e}roux C.,  Popping A.,  Milliard B.,  Deharveng J.-M.,   Frank
  S.,  2013b, \mn@doi [A\&A] {10.1051/0004-6361/201321154}, \href
  {http://adsabs.harvard.edu/abs/2013A%26A...556A.141Z} {556, A141}

\makeatother
\end{thebibliography}

%%%%%%%%%%%%%%%%%%%%%%%%%%%%%%%%%%%%%%%%%%%%%%%%%%

%%%%%%%%%%%%%%%%% APPENDICES %%%%%%%%%%%%%%%%%%%%%

%\appendix

%\section{Appendix name}\label{a:name}

%%%%%%%%%%%%%%%%%%%%%%%%%%%%%%%%%%%%%%%%%%%%%%%%%%

\section*{Supporting Information}\label{sec:supp}

Additional Supporting Information may be found in the online version
of this article and in \citet{Murphy:2018:UVESSQUADDR1}:\vspace{-0.5em}\newline

\noindent \textbf{DR1\_quasars\_master.csv.} Complete version of \Tref{t:cat}, incorporating the DLA information from \Tref{t:dla}.\vspace{-0.1em}\newline
\noindent \textbf{DR1\_DLAs.pdf.} Velocity plots of all 18 new DLAs reported in \Tref{t:dla}, similar to the example in \Fref{f:egDLA}.\vspace{-0.5em}\newline

\noindent Please note: Oxford University Press are not responsible for the
content or functionality of any supporting materials supplied by
the authors. Any queries (other than missing material) should be
directed to the corresponding author for the paper.

% Don't change these lines
\bsp	% typesetting comment
\label{lastpage}
\end{document}